\documentclass[12pt]{article}
\usepackage{amsmath,amssymb,epsfig,amsfonts}
\usepackage{graphicx,subfigure}
\usepackage[usenames, dvipsnames]{color}
\usepackage{multirow}
\usepackage[backref]{hyperref}
\usepackage[nosort]{cite}
\usepackage{lscape}
\usepackage{rotating}
\usepackage[normalem]{ulem} 

\usepackage{extarrows}

\usepackage{verbatim} 

\makeatletter

\newcommand{\be}{\begin{equation}}
\newcommand{\ee}{\end{equation}}
\newcommand{\ba}{\begin{aligned}}
\newcommand{\ea}{\end{aligned}}

\newenvironment{mathbox}
{\par\centering\begin{lrbox}{0}%
\begin{minipage}[c]{0.25\textwidth}}
{\end{minipage}\end{lrbox}%
\framebox[0.2\textwidth]{\usebox{0}}%
\par
\ignorespacesafterend}

\newenvironment{mathboxfat}
{\par\centering\begin{lrbox}{0}%
\begin{minipage}[c]{0.3\textwidth}}
{\end{minipage}\end{lrbox}%
\framebox[0.25\textwidth]{\usebox{0}}%
\par
\ignorespacesafterend}

\newcommand{\C}{\mathbb{C}}
\renewcommand{\P}{\mathbb{P}}

\newcommand{\nn}{\nonumber}
\newcommand{\bea}{\begin{eqnarray}}
\newcommand{\eea}{\end{eqnarray}}

\newcommand{\R}{{\mathbb R}}
\newcommand{\Z}{{\mathbb Z}}

\def\unit{{1\kern-.65ex {\rm l}}}
\def\1{{1\kern-.65ex {\rm l}}}

\newcount\hour \newcount\minute
\hour=\time \divide \hour by 60
\minute=\time
\def\now{%
\ifnum \hour<13
  \ifnum \hour=0 \advance \hour by 12 \number\hour:\else \number\hour:\fi%
     \ifnum \minute<10 0\fi%
     \number\minute%
\ A.M.%
\else \advance \hour by -12 \number\hour:%
  \ifnum \minute<10 0\fi%
  \number\minute%
  \ P.M.%
\fi%
}

\makeatother

\begin{document}

\baselineskip=18pt  
\numberwithin{equation}{section}  
\allowdisplaybreaks  


%
%


\thispagestyle{empty}

\vspace*{-2cm} 
\begin{flushright}
{\tt KCL-MTH-14-14} 
\end{flushright}

\vspace*{0.8cm} 
\begin{center}
{\Huge Box Graphs and Resolutions I}\\

 \vspace*{1.5cm}
{Andreas P. Braun and Sakura Sch\"afer-Nameki}\\

 \vspace*{1.0cm} 
 {\it Department of Mathematics, King's College, London \\
 The Strand, London WC2R 2LS, England }\\
 {\tt {gmail:$\,$ andreas.braun.physics, sakura.schafer.nameki}}\\

\vspace*{0.8cm}
\end{center}
\vspace*{.1cm}
\noindent 
Box graphs succinctly and comprehensively characterize singular fibers of elliptic fibrations in codimension two and three, as well as flop transitions connecting these, in terms of representation theoretic data.  
We develop a framework that provides a systematic map between a box graph and a crepant algebraic resolution of the singular elliptic fibration, thus allowing an explicit construction of the fibers from a singular Weierstrass or Tate model. The key tool is what we call a fiber face diagram, which shows the
relevant information of a (partial) toric triangulation and allows the inclusion of more general algebraic blowups. We shown that each such diagram defines a sequence of weighted algebraic blowups, thus providing a realization of the fiber defined by the box graph in terms of an explicit resolution. We show this correspondence explicitly for the case of $SU(5)$ by providing a map between box graphs and fiber faces, and thereby a sequence of algebraic resolutions of the Tate model, which realizes each of the box graphs. 

\newpage

\tableofcontents

\newpage

\section{Introduction}

Elliptic fibrations have a rich mathematical structure, which dating back to Kodaira and N\'eron's work \cite{Kodaira, Neron} on the classification of singular fibers has been in close connection with the theory of Lie algebras. 
Recently, this connection was been extended with a representation-theoretic characterization of singular fibers in higher codimension, in particular for three- and four-folds \cite{Hayashi:2014kca}. The inspiration for this work came from string theory and supersymmetric gauge theory, in particular the Coulomb branch phases of three-dimensional $N=2$ gauge theories. 
However the final result can be entirely presented in terms of geometry and representations of Lie algebras overlayed with a combinatorial structure, the so-called box graphs. The purpose of this paper is to complement this description of singular elliptic fibers with a direct resolution of singularity approach, and to develop a systematic way to construct the resolutions based on their description in terms of box graphs. 

Consider a singular elliptic fibration with two or three-dimensional base $B$. In codimension one, the singular fibers fall into the Kodaira-N\'eron classification, and for ADE type Lie algebras, the singular fibers are a collection of $\mathbb{P}^1$s intersecting in an affine ADE Dynkin diagram. The main interest in the present work and the motivation for the works \cite{Hayashi:2013lra, Hayashi:2014kca} is the extension of this to higher codimension fibers. Consider a singular Weierstrass (or Tate) model
\be
y^2 = x^3 + fx + g \,,
\ee
which describes the elliptic fibration. As is well known, the main advantage of this is that we do not need to specify the base, except for requiring that the sections $f, g$ (or the corresponding sections of the Tate model) exist.  
Then the discriminant of this equation characterizes the loci in the case where the fiber becomes singular. Let $z=0$ be the local description of a component of the discriminant $\Delta = 4 f^3 + 27 g^2$. I.e. $\Delta$ has an expansion $\Delta= \delta_0 z^{n_0} + \delta_1 z^{n_1} + \cdots$.  The vanishing order in $z$ of $(f, g, \Delta)$ determines the Kodaira type of the singular fiber above the codimension one locus $z=0$. Along special codimension two loci, $z=\delta_0=0$, the vanishing order of the discriminant increases, and thereby the singularity type enhances.

The box graphs provide answers to the following questions: for a fixed codimension one Kodaira singular fiber, 
what are the possible fiber types that can arise in codimension two and three. 
Secondly, how many distinct such fibers in codimension two and three are there, and how are these related through flop transitions. 
The Kodaira classification can be thought of as associating a Lie algebra $\mathfrak{g}$ (or affine Dynkin diagram) to the codimension one fibers. In the same spirit, the box graph supplements this with codimension two information, which is encoded in the representation-theoretic data associated to $\mathfrak{g}$. More precisely, the box graphs are sign (or color) decorated representation graphs. 
They give a succinct and elegant answer to these questions by characterizing the possible higher codimension fibers in terms of representation theoretic data alone. The box graphs determine the extremal generators of the cone of effective curves in codimension two and three, and flop transitions are implemented in terms of simple operations on the graph.

Box graphs are applicable to all Kodaira fibers in codimension one \cite{Hayashi:2014kca, CLSSN} and provide a framework to classify the fibers in higher codimension. One of the most studied examples is the case of $\mathfrak{su}(5)$, largely due to its relevance in F-theory compactifications, but also because it is one of the simplest examples which containss various interesting features of codimension two and three fibers. In this case the flop diagram was determined \cite{Hayashi:2013lra} in the map to the Coulomb branch of the three-dimensional $N=2$ gauge theory that describes low energy effective theory of the M-theory compactification on the resolved elliptic fibration \cite{Intriligator:1997pq, Aharony:1997bx, deBoer:1997kr, Grimm:2010ks, Grimm:2011fx} and confirmed from the box graphs in \cite{Hayashi:2014kca}. 

This simple description in terms of box graphs is in stark contrast to the process of explicitly constructing crepant resolutions of singular fibers for elliptic three- and four-folds. The starting point for this process is the singular Weierstrass or Tate model and the resolutions are either based on toric \cite{Bershadsky:1996nh, Krause:2011xj} or algebraic blowups \cite{MS, EY, Lawrie:2012gg}. One of the most tantalizing issues in 
explicit resolutions of the singular geometry is that flops are entirely obscured, or at best only known in a subclass of resolutions. In \cite{Hayashi:2013lra} the case of $\mathfrak{su}(5)$ was understood and all phases and resolutions were obtained either directly by algebraic or toric blowups, or they were shown to arise from these by flop transitions.

The concise and representation-theoretic description of singular fibers in terms of box graphs is highly suggestive of the existence of a more unified, elegant approach to resolutions of singular elliptic fibrations. The goal of this paper and of the followup \cite{ABSSN} is to develop resolution methods for singular elliptic fibrations which provide an explicit map between a given box graph and an associated resolution of the singular fibration.  

The framework that we propose is a hybrid between toric resolutions\footnote{I.e. resolutions obtained by simply refining the
fan of the toric ambient space our Tate model is embedded in.} and algebraic blowups: we use  partial toric triangulations, represented in terms of {\it fiber face diagrams}, which in turn determine a resolution sequence of weighted projective blowups. The various subcases that fall into this framework are:
\begin{itemize}
\item Standard toric triangulations, which have a description in terms of weighted blowups as is known from e.g. \cite{Reid:young1987}\footnote{There are exceptions to this, which however do not feature in the case of $SU(5)$ studied in this paper, but will be important in \cite{ABSSN}.}
\item Standard algebraic resolutions, which correspond to the specialization to unit weights 
\item Algebraic resolutions leading to a realization of the fiber as a complete intersection, which appeared already in the resolution studied in \cite{EY, MS, Hayashi:2013lra}.
\item Determinantal blowups.
\end{itemize}
Our proposal is to use the top \cite{Candelas:1996su,Perevalov:1997vw} corresponding to a degenerate fiber as an organizing tool for weighted blowups, which realize the different box graphs, or equivalently Coulomb phases.  This realization by direct blowups guarantees in particular projectivity of the resolved space. 
Each phase or box graph can be mapped  to a resolution by computing the splitting of fiber components over codimension two loci in the base. Even though it is not possible to obtain all box graphs by a triangulation of the top, we can 
use partial triangulations to map out the entire network of the corresponding resolutions. Such partial triangulations correspond to only partial resolutions, after which
singular loci are still present. We may then continue the resolution process in ways which can not be obtained through straightforward 
triangulation of the top, e.g. turning the Tate form into a complete intersection. Keeping this in mind, we hence display the partial triangulation
of the top which is relevant to obtain each phase. The main advantage to this way of or organizing the resolutions is that it is systematic and is amenable to generalization \cite{ABSSN}. 

In all but one\footnote{In fact, there is another resolution, which corresponds to inverting the ordering of the simple roots, and thereby  the fiber components, so this really corresponds to two resolutions.} of the box graphs/phases for $\mathfrak{su}(5)$, the associated resolutions of the  Tate form are given as a hypersurface\footnote{As is common in the literature on elliptic fibrations, we hereby mean that the fiber is embedded as a hypersurface into a projective space, not necessary the full fibration, as the base remains unspecified. } or complete intersection of codimension two. In the remaining case, we need to blow up along a divisor realized as a determinantal variety. This turns the Tate model into a non-complete intersection. 

In summary, we propose the following correspondence between box graphs and algebraic resolutions of singular elliptic fibers, via fiber face diagrams: 
\be
\begin{array}{ccccc}
\begin{mathbox}
\begin{center}
${\, }$\\
Box Graphs\\
${\, }$\\
\includegraphics[width=1.5cm]{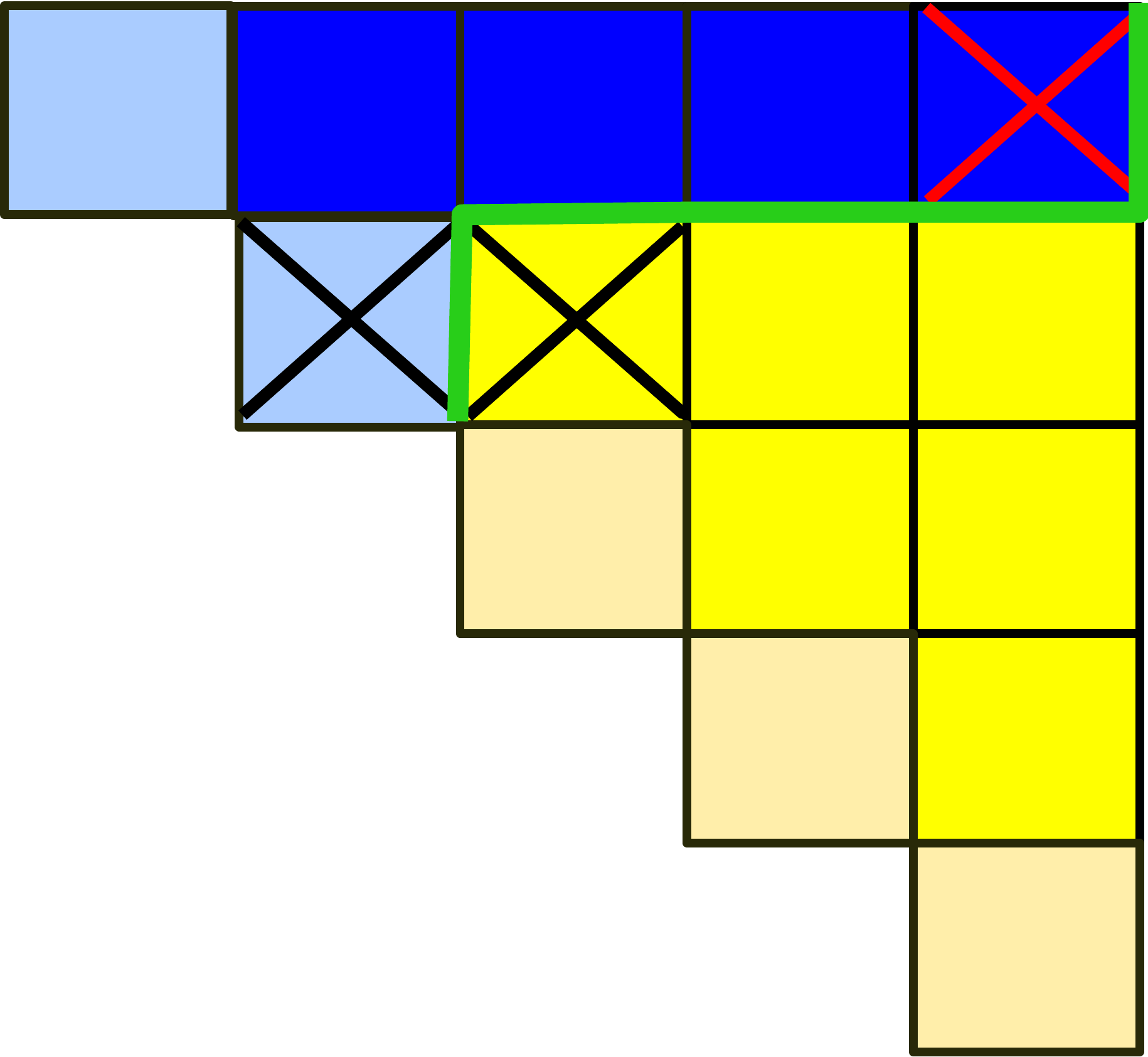}
${\, }$\\
${\, }$\\
\end{center}
\end{mathbox} &\rightarrow&
 \begin{mathbox}
\begin{center}
${\, }$\\
Fiber Faces\\
${\, }$\\
\includegraphics[width=1.5cm]{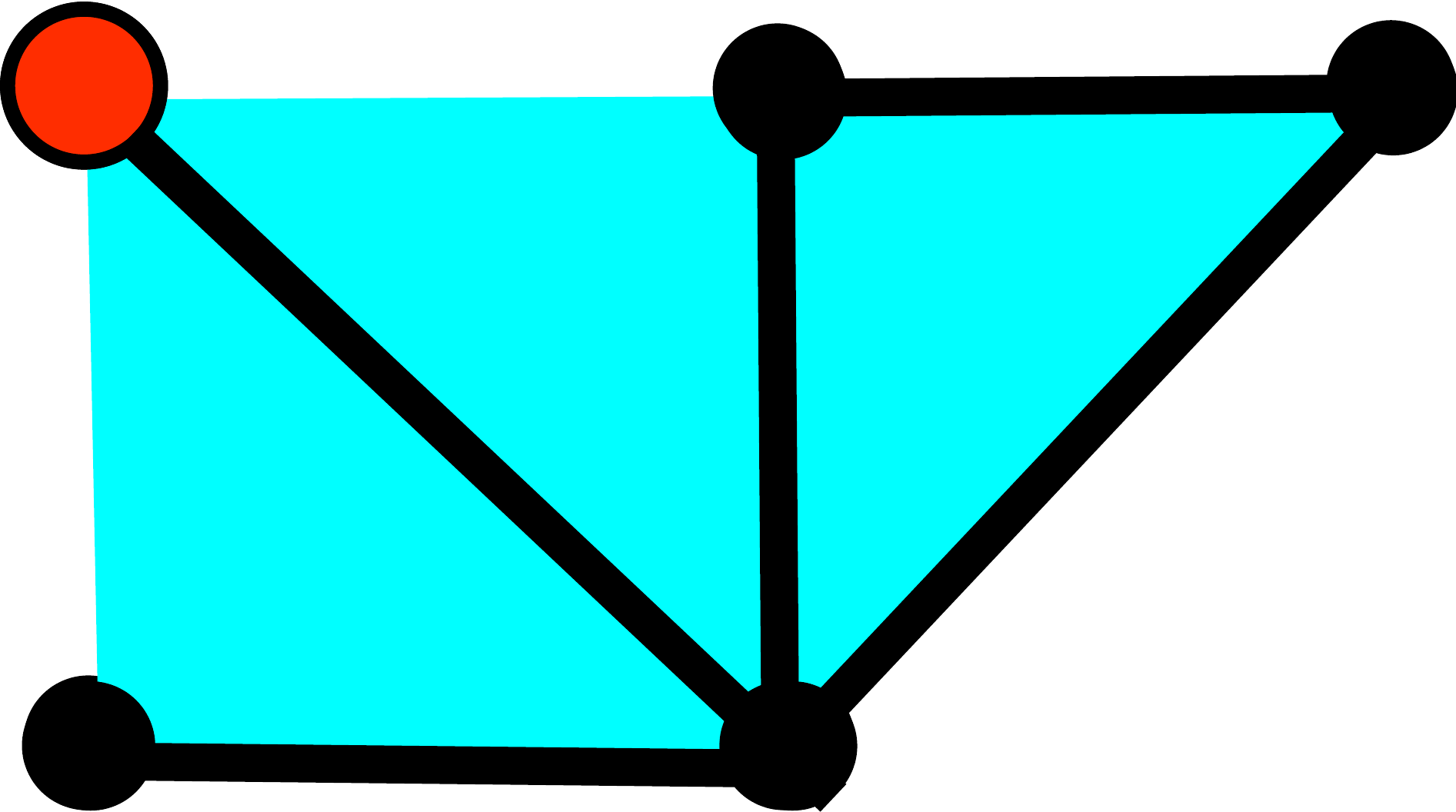}
${\, }$\\
${\, }$\\
 \end{center}
\end{mathbox}
&\rightarrow&
\begin{mathboxfat}
\begin{center}
Algebraic Resolution:\\
Weighted blowups 
\end{center}
\end{mathboxfat}
\end{array}
\ee
The box graphs determine  the codimension two fibers, or equivalently Coulomb branch phases. From the splitting of the fibers in codimension two we determine an associated fiber face diagram, which is based on the top of the fiber in codimension one. This in turn determines a sequence of algebraic resolutions of the Tate form. 
In the present paper we develop this direct correspondence for $\mathfrak{su}(5)$, with codimension two fibers associated to the representations ${\bf 5}$ and ${\bf 10}$, construct the fiber face diagrams, and associated associated weighted blowups. Through direct comparison of the fibers in codimension two with the box graph we establish the correspondence. Finally, it is possible to also map all the flops into flops of the resolved geometries, and both networks are in agreement. 


The plan of this paper is as follows. Section \ref{sec:Box} is a lightning review of box graphs, with a focus on  the $\mathfrak{su}(5)$ case.  
In section \ref{sect:toricbu} we discuss crepant weighted blowups and how to systematically determine these for a given singularity. 
In section \ref{sec:SU5Blowups} we discuss the precise correspondence between triangulations, fiber faces and weighted blowups for $SU(5)$. Finally in section \ref{sec:DetBlowup} we discuss the determinantal blowups. The main result is table \ref{tab:THETABLE}. Here, the correspondence is succinctly summarized for all cases, as well as the networks of flop transitions in box graph and fiber face presentation as given in figures \ref{fig:SU5AF} and \ref{fig:SU5AFRes}.

{\bf Note added:}\\
 As we were completing this paper, another work \cite{Esole:2014hya} appeared which claims to also construct all the $\mathfrak{su}(5)$ resolutions, based on the earlier work on $\mathfrak{su}(n), n=2,3,4$ \cite{Esole:2014bka}. In {\tt v2} of \cite{Esole:2014hya} it is erroneously claimed 
 that the resolutions in the present
paper are restricted to ``the special case of singular Calabi-Yau
hypersurfaces in compact toric varieties".  The crepant resolutions we
construct can be applied to any singular elliptic fibration for which
the fiber is embedded in $\mathbb{P}^{123}$.

\section{Box Graphs and Singular Fibers}
\label{sec:Box}

\subsection{Box graph primer}
\label{sec:BoxPrim}
The main result of \cite{Hayashi:2014kca} is the chacracterization of singular fibers in higher codimension of an elliptic fibration in terms of representation theoretic objects, the box graphs. The goal of this paper is to develop a precise map between explicit resolutions of singular fibrations and the data describing singular fibers in higher codimension that is encoded in the box graphs.
We will start with a  brief primer on how to use box graphs to determine the codimension two and three fibers. 

Consider simple Lie algebras $\mathfrak{g}\subset \widetilde{\mathfrak{g}}$, 
 and let ${\bf R}$ be a representation of $\mathfrak{g}$, with weights $\lambda_i$, $i= 1, \cdots, d=\hbox{dim} ({\bf R})$, such that the adjoint of $\widetilde{\mathfrak{g}}$ decomposes as\footnote{More generally, the commutant can be non-abelian, however this case will not be relevant in the  present paper, so we refer the reader to \cite{Hayashi:2014kca, CLSSN} for details on the more general case, in particular for the definition of box graphs beyond $\mathfrak{su}(n)$.  }
\be\ba\label{Gtilde}
\widetilde{\mathfrak{g}} &\quad\rightarrow\quad \mathfrak{g} \oplus \mathfrak{u}(1)\cr
\hbox{adj}(\widetilde{\mathfrak{g}}) &\quad \rightarrow \quad  \hbox{adj}(\mathfrak{g}) \oplus \hbox{adj}(\mathfrak{u}(1)) \oplus {\bf R}_+ \oplus {\bf R}_- \,,
\ea
\ee
For the present paper, the case of interest is $\mathfrak{g}=\mathfrak{su}(n)$, and ${\bf R}= {\bf n}$ or ${\Lambda^2 {\bf n}}$, in which case $\widetilde{\mathfrak{g}} = \mathfrak{su}(6)$ and $\mathfrak{so}(10)$, respectively. The representation graphs, including the action of the simple roots, are shown in figure \ref{fig:SU5RepGraphs}.  In this case the weights will be denoted by $L_i$, ${i=1, \cdots, n}$, and $L_i + L_j$, ${i<j}$, respectively, with the tracelessness condition
\be
\sum_{i=1}^n L_i =0 \,.
\ee
A box graph  for the pair ($\mathfrak{su}(n), {\bf n})$   is  a sign (color) decorated representation graph of ${\bf n}$, i.e. a map
\be
(\lambda_1, \cdots, \lambda_d)  \quad \rightarrow \quad (\epsilon_1 , \cdots, \epsilon_d), \qquad \epsilon_i =\pm \,, 
\ee
which  satisfies the following two conditions:
\begin{itemize}
\item Flow rules:  \\
If $\epsilon_i= +$, then $\epsilon_{j}=+$ for all $j<i$. Likewise, $\epsilon_i=-$, then $\epsilon_j =-$ for all $j>i$:
\begin{center}
\includegraphics*[width =1.5cm]{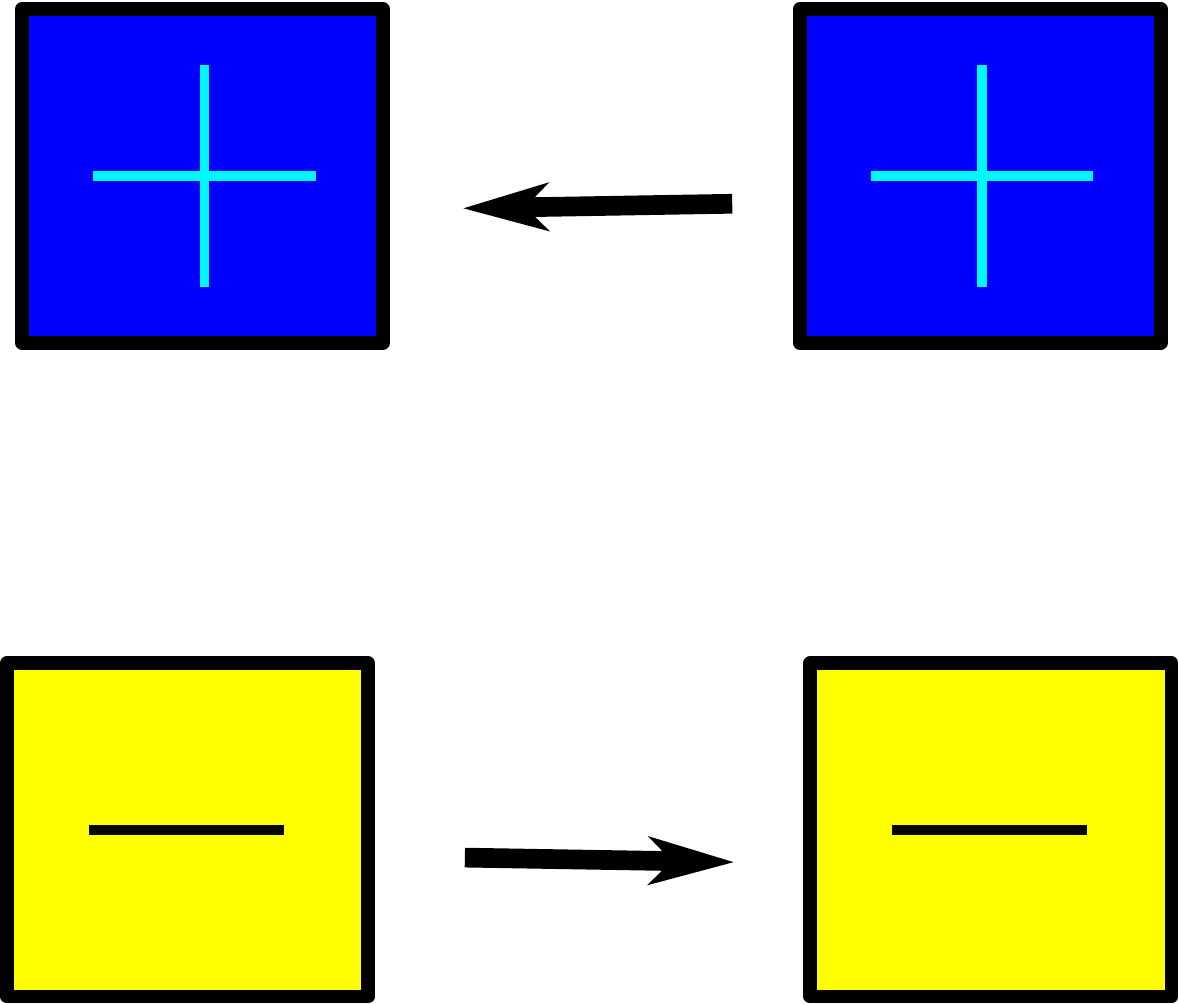}  
\end{center}
\item Diagonal condition: \\
The signs $\epsilon_i$ cannot all be the same. This follows from the fact that $\sum L_i =0 $ for $\mathfrak{su}(n)$, and thus the trace should not have a definite sign. 
\end{itemize}
A box graph for $(\mathfrak{su}(n), \Lambda^2 {\bf n})$  is again a sign-decoration or coloring of the representation graph of $\Lambda^2{\bf n}$, with weights $L_{i,j}= L_i + L_j$, $i<j$
\be
(L_{i,j}) \quad \rightarrow\quad (\epsilon_{i,j}) =\pm  \,,\qquad i<j \,,
\ee
again satisfying the constraints:
\begin{itemize}
\item Flow rules:  \\
If $\epsilon_{i, j}= +$, then $\epsilon_{k, l}=+$ for all $k<i$ and $l<j$, i.e. `` + signs flow up and to the left".
Likewise if $\epsilon_{i,j}=-$, then $\epsilon_{k, l}=-$ for all $k>i$ and $l>j$, i.e. ``- signs flow down and to the right".  
\begin{center}
\includegraphics*[width =4cm]{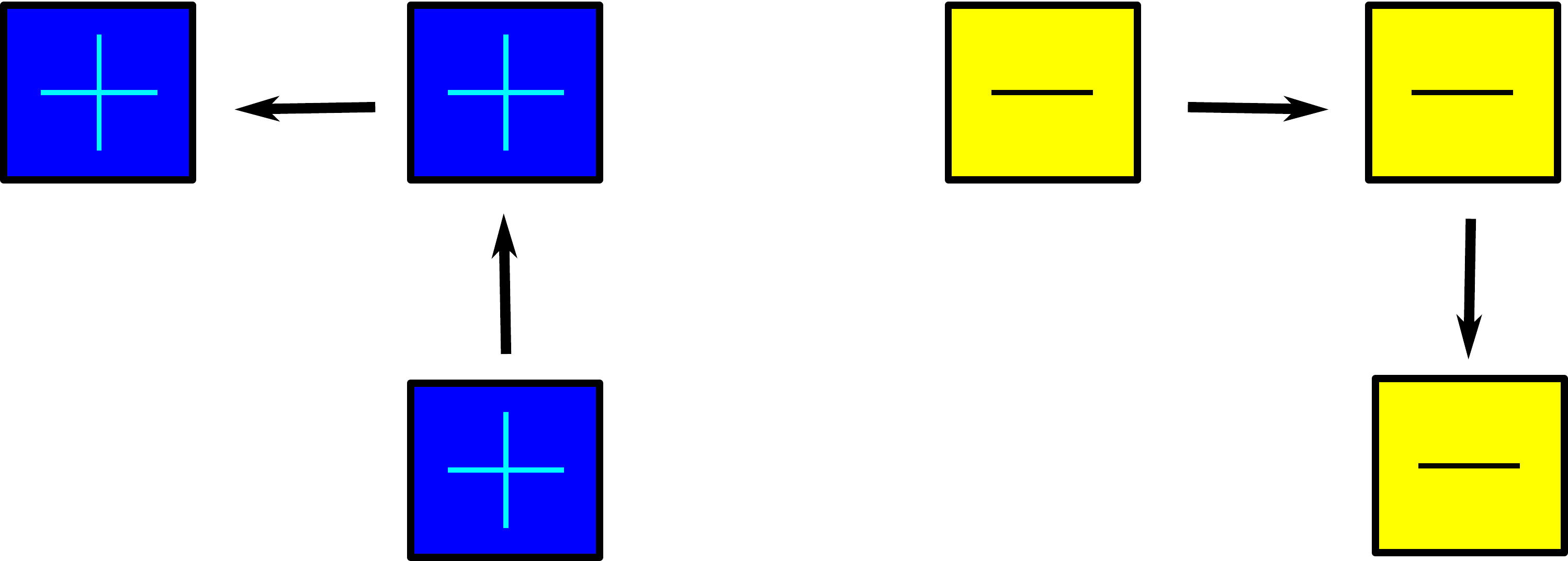}
\end{center}

\item Diagonal condition: \\
The signs along the diagonals (defined below) cannot all be the same. And example is shown in figure \ref{fig:SU5RepGraphs}. This is again related to the trace, and differentiates  between $\mathfrak{su}(n)$ and $\mathfrak{u}(n)$ box graphs:
\be\ba\label{Diagos}
\hbox{For $n=2k$:} &\qquad (\epsilon_{1, 2k}, \epsilon_{2, 2k-1}, \cdots, \epsilon_{k, k+1} ) \not= (+ , \cdots, +), (-, \cdots, -)\cr
\hbox{For $n=2k+1$:}&\qquad (\epsilon_{1, 2k+1}, \epsilon_{2, 2k}, \cdots, \epsilon_{k-1, k+3}, \epsilon_{k, k+1}, \epsilon_{k+1, k+2})
\ea\ee
\end{itemize}
The box graphs can equivalently be described in terms of the convex path, that separates the $+$ and $-$ sign boxes. For $\mathfrak{su}(n)$, this path has to cross the diagonals (\ref{Diagos}), and therefore is called an {\it anti-Dyck path}. 

Each box graph corresponds to a small resolution of an elliptic fibration with codimension one singular fiber specified by the Lie algebra $\mathfrak{g}$ via Kodaira's classification. 
Here, we will summarize the rules for how to determine the splitting of the codimension one fiber into the codimension two fiber, as well as the intersections of the fiber components. Let us denote the curves associated to the simple roots $\alpha_i$ and weights $\lambda$ by
\be
F_i \quad \leftrightarrow \quad \alpha_i = L_i- L_{i+1} \,,\qquad 
C_{\lambda_i}^\pm \quad \leftrightarrow \quad \lambda_i \quad \hbox{with} \quad \epsilon_i = \pm \,.
\ee
The initial fiber is given by $I_n$, where the intersection matrix between $F_i$ and the curve associated to the zero section $F_0 = -\sum F_i$, is given by the affine $\mathfrak{su}(n)$ Cartan matrix. 
Given a box graph, we can read off which curves $F_i$ split along the codimension two loci, and secondly, what the intersections of the irreducible fiber component are:

\begin{figure}
\centering
\includegraphics[width=10cm]{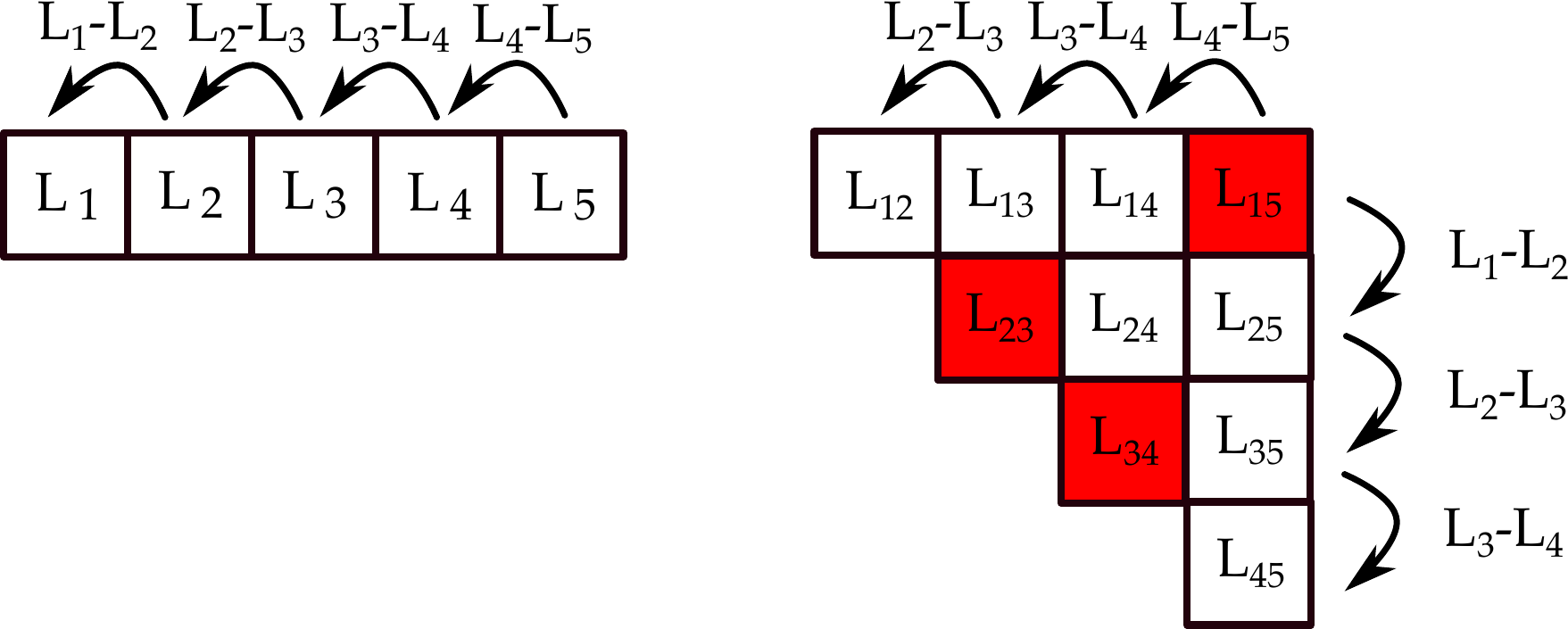}
\caption{Representation graphs for ${\bf 5}$ and ${\bf 10}$ of $\mathfrak{su}(5)$, with the action of the simple roots $L_i- L_{i+1}$ on the diagram as shown along the edges. In the representation graph for ${\bf 10}$, the red boxes correspond to the `diagonal' (\ref{Diagos}), i.e. the signs of these three boxes cannot be the same in an $\mathfrak{su}(5)$ box graph. }
\label{fig:SU5RepGraphs}. 
\end{figure}


\begin{itemize}
\item Fiber splitting rules:\\
If adding the simple root $\alpha_i$ crosses from a $+$ to $-$ box (i.e. it crosses the anti-Dyck path) then the associated curve $F_i$ splits. If not, then $F_i$ remains irreducible. 
\item Extremal generators: \\ 
The extremal generators of the cone of effective curves above the codimension two locus that the box graph describes are the irreducible $F_i$, as well as the {\it extremal curves}, which are defined as follows: a curve $C_{\lambda}$ is extremal, 
if changing the sign of the box associated to $\lambda$ maps the graph to another decorated representation graph, that satisfies the flow rules. These extremal curves, which always lie along the anti-Dyck path, will be marked by an $X$ in the box graph. An extremal curve cannot necessarily be flopped (sign changed), as this might violate the diagonal condition. If it can be flopped, it will be marked by a black $X$, otherwise by a red $X$. 
\item Intersections:\\
 The extremal curves $C_\lambda^\pm$ intersect the irreducible $F_i$ by $\pm1$ if adding the corresponding root to $\lambda$ retains/changes the sign.
 We define the intersection with the (representation-theoretically prefered) sign convention, where $D_i$ is the divisor dual to the curve $F_i$
 \be
 F_i\cdot F_i := - \# D_i\cap F_i = 2   \,.
 \ee

\end{itemize}


\subsection{Box graphs and singular fibers for $\mathfrak{su}(5)$}

For the fundamental representation ${\bf 5}$, the box graphs are based on the representation graph shown in figure \ref{fig:SU5RepGraphs}.
Here, $L_i$ are the weights, and $L_i-L_{i+1}$ the simple roots, which act between the weights. 
Similarly, for ${\bf 10}$ the representation graph can be written in terms of the weights $L_{i, j} = L_{i}+ L_j$, with $i<j$, $i, j= 1, \cdots, 5$, and  the simple roots act as indicated in figure \ref{fig:SU5RepGraphs}.

\begin{figure}
\centering
\includegraphics[width=4cm]{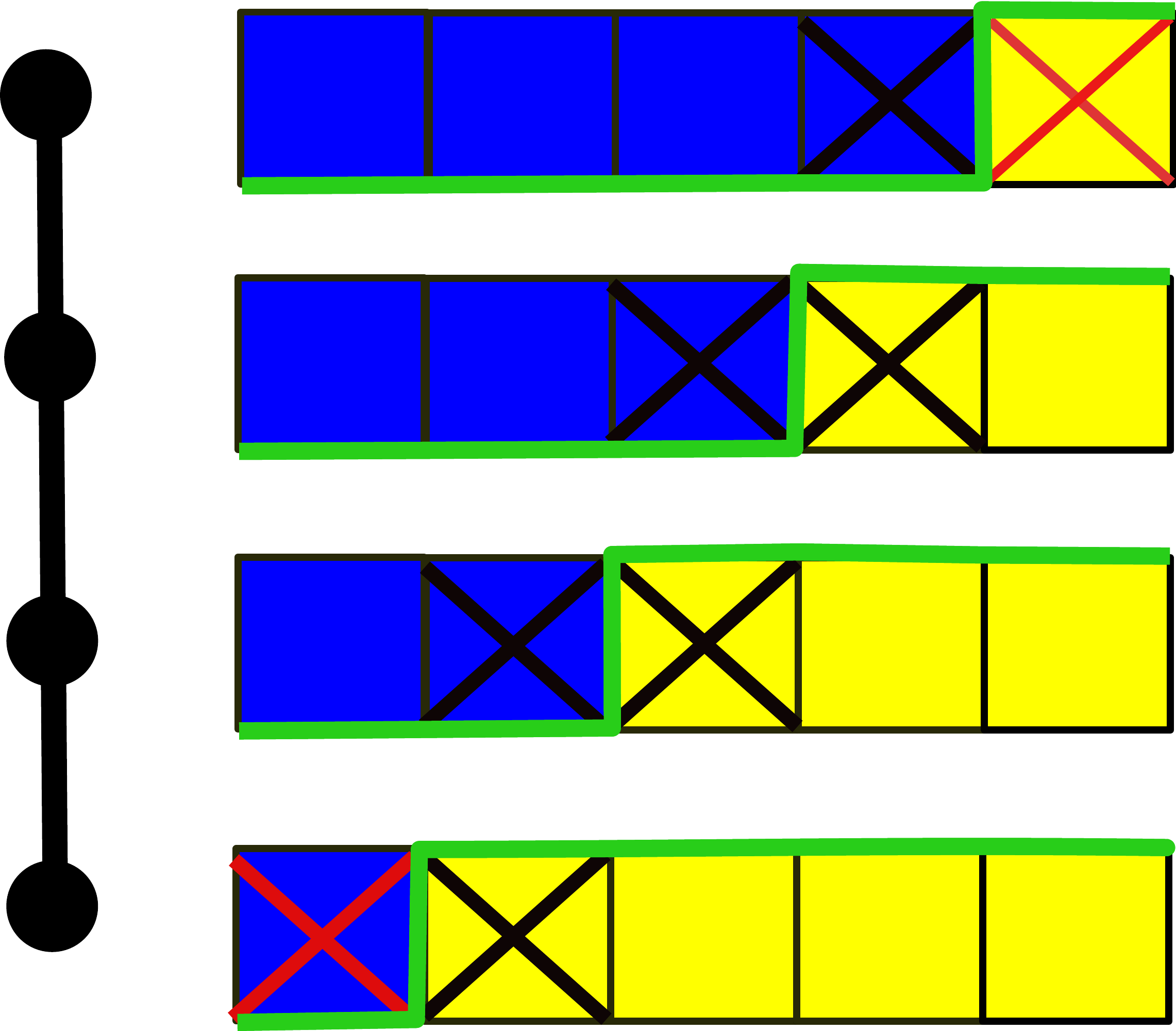}
\caption{Box graphs for $\mathfrak{su}(5)$ with ${\bf 5}$ representation, on the left the corresponding flop diagram is shown. The extremal generators, which in the geometry correspond to the curves that can be flopped, are marked with a black $X$, whereas red $X$'s indicate cone generators which cannot be flopped as they would yield $\mathfrak{u}(5)$ phases. The green line marks the anti-Dyck path.}
\label{fig:SU5F}
\end{figure}

\begin{figure}
\centering
\includegraphics[width=7cm]{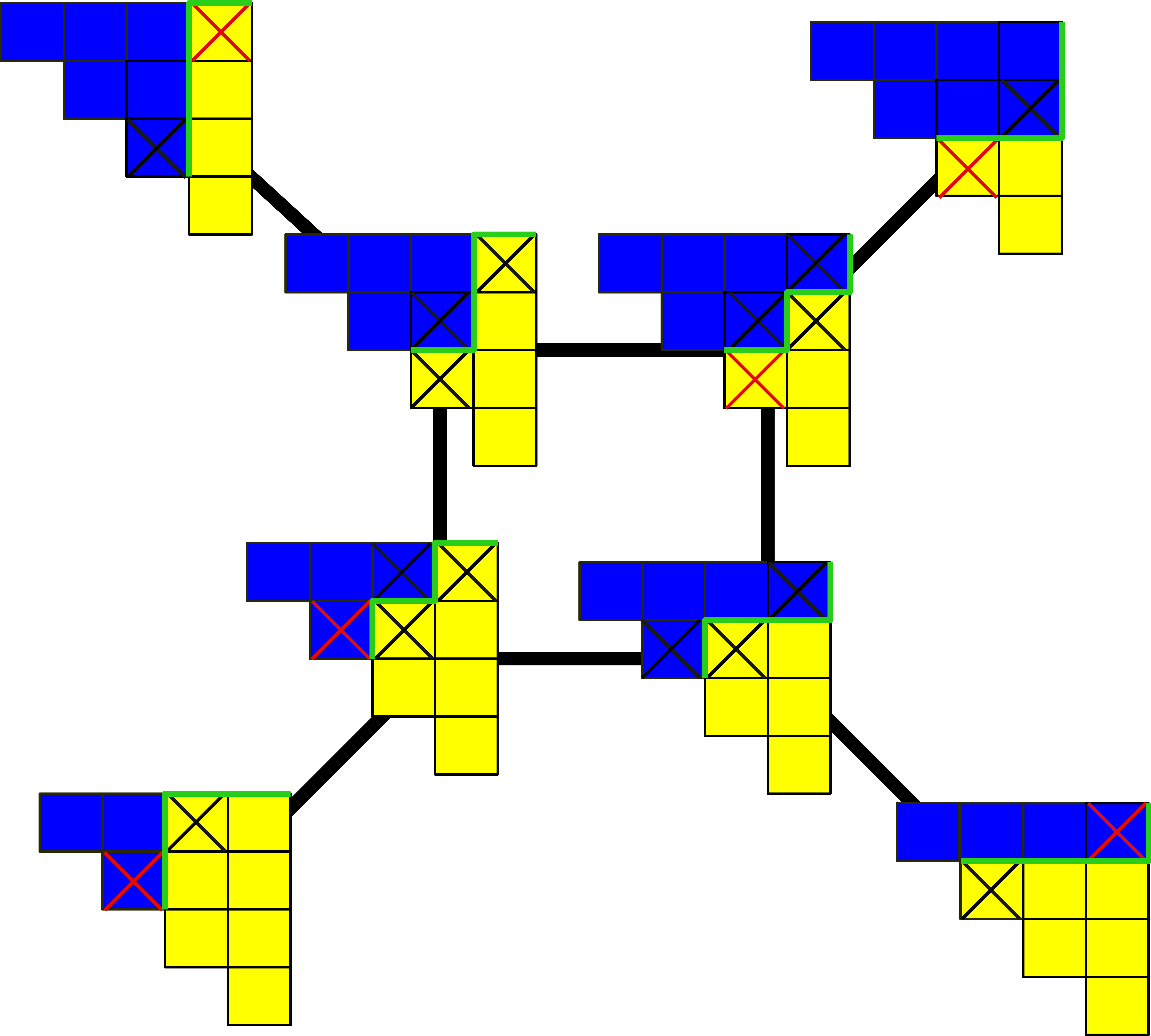}
\caption{Box graphs for $\mathfrak{su}(5)$ with ${\bf 10}$ representation. Two box graphs that are connected by a black line can be flopped into each other. The extremal generators, which in the geometry correspond to the curves that can be flopped are marked with a black $X$, whereas red $X$'s indicate cone generators which cannot be flopped as they would yield $\mathfrak{u}(5)$ phases. }
\label{fig:SU5A}
\end{figure}

The box graphs for $\mathfrak{su}(5)$ with fundamental ${\bf 5}$ and/or anti-symmetric ${\bf 10}$ representation, which characterize the fibers in codimension two and three of the elliptic fibration with $I_5$ Kodaira fiber in codimension one, were determined in \cite{Hayashi:2014kca}. 
The box graphs for each of these situations are shown in figures \ref{fig:SU5F}, \ref{fig:SU5A} and \ref{fig:SU5AF}, respectively. 
The main result in \cite{Hayashi:2014kca, CLSSN} is that the box graphs determine the complete set of small resolutions, which is characterized by the fibers in codimension two and three. 


\subsubsection{Singular fibers for ${\bf 5}$ Representation}

The possible box graphs are shown in figure \ref{fig:SU5F}. We denote the curves with a $\pm$ sign associated to the weight $L_i$ by $C_i^\pm$. 
It is clear that these are all possibilities that satisfy the flow rule and diagonal condition. In each diagram there is exactly one simple root that splits by the rules specified in section \ref{sec:BoxPrim}. 
For instance, in the first box graph, the blue (+) and yellow (-) separation is between $L_4$ and $L_5$, i.e. adding $\alpha_4= L_4- L_5$ changes the sign, and thus $F_4$ splits into $C_4^++ C_5^-$.
The resulting splittings, extremal generators of the cone of effective curves, and the new intersections are as follows, and give rise to $I_6$ fibers in all cases, as shown in (\ref{5Split}).
\be\label{5Split}
\begin{array}{|c|c|c|c|c|}\hline
\# & \hbox{Box Graph} & \hbox{Splitting} & \hbox{Generators}& \hbox{Intersections}\cr\hline
&&&&\cr
\hbox{I}&\includegraphics[width=2cm]{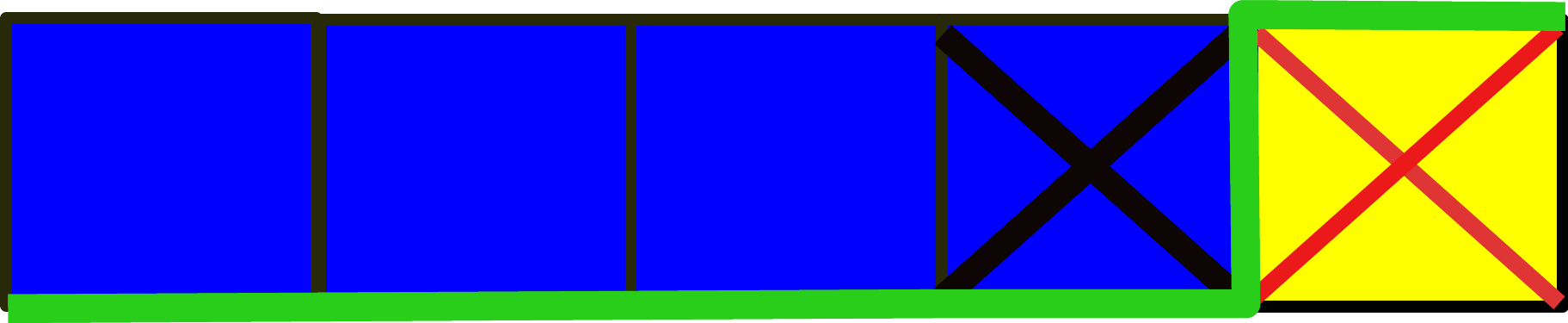} 
&  F_4 \rightarrow C_{4}^+ +  C_5^- 
& \{F_1, F_2, F_3, C_{4}^+ ,  C_5^-  \}
&  C_4^+ \cdot F_4= C_4^+ \cdot C_5^- =C_5^- \cdot F_0= -1 
\cr
&&&&\cr
\hbox{II}&\includegraphics[width=2cm]{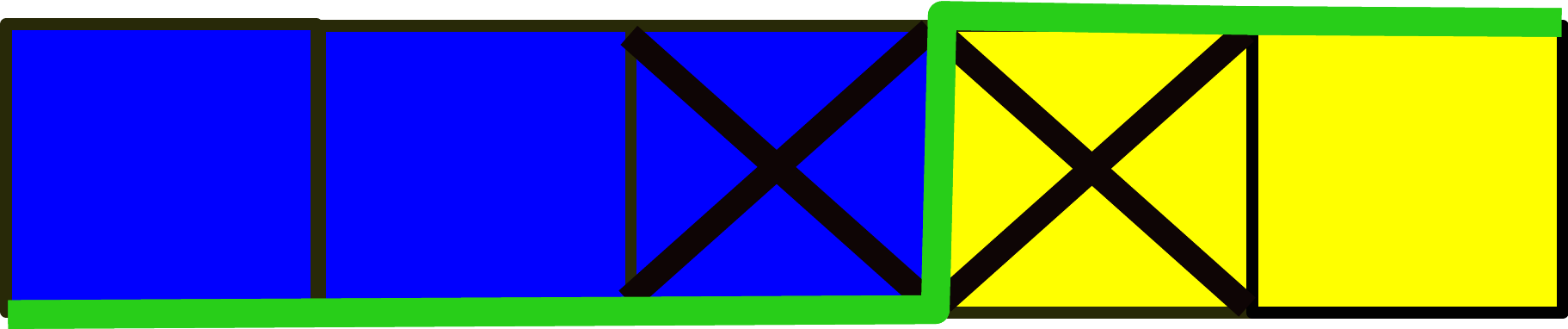} 
&F_3 \rightarrow C_{3}^+ +  C_4^-  
& \{F_1, F_2,  C_{3}^+ ,  C_4^-, F_4  \}
& C_3^+ \cdot F_3= C_3^+ \cdot C_4^- =C_4^- \cdot F_4= -1 
\cr
&&&&\cr
\hbox{III} & \includegraphics[width=2cm]{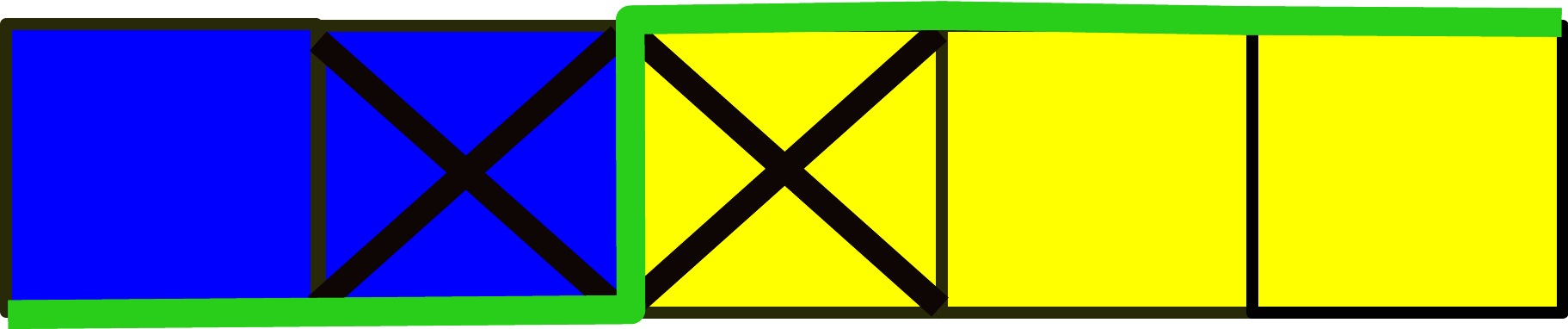} 
&F_2 \rightarrow C_{2}^+ +  C_3^-  
&  \{F_1,   C_{2}^+ ,  C_3^-,  F_3, F_4  \}
& C_2^+ \cdot F_3= C_2^+ \cdot C_3^- =C_3^- \cdot F_3= -1 
\cr
&&&&\cr
\hbox{IV}& \includegraphics[width=2cm]{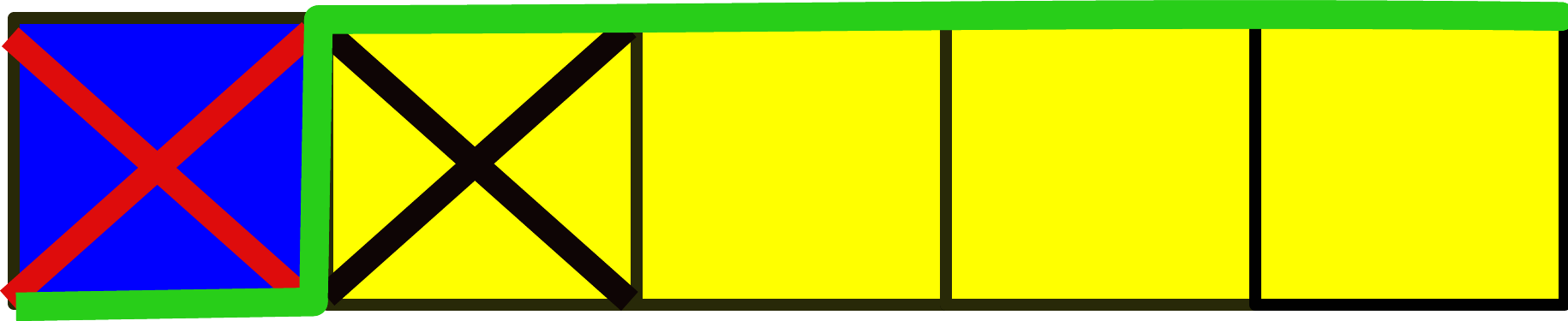} 
&F_1 \rightarrow C_{1}^+ +  C_2^-  
& \{  C_{1}^+ ,  C_2^-,  F_2, F_3, F_4  \}
&C_1^+ \cdot F_0= C_1^+ \cdot C_2^- =C_2^- \cdot F_2= -1 
 \cr
&&&&\cr\hline
\end{array}
\ee
Note that for each resolution, there is one simple roots that split into two weights $C_{i}^+$ and $C_{i+1}^-$, which are marked with $X$ in the box graphs. These intersect each other transversally, and with the remaining irreducible roots to form an $I_6$ Kodaira fiber in codimension two. 


\subsubsection{Singular fibers for ${\bf 10}$ Representation}\label{sect:singfib10}

The fibers of the  ${\bf 10}$ representation are obtained similarly from the box graphs in figure \ref{fig:SU5A}.
There is a $\mathbb{Z}_2$ symmetry that corresponds to reversal of the ordering of simple roots of $\mathfrak{su}(5)$, so that we only need to discuss half of the box graphs. The resulting fibers are all $I_1^*$, consistent with the local enhancement to $\mathfrak{so}(10)$, and the splittings produce the correct multiplicities: 
\be
\label{10Split}
\begin{array}{|c|c|c|c|c|}\hline
\# &\hbox{Box Graph} & \hbox{Splitting} & \hbox{Generators}& \hbox{Intersections}\cr\hline
&&&&\cr
4&  \multirow{2}{*}{\includegraphics[width=2cm]{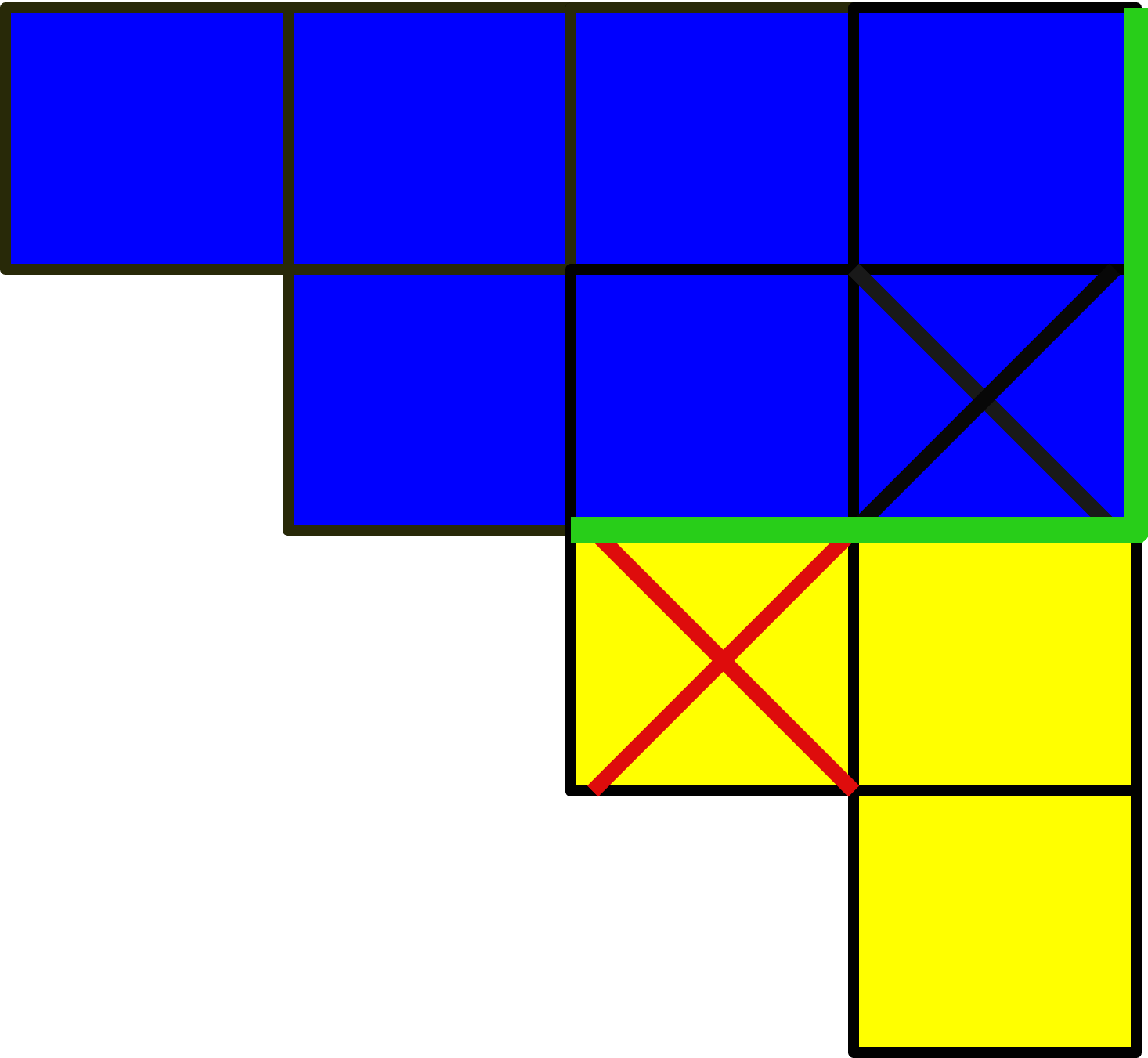}}
&  F_2 \rightarrow C_{2,5}^+ +  C_{3,4}^- + F_4 
& \{F_1,  F_3, F_4  ,  C_{2,5}^+ ,  C_{3,4}^-  \}
&\multirow{2}{*}{\includegraphics[width=3cm]{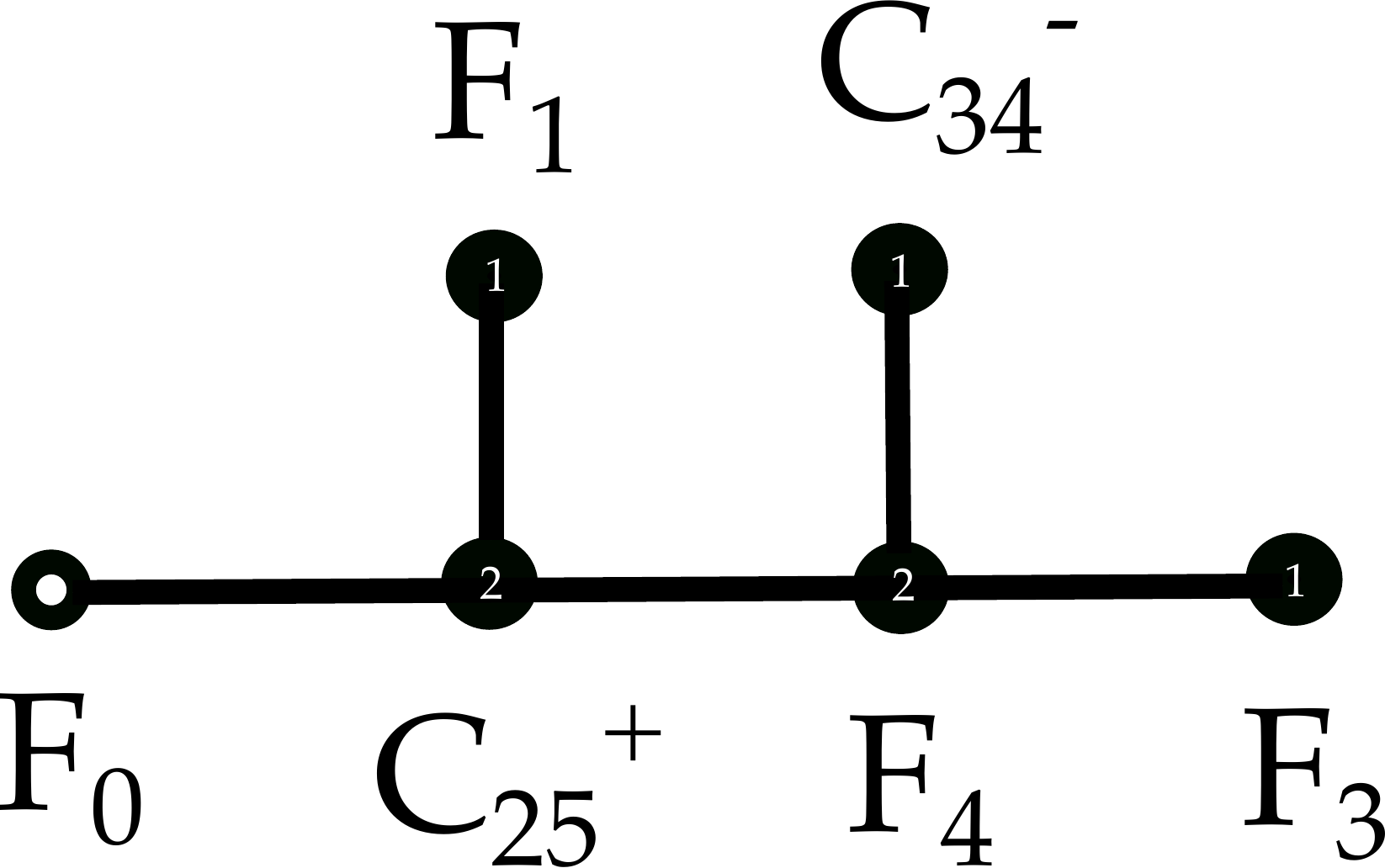}}
\cr
&& F_0 \rightarrow C_{2,5}^+ + \tilde{F}_0&
&\cr
&&\tilde{F}_0= C_{12}^+
&&\cr
&&&&
\cr\hline
&&&&\cr
7&  \multirow{2}{*}{\includegraphics[width=2cm]{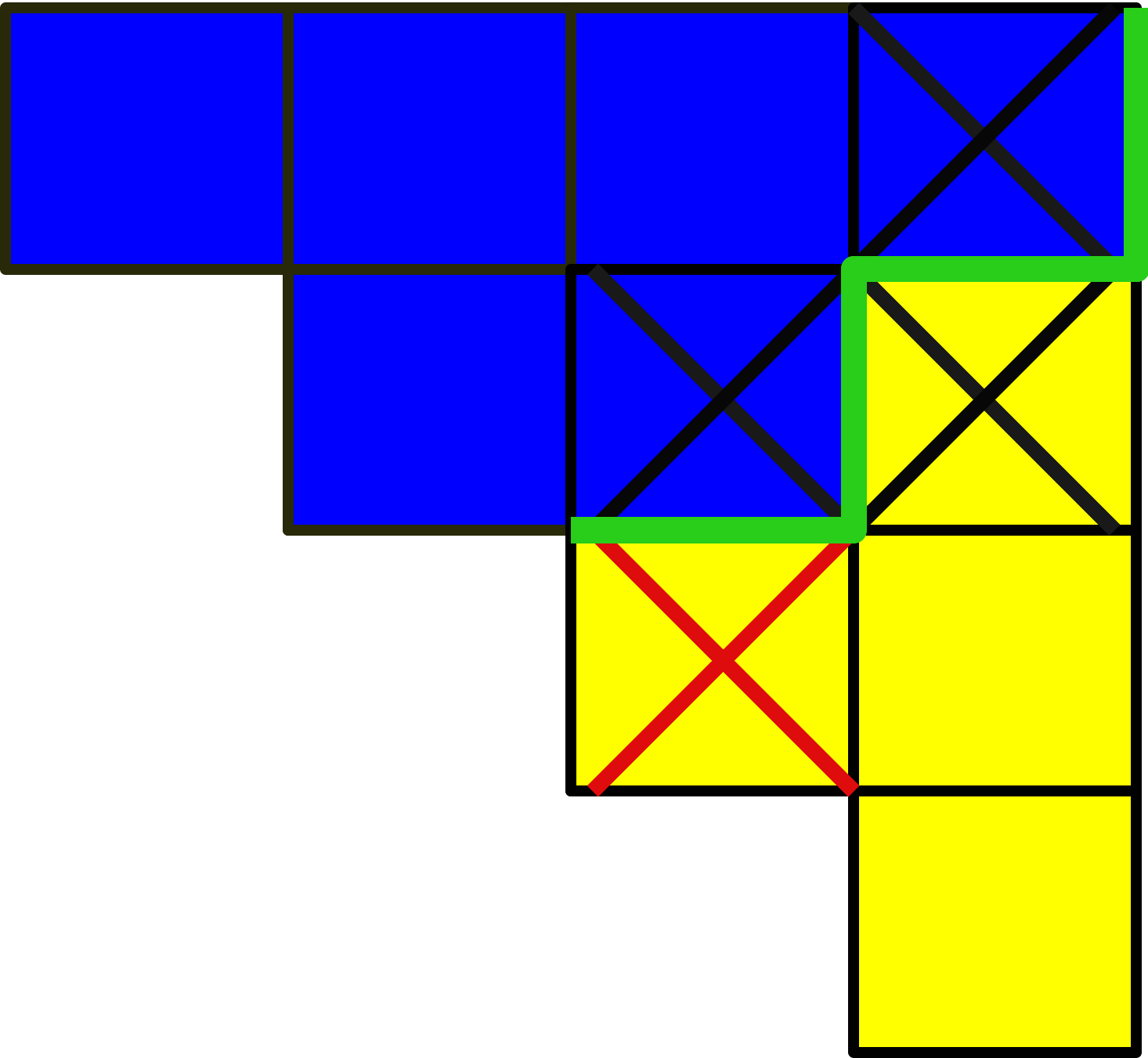} }
&F_1 \rightarrow C_{1,5}^+ +  C_{2,5}^-  
& \{F_3, C_{3,4}^-,  C_{2,4}^+ ,  C_{2,5}^-, C_{1,5}^+  \}
& \multirow{2}{*}{\includegraphics[width=3cm]{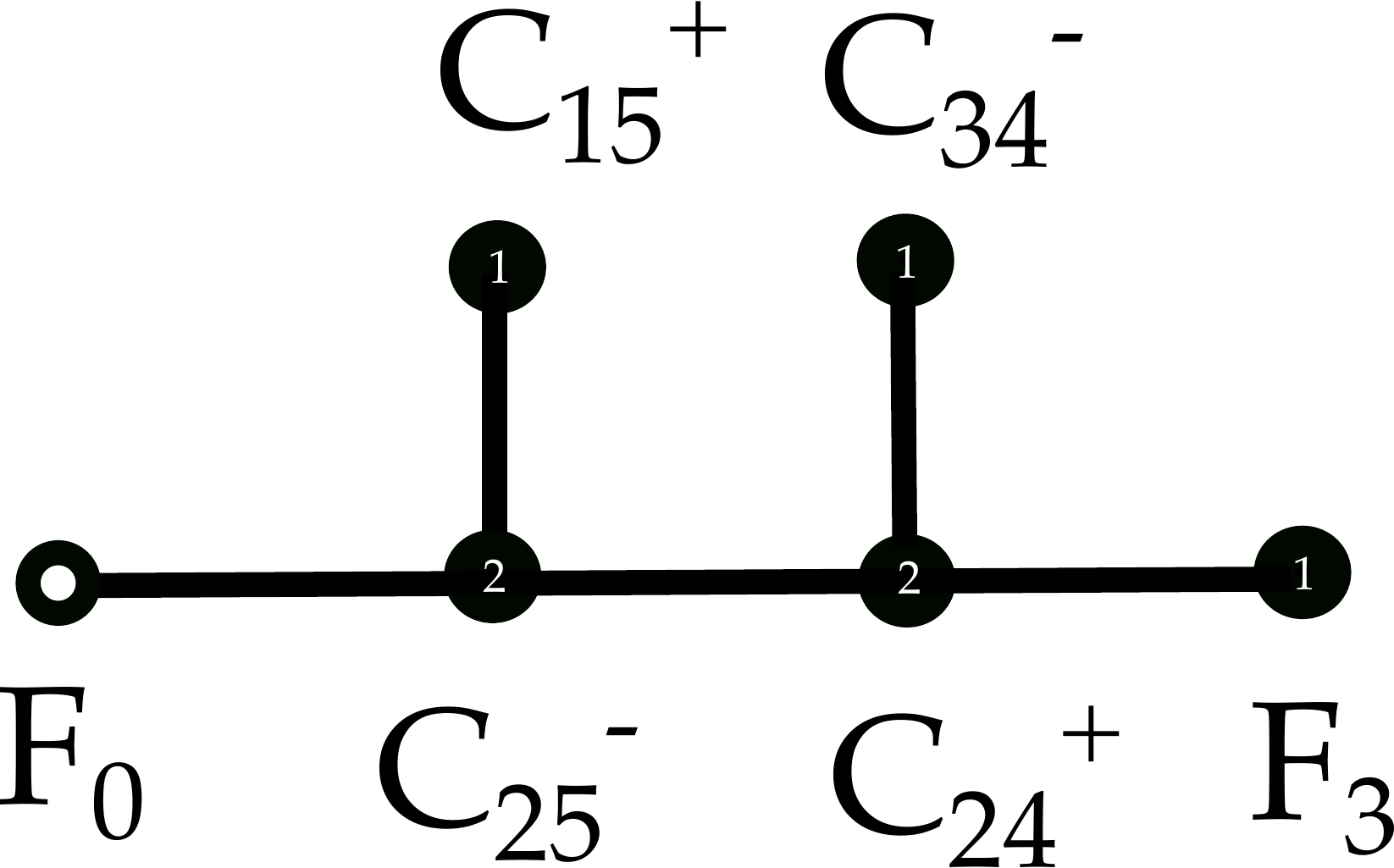}}
\cr
&& F_2 \rightarrow C_{2,4}^+ +  C_{3,4}^-  
&&\cr
&& F_4 \rightarrow C_{2,4}^+ +  C_{2,5}^-  
&&\cr
&&&&\cr\hline
&&&&\cr
 9& \multirow{2}{*}{\includegraphics[width=2cm]{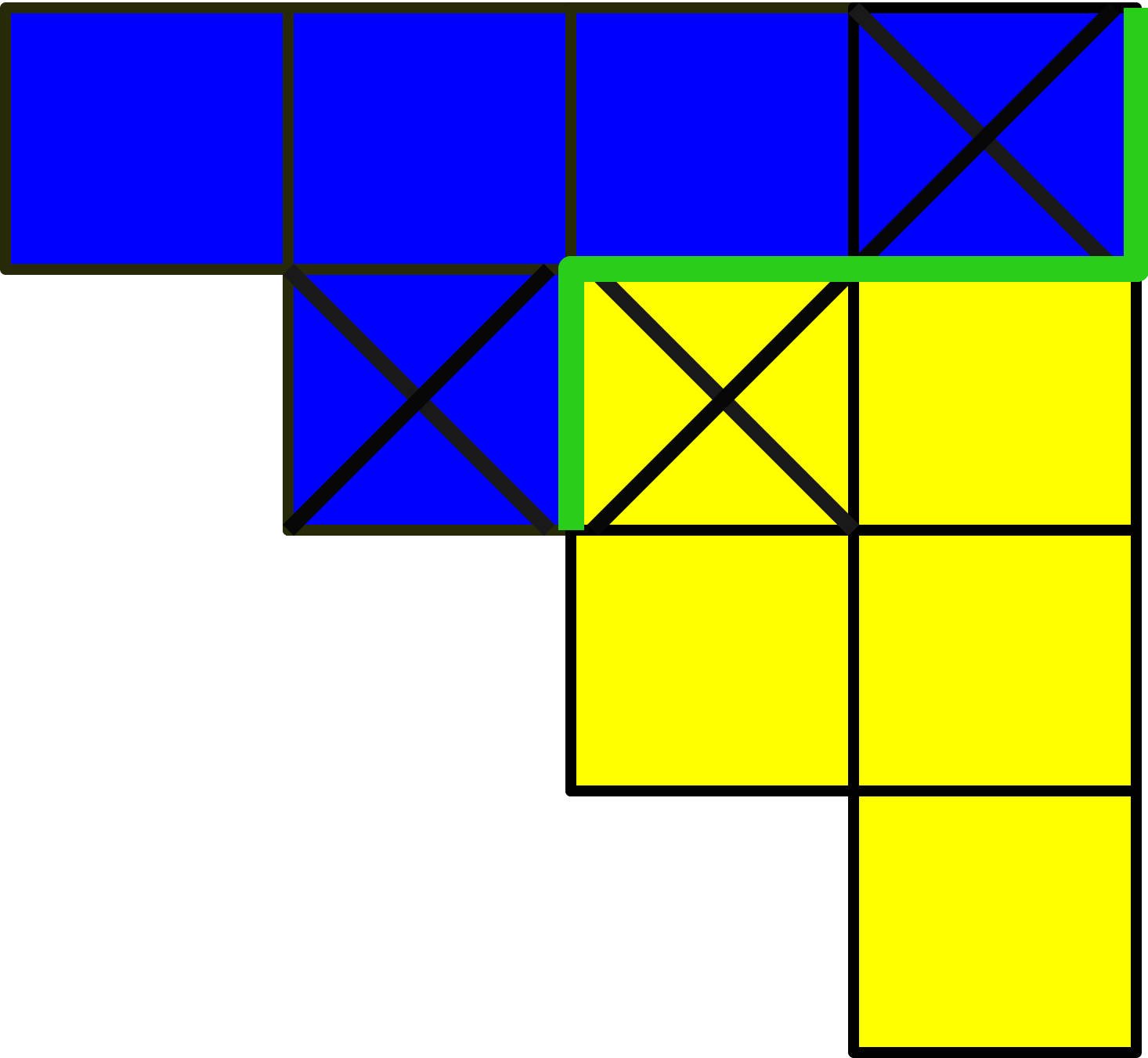} }
&F_1 \rightarrow C_{1,5}^+ + F_4 + C_{2,4}^-  
&  \{ C_{1,5}^+, F_2,  C_{2,4}^-, F_4,C_{2,3}^+ \}
& \multirow{2}{*}{\includegraphics[width=3cm]{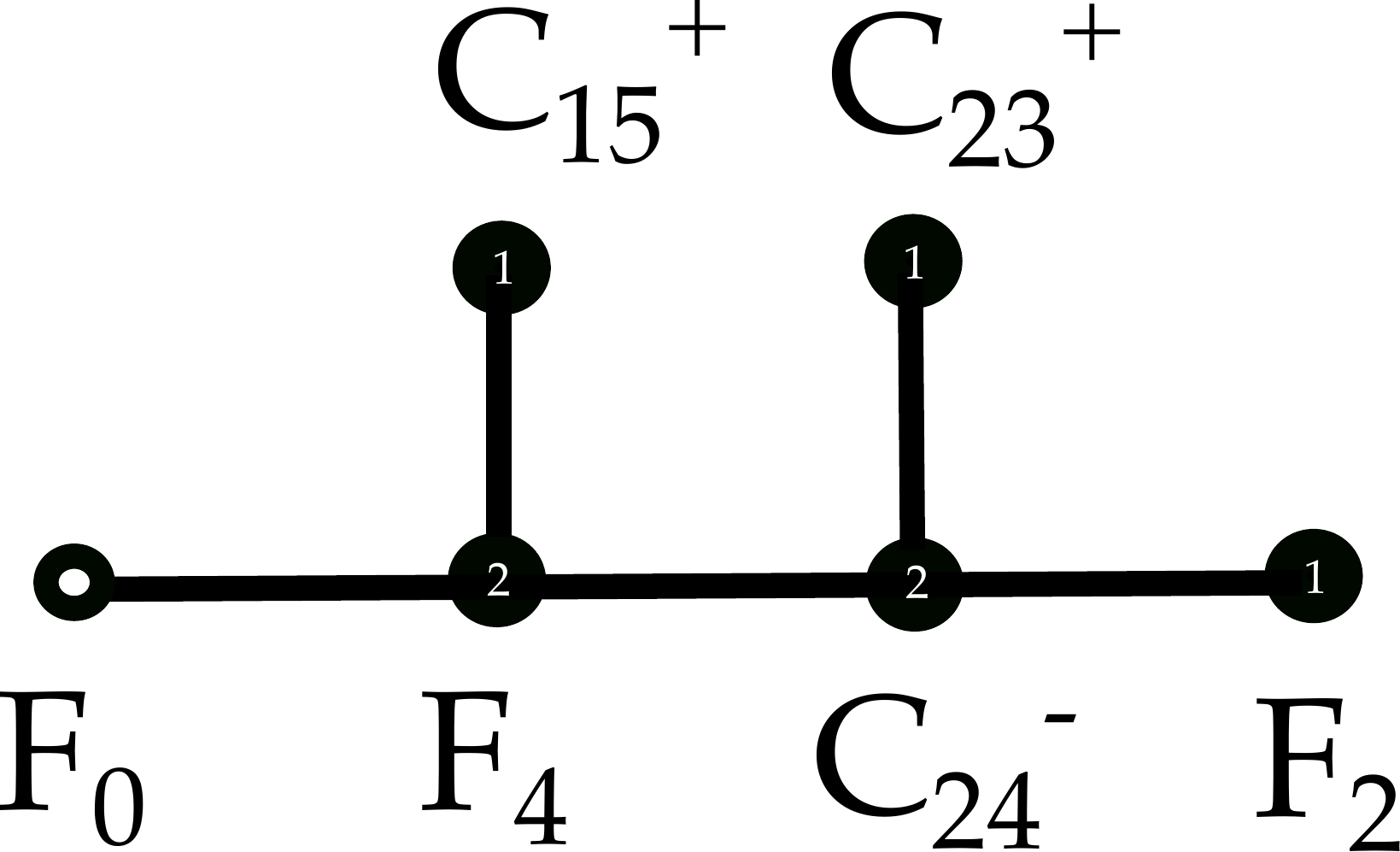}}
\cr
&&
F_3 \rightarrow C_{2,3}^+ + C_{2,4}^- 
&&\cr
&&&&\cr
&&&&\cr\hline
&&&&\cr
 11& \multirow{2}{*}{\includegraphics[width=2cm]{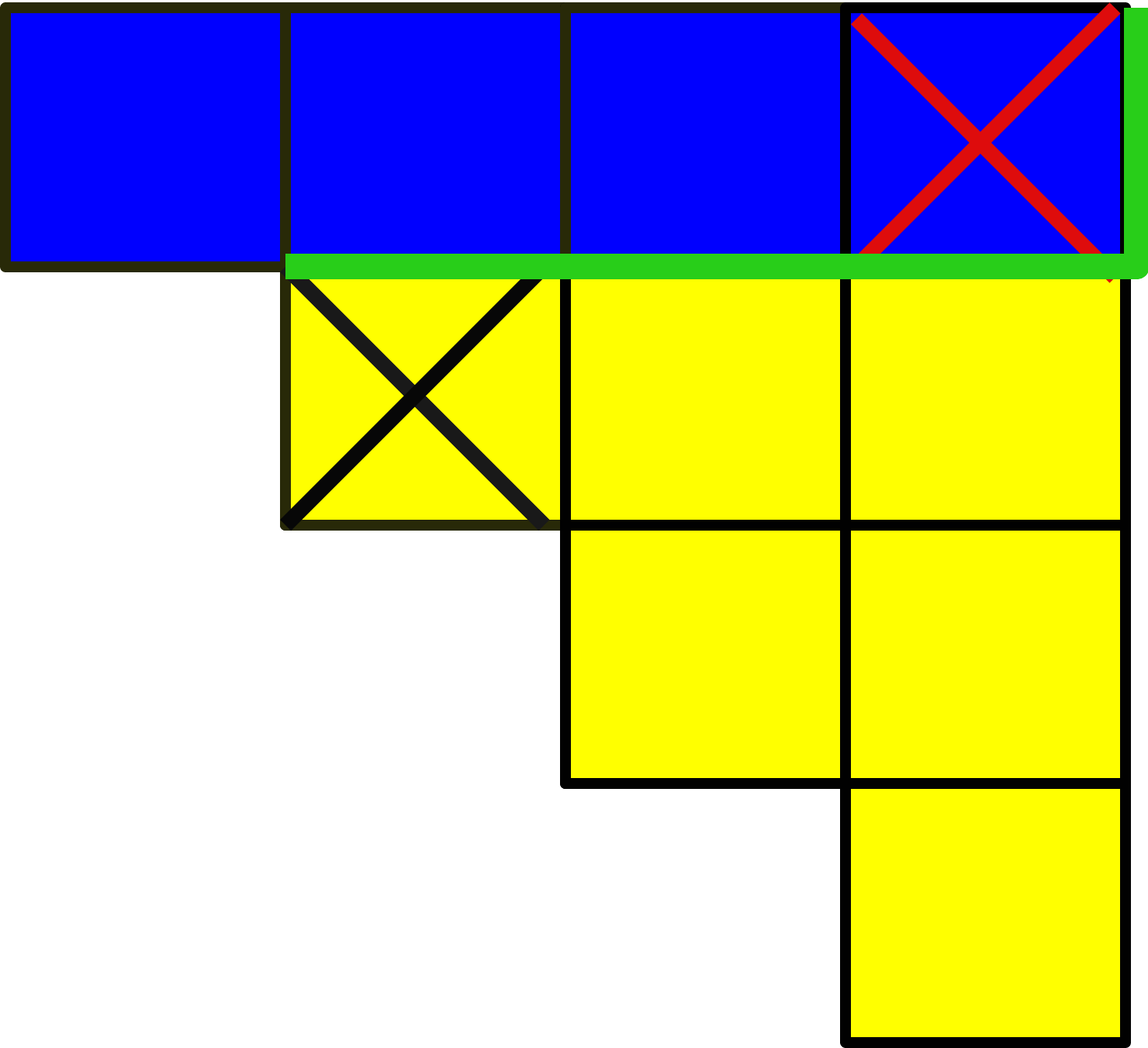} }
&F_1 \rightarrow   C_{1,5}^+ + F_4 + F_3 + C_{2,3}^- 
& \{ F_2, C_{2,3}^-, F_3, F_4, C_{1,5}^+  \}
& \multirow{2}{*}{\includegraphics[width=3cm]{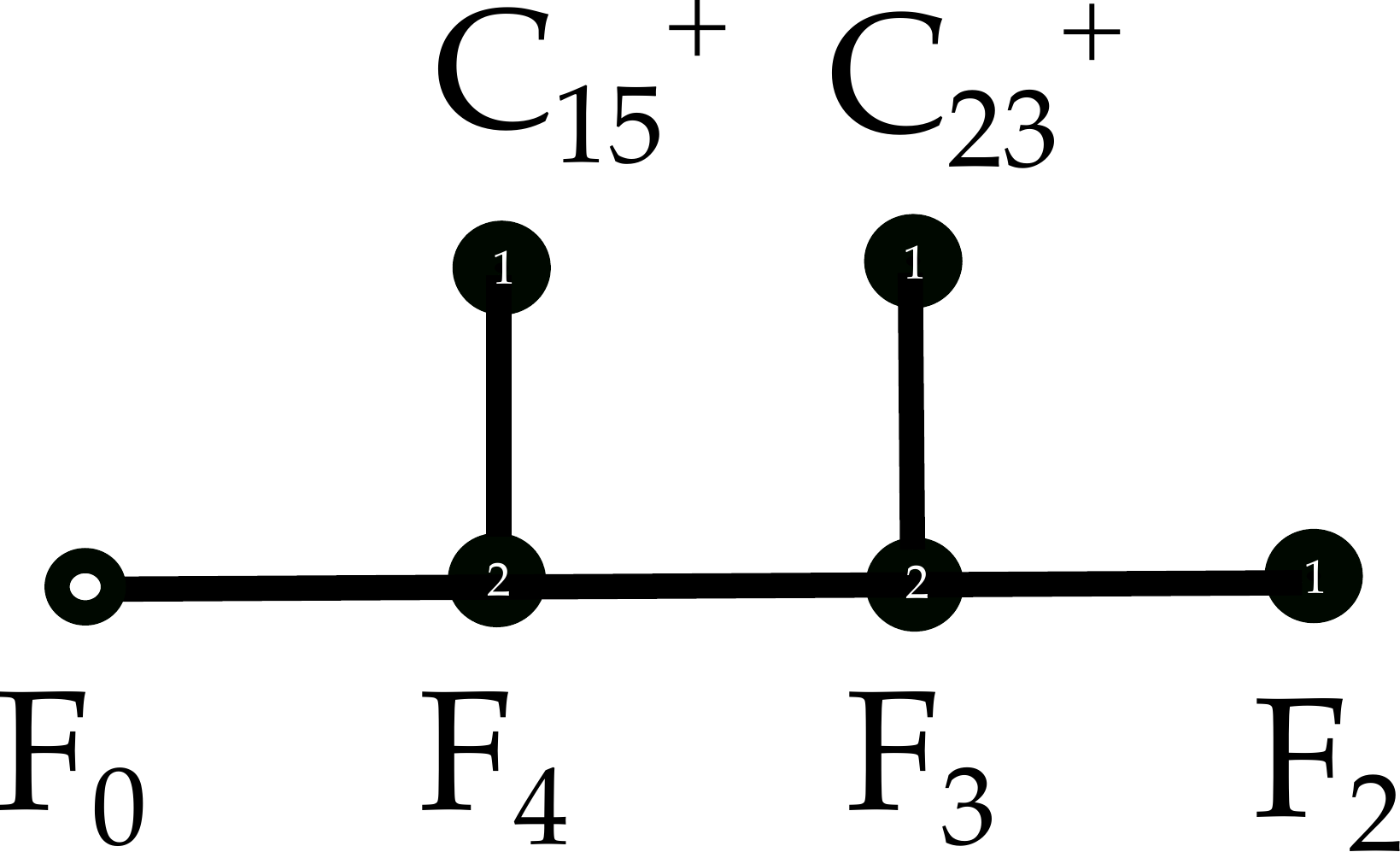}}
 \cr
 &&&&\cr
&&&&\cr
&&&&\cr\hline
\end{array}
\ee

\begin{figure}
\centering
\includegraphics[width=14cm]{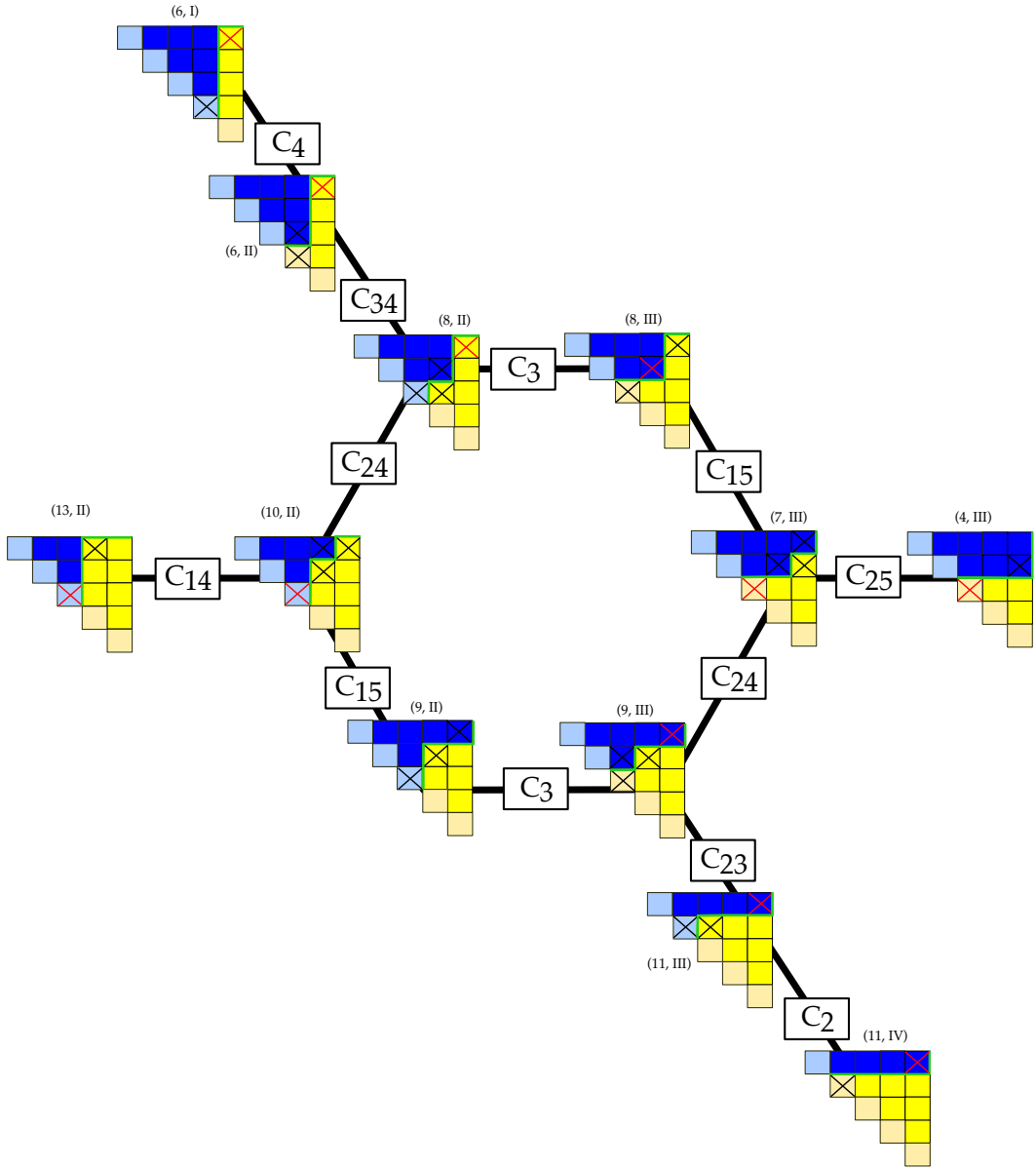}
\caption{Box graphs for $\mathfrak{su}(5)$ with both ${\bf 5}$ and ${\bf 10}$ representation. The extremal generators, which in the geometry correspond to the curves that can be flopped, are marked with a black $X$, whereas red $X$'s indicate cone generators, which cannot be flopped as they would yield $\mathfrak{u}(5)$ phases. The lines connecting the box graphs are labeled by the curve that is being flopped, i.e. $C_i$ ($C_{ij}$) corresponds to flopping a ${\bf 5}$ (${\bf 10}$)curve.  \label{fig:SU5AF}}
\end{figure}

\subsection{Combined box graphs and flops}

The possible combined box graphs are obtained by consistently combining the ones from ${\bf 5}$ and ${\bf 10}$, which turns out to be equivalent to consistent $\mathfrak{su}(6)$ box graphs with the ${\bf 15}$ representation \cite{Hayashi:2014kca}. This structure encodes also codimension three information, as was shown there, and allows to compute all possibly non-Kodaira fibers along the $\mathfrak{e}_6$ enhacement loci. 
The combined flop graph is shown in figure \ref{fig:SU5AF}. Each box graph is combined from one ${\bf 10}$ and one ${\bf 5}$ box grahs, carrying labels  ({\it arabic}, {\it roman}), and the combined resolved geometry has to exhibit both types of splittings, as determined in (\ref{5Split}) and (\ref{10Split}). 

The flops are either with respect to curves corresponding to ${\bf 5}$, or to ${\bf 10}$ weights. This again is easily read off from the flop network figure \ref{fig:SU5AF}: if two box graphs are connected, they differ by either their arabic or roman numeral. Correspondingly, a ${\bf 10}$ or ${\bf 5}$ curve is flopped. We labeled all connecting lines with the curves that are being flopped. In figure \ref{fig:SU5AF}, each connecting line is labeled by the curve, $C_i$ or $C_{ij}$, that is being flopped.


\section{Crepant weighted blowups}
\label{sect:toricbu}

In this section we explain how to determine weighted blowups that give rise to crepant resolutions. One of the organizational tools is to use the connection between toric triangulations, which we define to be toric resolutions, based on  fine triangulations of polytopes\footnote{These are resolutions that are commonly referred to as toric resolutions, for instance in the context of triangulations of tops and polytopes. However, we will consider more general toric resolutions, that do not directly correspond to such triangulations, but to more general refinements of cones. As the resolved are projective, which follows form the direct blowup procedures, one can of course construct an extended polytope whose triangulation yields the resolution. However, for a systematic analysis of all possible crepant resolutions, our approach is more efficient.}, and weighted blowups. 
Such toric triangulations form a strict subclass of possible crepant resolutions. 
However, the way we will characterize these will be generalized and extended to resolutions, do not necessarily arise from a (fine) triangulation.
These generalizations will be discussed for $SU(5)$ in the next sections and in general in \cite{ABSSN}.
Basic definitions and facts from toric geometry (as well as an explanation of our notation) are contained in appendix \ref{sect:toric101}.

\subsection{Cones and Toric Resolutions}

\begin{figure}
\begin{center}
\includegraphics[width=8cm]{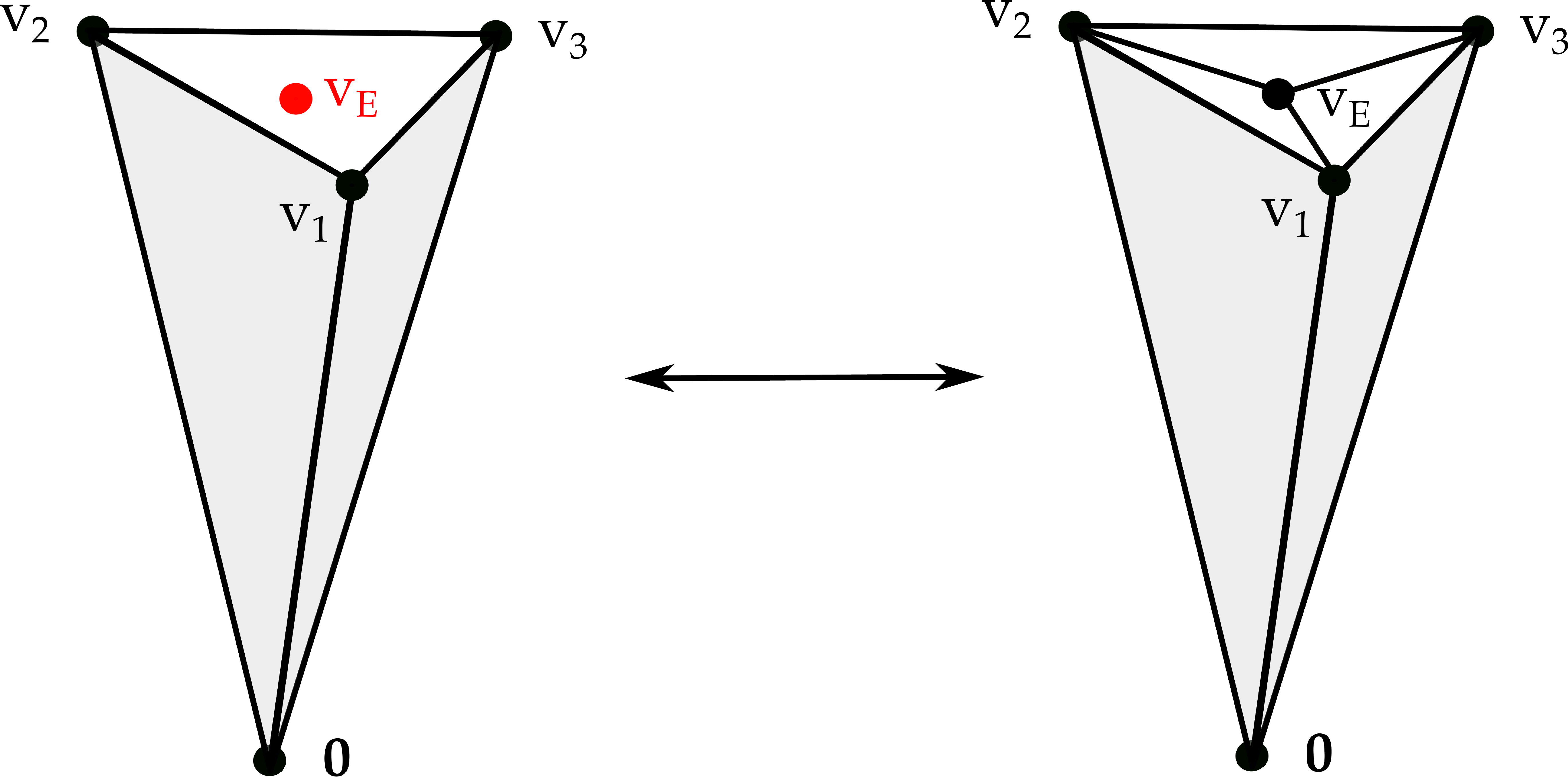} 
\end{center}
\caption{\label{fig:3dfansubdiv}The subdivision of a three-dimensional cone $\sigma$ by introducing a new one-dimensional cone in its interior. 
On the left, a (simplicial) three-dimensional cone generated by the three lattice vectors $v_1, v_2$ and $v_3$ is displayed. We have also
included the lattice point $v_E$ we wish to use for the subdivision, which is drawn in red. Note that this point does not need to lie 
on the hyperplane supporting $v_1, v_2$ and $v_3$. The refinement of $\sigma$ including $v_E$ is shown on the right. This refinement
introduces three three-dimensional cones, three two-dimensional cones and the one-dimensional cone generated by $\sigma$. This figure 
can also be used as an example of a toric blowdown: if we have three three-dimensional cones sitting in a fan as shown on the right,
we can blow down the coordinate corresponding to the lattice vector in the interior. This will eliminate three two-dimensional cones
and glue three three-dimensional cones into a single one. Note that the combinatorics will be more complicated if $\sigma$ is a three-dimensional
cone in a fan of four dimensions or more.}
\end{figure}

Consider a toric variety described in terms of a fan. For any given fan $\Sigma$, we may consider a refinement $\Sigma'$
in which we consistently subdivide cones. By construction, there is a projection $\pi : \Sigma' \rightarrow \Sigma$ such that 
any cone of $\Sigma'$ is mapped to a single cone of $\Sigma$. Hence there is an associated toric morphism which gives rise to 
a proper birational map $T_{\Sigma'} \rightarrow T_\Sigma$, i.e. we may think of a refinement of a fan as
a (generalized) blowup and a fusing of appropriate cones as a blowdown. A simple refinement of a 3-dimensional cone is 
shown in figure \ref{fig:3dfansubdiv}.

Let us now consider an algebraic subvariety $X$ of a toric variety $T_\Sigma$. $X$ has singularities if singularities of $T_\Sigma$ meet $X$ or the
defining equations of $X$ are not transversal. We can try to (partially) resolve such singularities by refining the cones of the fan
$\Sigma$, which is what we will discuss in the following. Consider a singularity of $X$ coming from the non-transversality\footnote{To check this, 
we have to go to a patch where we can use a set of affine coordinates.}  of one of its defining
equations
\begin{equation}
 P(z_1,\cdots z_n)=0 \, ,
\end{equation}
along a locus $z_1 = \cdots z_k = 0$. This singularity has the codimension $k-1$. We can now easily describe blowups along this 
locus by a toric morphism of the ambient space. As $z_1 = \cdots z_k = 0$ are allowed to vanish simultaneously by assumption,
they must share a common cone $\sigma = \langle v_1,\cdots, v_k\rangle$. Hence we want to refine the cone $\sigma$ by introducing
a new one-dimensional cone with generator $v_E$ and appropriate higher-dimensional cones. In the simplest case, where $v_E$ is
in the interiour of an $n$-dimensional cone, $\sigma = \langle v_1,\cdots, v_n\rangle$ is subdivided in to
\be\label{SubDiv}
\langle v_1 ,\cdots,v_{n-1},v_E \rangle, 
\langle v_1 ,\cdots,v_{n-2}, v_{E},v_n \rangle, 
\cdots, 
\langle v_E,v_2 , \cdots,v_n\rangle\,.
\ee
This means that the Stanley-Reisner ideal now contains $z_1 \cdots z_n$, or written in terms of projective relations more commonly used for algebraic resolutions, $[z_1,\cdots, z_n]$. We have shown two elementary examples of such subdivisions
in figures \ref{fig:3dfansubdiv} and \ref{fig:2dfansubdiv}. We will frequently be interested in displaying such cones and their subdivisions, for which we will introduce cone diagrams. 

\subsection{Cone diagrams}
\label{sec:ConeDiag}

\begin{figure}
\begin{center}
\includegraphics[width=12cm]{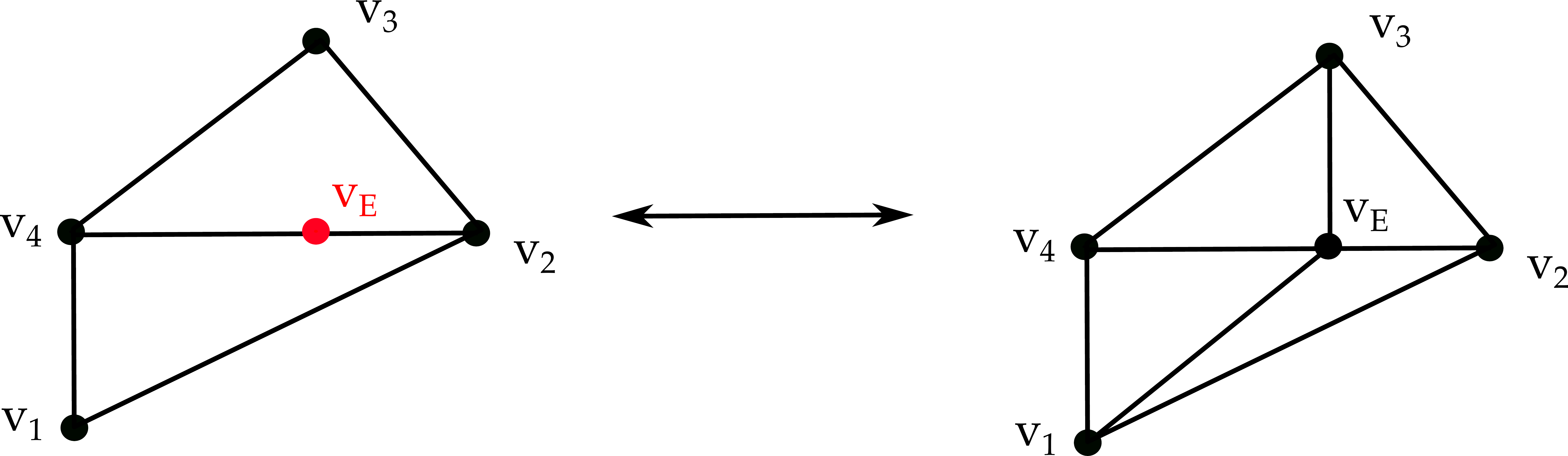} 
\end{center}
\caption{\label{fig:2dfansubdiv}The subdivision of two three-dimensional cones by introducing a new one-dimensional cone interior to their intersection. 
Contrary to figure \ref{fig:3dfansubdiv}, we are using a projection in which $p$-dimensional cones are mapped to $(p-1)$ - dimensional simplices, keeping
their combinatoric intact. We refer to such illustrations as cone diagrams. In the specific example shown, we wish to introduce a new one-dimensional cone
generated by $v_E$, which sits in the interior of a two-dimensional cone generated by $v_2$ and $v_4$. If these cones are part of a three-dimensional fan, this kind of blowup will subdivide each of the two adjacent cones in two. Starting from the fan on the right, it is possible to blow down $v_E$. Note that
this will necessarily take us back to the figure on the left as the resulting cones after the blowdown must be strongly convex. {Note that the four
vertices $v_1 \cdots v_4$ are not necessarily on a hyperplane.}}
\end{figure}

We now define cone diagrams, which are one of the tools that we will use to systematially describe resolutions of singular fibrations. 
Instead of depicting the entire cone of a fan, as in figure \ref{fig:3dfansubdiv}, we will consider diagrams, such as the one shown in figure \ref{fig:2dfansubdiv}, which are more convenient visualizations using a projection. This is done such that the relevant combinatorics is kept intact\footnote{Note that the two figures correspond to different situations.} and the relative locations of the various cones are faithfully represented. 
We will call such pictorial representations of one or several cones {\it cone diagrams} \footnote{These are different from the toric diagrams used in the
litarature to describe toric varieties which are Calabi-Yau manifolds at the same time.}. We will use cone diagrams to describe partial triangulations (or resolutions), and most importantly, to characterize which triangulations (or crepant resolutions) can still be applied to further resolve the geometry. {It is important to keep in mind that such a representation is not possible for any collection of cones in fan. In the situations we encounter, however, this presentation allows
us to restrict ourselves to the salient information.}

\subsection{Toric Resolutions as Weighted Blowups}

Let us return to subdivisions such as the one displayed in figure \ref{fig:torictop}. We will now realize this toric resolution in terms of a weighted blowup in the coordinates $z_i$. For a cone $\sigma$ and a point $v_E$ in the inside of $\sigma$, we can write 
\begin{equation}\label{genrelations}
 \sum_i a_i v_i = a_E \rho_E \,,
\end{equation}
where $\rho_E$ is the one-dimensional cone associated to $v_E$.
In the toric variety corresponding to the refined fan $\Sigma'$, we have a new homogeneous coordinate $z_E$. 
Due to the above relation, there is also a new $\C^*$ action with the weights
\begin{equation}\label{eq:scalingE}
 \begin{array}{cccc}
 z_1 & \cdots & z_k & z_E \\
 \hline 
 a_1 & \cdots & a_k & -a_E  
 \end{array} \, .
\end{equation}
Depending on the details at hand, this will reproduce customary algebraic blowups, but also naturally includes cases 
with non-trivial weights, see \cite{Reid:young1987} for a classic exposition. For such a weighted blowup we will use the notation
\be
([z_1,a_1],[z_2,a_2],\cdots,[z_k,a_k]; [z_E,a_E]) \,.
\ee
In these cases, the ambient space can potentially become singular.
The power of describing these data in terms of a fan is that it is easy to trace the fate of the singularity as we are blowing up
and determine the singular strata of the ambient space $\Sigma'$.

Whereas a refinement of cones in a fan $\Sigma$ can also be conveniently captured in terms of projective relations, the situation
is more subtle for blowdowns. Here, the language of fans allows us to determine when such a blowdown can be carried out at the level of
the ambient space: we need to be able to consistently eliminate cones from the fan and/or glue cones together: we have to make
sure that all resulting cones are strongly convex and can be collected into a fan. See figures \ref{fig:3dfansubdiv} and \ref{fig:2dfansubdiv}
for examples.

\subsection{Crepant weighted blowups}

In this paper, we are interested in crepant resolutions, so that we only want to consider (partial) resolutions 
keeping the canonical class invariant. The anticanonical bundle of a toric variety is 
\begin{equation}
-K_{X_\Sigma} = \sum_i D_i \, ,
\end{equation}
where the sum goes over all one-dimensional cones in $\Sigma$, i.e. all toric divisors. If we perform a blowup associated with 
a refinement $\Sigma' \rightarrow \Sigma$ which introduces a single one-dimensional cone with generator $v_E$, the anticanonical 
class of $T_\Sigma$ hence receives the contribution
\begin{equation}
\delta K = (a_E - \sum_i a_i)D_E \, .
\end{equation}
This tells us that the above only is a crepant (partial) resolution of $X$ if its class after proper transform is $-K_{X_\Sigma} - \delta K$. 
In other words, the proper transform must allow us to `divide out' the right power of the exceptional coordinate $z_E$ to make $P(z_i)$
aquire the weight $(-a_E + \sum_i a_i)$ under the $\C^*$ action \eqref{eq:scalingE}.

Of course, it is an option to check the above condition case by case for any sequence of weighted blowups. Here, we are going to use
a more elegant method. Assume that the singularity we want to resolve is captured by an equation\footnote{This does not mean that the 
manifold in question is a hypersurface in an ambient projective (or toric) space or a Calabi-Yau variety, as it may e.g. be defined by
a complete intersection involving \eqref{eq:monos}.}
\begin{equation}\label{eq:monos}
\sum_j c_j \prod_i z_i^{\langle v_i,m_j\rangle+1} = 0 \, ,
\end{equation}
where $v_i$ are generators of a fan and the $m_j$ are a set of lattice points in the $M$ lattice. The singularities we are interested in,
which arise in singular Tate models, are of this type. We will describe the singular Tate model in this language in the next section.

A weighted blowup sends $z_i \rightarrow z_i z_E^{a_1/a_E}$. In order for such a blowup to be crepant, \eqref{eq:monos} must be divided by
$z_E^{(-a_E + \sum_i a_i)/a_E}$ when doing the proper transform. Using \eqref{genrelations}, an arbitrary monomial in \eqref{eq:monos} is 
then turned into
\begin{equation}
\ba\label{eq:monoscrep}
 &z_E^{(a_E - \sum_i a_i)/a_E} \prod_i z_i^{\langle v_i,m_j\rangle+1} z_E ^{\tfrac{1}{a_E}(a_i \langle v_i,m_j\rangle + a_i)} \cr
  =\, &z_E^{(a_E - \sum_i a_i)/a_E} z_E ^{\langle v_E,m_j\rangle + \sum_i a_i/a_E} \prod_i z_i^{\langle v_i,m_j\rangle+1} \cr
  =\,& z_E^{\langle v_E,m_j\rangle + 1 }  \prod_i z_i^{\langle v_i,m_j\rangle+1} \, ,
\ea
\end{equation}
i.e. we simply need to use \eqref{eq:monos} for the new coordinate $z_E$ as well. Note, however, that \eqref{eq:monoscrep}
is holomorphic if and only if 
\begin{equation}\label{constrv_E}
 \langle v_E , m_j \rangle \geq -1 \quad\quad \mbox{for all}\quad m_j \, .
\end{equation}
Hence only blowups related to the introduction of new generators $v_E$ satisfying the above relation can be crepant.
For a given singularity, this will single out a finite number of crepant weighted blowups. 
After performing such a weighted blowup (cone refinement), the set $m_j$ of monomials is not 
changed, i.e. at every step of a sequence of blowups we find the same condition \eqref{constrv_E} for the next step.
We hence learn that we can only use weighted blowups originating from the set of $v_E$ satisfying 
\eqref{constrv_E} in any step of a sequence of blowups.

Note that even though we have used toric language, the result stands on its own. We may completely discard all of the 
toric language at this point and merely proceed to carry out the weighted blowups we have found. We will however, continue
to use the diagrams associated with the fan spanned by the $v$, as these conveniently encode the projective relations
(i.e. the SR ideal) of the ambient space coordinates.

\label{subsec:CI}
In the discussion above, we have assumed that the locus we want to blow up can be described by the vanishing of a set of 
homogeneous coordinates of the ambient space. The above discussion is still applicable, however, if we appropriately
enlarge the dimension of the ambient space we are working with.

Let us give a schematic example and describe the blowup of a hypersurface $X$ given by $P=0$ in a toric variety along the locus 
\be
z_1= \cdots= z_k = \phi(z_i) = 0\,,
\ee
for some homogeneous polynomial $\phi(z_i)$. The trick is to introduce another coordinate $z_\phi$ which lifts $\phi(z_i)$ to a coordinate
of the ambient space. We hence ask this new coordinate to fulfill the equation $z_\phi = \phi(z_i)$, which by homogeneity also uniquely fixes the 
weights of $z_\phi$. After fixing $v_\phi$ we lift the generators $v_i$ of one-dimensional cones in $\Sigma$ to $v_i^\sharp$ in $n+1$ 
dimensions such that the scaling relations involving $z_\phi$ are reproduced. The lift of the fan $\Sigma$,  $\Sigma^\sharp$, is then obtained
as follows. For every $p$-dimensional cone of $\Sigma$ we add $z_\phi$ as an extra vertex, turning it into a $p+1$-dimensional cone of the lifted 
fan $\Sigma^\sharp$. We have now increased the dimension of the ambient space by one and gained a further equation. 
In particular, we have managed to place the locus we intend to blow up along the intersection
$z_1, \cdots, z_k = z_\phi= 0$ of toric divisors. We can now perform a blowup by introducing a new generator $v_E$ and 
subdividing the cone $\langle v_1, v_2,\cdots, v_k,v_\phi \rangle$ appropriately. The resolved complete intersection is then given by two
equations of the form
\begin{equation}
\ba
z_\phi z_E &= \phi(z_i,z_E) \cr
 P(z_i,z_E,z_\phi) &= 0 \, .
\ea
\end{equation}
The description as a complete intersection was redundant before this blowup (we could simply solve the equation of $z_\phi$ and discard this coordinate), 
however becomes non-trivial after the blowup. Resolutions of this type will form another subclass of algebraic resolutions that are necessary in 
order to construct all possible small resolutions. 

If we restrict ourselves to the case of Calabi-Yau varieties, the above discussion boils down to the reflexive polytopes 
of \cite{Batyrev94dualpolyhedra,Borisov93,BatyrevBorisov}. 

\subsection{Flops}

\begin{figure}
\begin{center}
\includegraphics[width=10cm]{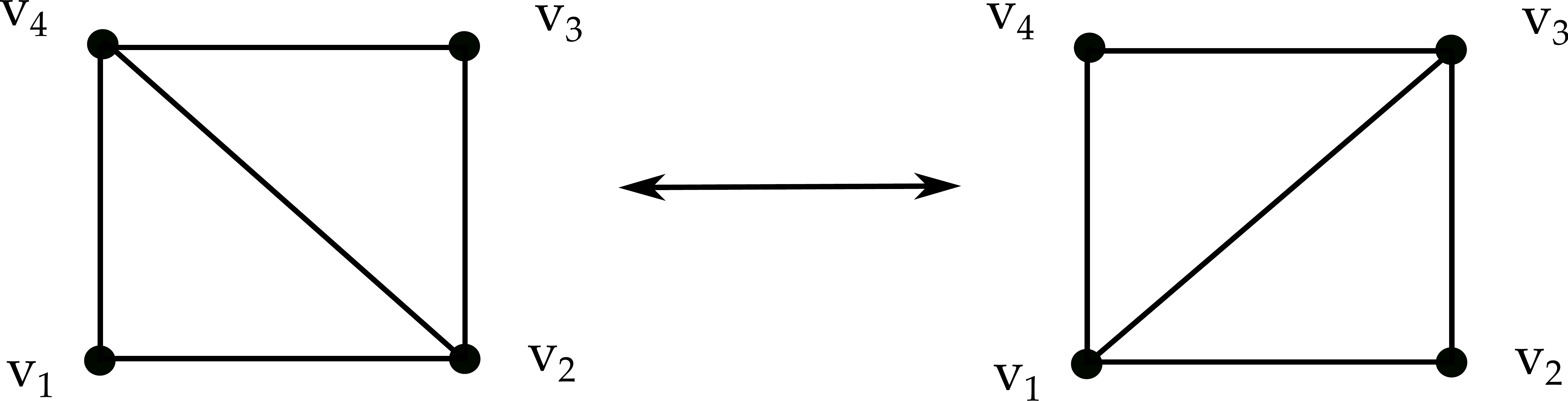} 
\end{center}\caption{\label{fig:toricflop}Two three-dimensional cones for which a flop can be realized at the level of the ambient space.
As done already in figure \ref{fig:2dfansubdiv}, we only show a projection containing the relevant combinatorics. The flop consists of
changing the way the non-simplicial cone spanned by all four generators is subdivided into two simplicial cones, i.e. we either separate
$D_1$ and $D_3$, or $D_2$ and $D_4$. As is well-known, the intermediate singular ambient space in which we only have a single non-simplicial
cone, has Weil divisors which are not Cartier. The exceptional $\P^1$'s appearing in the two possible
small resolutions correspond to the interior one-simplex (two-dimensional cone) in the figure. Note that the four generators $v_i$ 
need not be on a hyperplane in $N\otimes\R$.}
\end{figure}

We now turn to a discussion of flops in the toric context. A flop is realized by blowing down a
subvariety of codimension two of $X$ and resolving to a different manifold $\tilde{X}$. In toric
geometry, such objects correspond to two-dimensional cones\footnote{{Here, we of course assume that the corresponding divisors
do indeed meet on the embedded manifold.}}. To see if we can do a flop, we hence have to 
ask if we can consistently remove a two-dimensional cone from $\Sigma'$ and replace it with a different one.
The prototypical example is shown in figure \ref{fig:toricflop}. We will encounter more complicated examples 
in the rest of this paper.

\section{Fiber Faces and weighted blowups for $SU(5)$ }
\label{sec:SU5Blowups}

In this section we will apply the general insights obtained in the last section, and construct an explicit algebraic resolution sequence
for each box graph of $\mathfrak{su}(5)$ with both ${\bf 5}$ and ${\bf 10}$ representation. 

\subsection{Top Cone and Fiber Faces}\label{eq:ptforbus}
\label{sec:TCFF}

A singular Weierstrass model with a fiber of type $I_5$ over $S = \{ \zeta_0 = 0 \}$ is best consumed in Tate form \cite{Bershadsky:1996nh, Katz:2011qp} 
\be
\label{TateSU5}
 y^2 + b_1 w x y + b_3\zeta_0^2  w^3 y = x^3 + b_2 \zeta_0 w^2 x^2 + b_4 \zeta_0^3 w^4 x + b_6 \zeta_0^5 w^6 \,.
\ee
The above equation embeds the elliptic fiber into the weighted projective space $\P_{123}$ with homogeneous coordinates
$(w,x,y)$ for every point in the base. We can make contact with the techniques reviewed in the last section 
by the following construction, which borrows from the idea of tops first introduced in \cite{Candelas:1996su,Perevalov:1997vw}
and has been widely adapted in the literature on F-theory\footnote{In particular, see  \cite{Collinucci:2010gz} for a recent paper, which discusses this in a similar spirit to ours.}

We first introduce the vectors
\be
\ba
p_x = \left(\begin{array}{c}
       -1 \\ 0 
      \end{array}
\right)\, , \, \,
p_y = \left(\begin{array}{c}
       0 \\ -1 
      \end{array}
\right)\, , \,\,
p_w = \left(\begin{array}{c}
       2 \\ 3 
      \end{array}
\right)\,,
\ea
\ee
and construct a fan from the cones $\langle p_x, p_y \rangle$, $\langle p_x, p_w \rangle$ and $\langle p_w, p_y \rangle$. The corresponding toric
variety is the weighted projective space $\P_{123}$ and we can think of both $\P_{123}$ and the elliptic curve \eqref{TateSU5}
as being fibered over the base. In order to be able to resolve the $I_5$ fiber in \eqref{TateSU5} using toric methods, we have
to introduce a toric coordinate corresponding to $\zeta_0$. We use the generators
\be
\ba\label{eq:toppts}
v_x = \left(\begin{array}{c}
       -1 \\ 0 \\ 0
      \end{array}
\right)\, , \, \,
v_y = \left(\begin{array}{c}
       0 \\ -1 \\ 0
      \end{array}
\right)\, , \,\,
v_w = \left(\begin{array}{c}
       2 \\ 3 \\ 0
      \end{array}
\right)\,,  \,\,
v_{\zeta_0} = \left(\begin{array}{c}
       2 \\ 3 \\ 1
      \end{array}
\right)\, ,
\ea
\ee
and construct a fan $\Sigma_{I_5}$ from the cones $\langle v_x, v_y, v_{\zeta_0} \rangle$, $\langle v_x, v_w, v_{\zeta_0} \rangle$
and $\langle v_w, v_y, v_{\zeta_0} \rangle$. The power of this construction is that \eqref{TateSU5} 
captures the behaviour of the elliptic fiber and allows us to find resolutions without having to explicitely specify the base. 
This works as follows. We describe \eqref{TateSU5} along the lines of \eqref{eq:monos} by assigning a vector 
$m_j$ in the $M$-lattice to each monomial in \eqref{TateSU5} by:
\be\label{eq:mpolyI5}
\begin{array}{c|c}
\hbox{Monomial} & \hbox{Vector} \\ \hline
y^{2}  &\left(1,\,-1,\,0\right) \\
x y & \left(0,\,0,\,-1\right)\\
y \zeta_{0}^{2} & \left(1,\,0,\,-1\right)\\
x^{3} & \left(-2,\,1,\,0\right)\\
x^{2}\zeta_{0} & \left(-1,\,1,\,-1\right)\\
x \zeta_{0}^{3} & \left(0,\,1,\,-1\right)\\
\zeta_{0}^{5} & \left(1,\,1,\,-1\right)
\end{array} 
\ee

\begin{figure}
\begin{center}
\includegraphics[width=10cm]{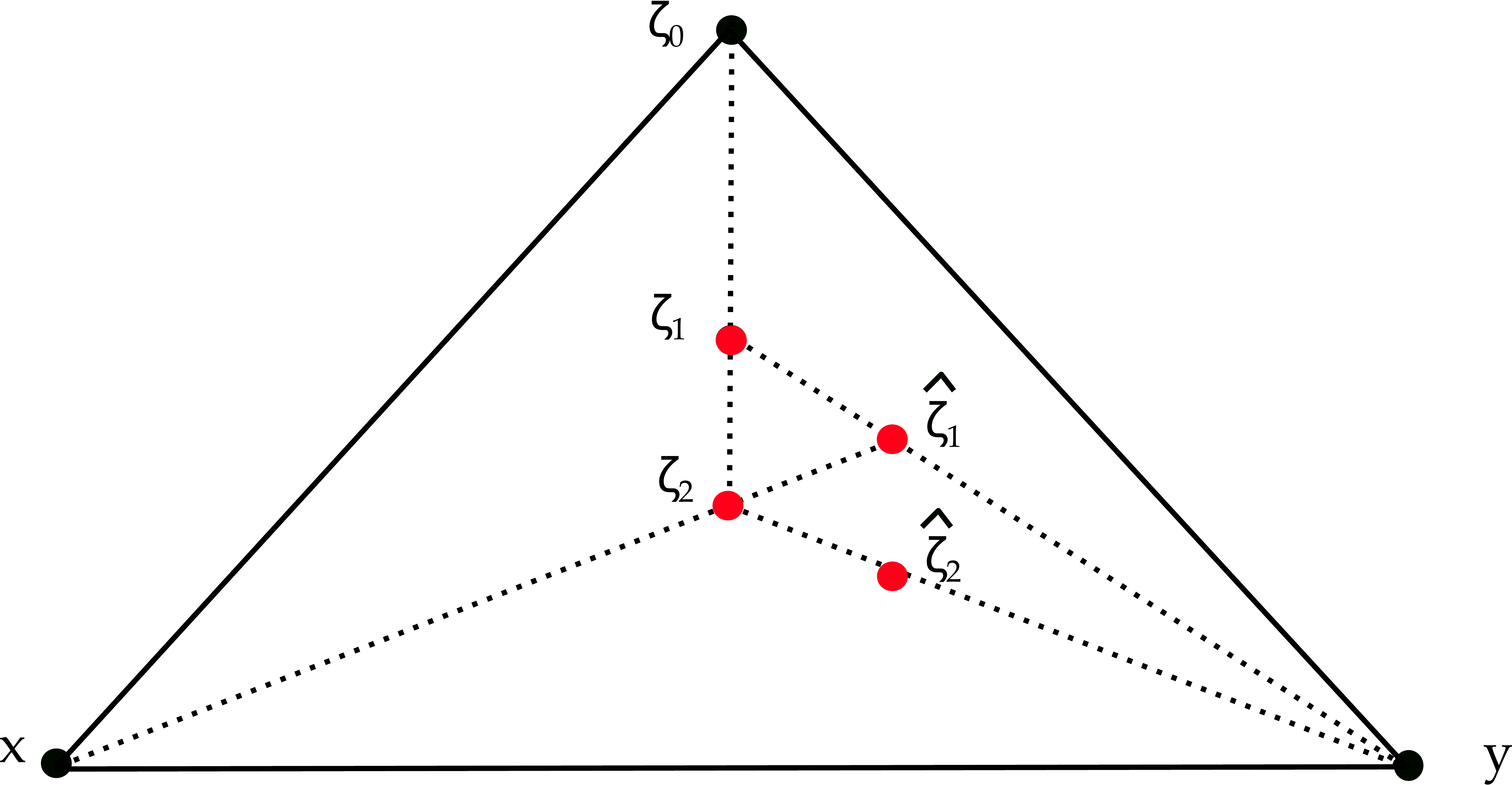} 
\end{center}\caption{\label{fig:torictop}
A cone diagram showing the top cone $\langle x,y,\zeta_0\rangle$.
Refinements of this cone corresponding to crepant resolutions of the singular locus in \eqref{TateSU5} must necessarily employ the 
four points shown in red. We have also displayed their relative positions: if three points lie along a dottet line, the middle 
one is located in the cone spanned by the two ones at the ends. }
\end{figure}

The singularity in \eqref{TateSU5} is located at $x=y=\zeta_0=0$. We hence want to refine the cone $\langle x,y,\zeta_0\rangle$ in such
a way as to resolve the singularity crepantly. As this cone plays a key role we will be refered to it 
as the {\it top cone}. 

As we have discussed in section \ref{sect:toricbu}, 
to refine this top cone, we have to demand that the generators $v_i$ of one-dimensional cones introduced in the refinement process satisfy  
\begin{equation}
\langle v_i , m_j \rangle  \geq -1 
\end{equation}
for all vectors $m_j$ in \eqref{eq:mpolyI5}. 
In the cone $\langle x,y,\zeta_0\rangle$, there are exactly four such lattice vectors given by
\be
\ba\label{eq:viI5}
v_{\zeta_1} = \left(\begin{array}{c}
       1 \\ 2 \\ 1
      \end{array}
\right)\, , \, \,
v_{\zeta_2} = \left(\begin{array}{c}
       0 \\ 1 \\ 1
      \end{array}
\right)\, , \,\,
v_{\hat{\zeta}_1} = \left(\begin{array}{c}
       1 \\ 1 \\ 1
      \end{array}
\right)\,, \,\,
v_{\hat{\zeta}_2} = \left(\begin{array}{c}
       0 \\ 0 \\ 1
      \end{array}
\right)\, .
\ea
\ee
Together with $v_x, v_y, v_{\zeta_0}$ and $v_w$, these points span the well-known $SU(5)$ top {for $\P_{123}$ fiber embeddings.

We have hence shown that any refinement of the cone $\langle x,y,\zeta_0\rangle$, which introduces a one-dimensional cone
generated by any one of the four lattice vectors above, will induce a crepant blowup of \eqref{TateSU5}. A projection, which we call 
{\it cone diagram} (introduced in section \ref{sec:ConeDiag}) showing the location of these four lattice points in the cone $\langle x,y,\zeta_0\rangle$ is shown in figure \ref{fig:torictop}. As we are only interested in refining the cone $\langle x,y,\zeta_0\rangle$ when resolving \eqref{TateSU5}, we can ignore $v_w$ and the coordinate $w$ in the following. 

Furthermore, it is sufficient to only consider the subdivisions of the {\it fiber face}, which we define as the point configuration comprising $v_{\zeta_0}$ and the points (\ref{eq:viI5}). This point configuration appears as the integral points on a face of the top generated by \eqref{eq:toppts}
and \eqref{eq:viI5}. It is precisely the face showing the components of the reducible $I_5$ fiber. As a triangulation of such a face uniquely
fixes a subdivision of the cone $\langle x,y,\zeta_0\rangle$, it provides are more condensed way of presenting this information. 

In the following we will provide a map between box graphs and triangulations of the fiber face. As with the triangulations of the top cones, black (red) points correspond to points that have (not) been used in a triangulation. Black lines connecting points correspond to actual triangulations, whereas black lines connecting to red points correspond to triangulations involving $v_x$ or $v_y$. An example corresponding to a single triangulation for the $SU(5)$ top is shown in figure \ref{fig:StartingFibFac}.

\subsection{Starting resolutions}

\begin{figure}
\begin{center}
\includegraphics[width=7cm]{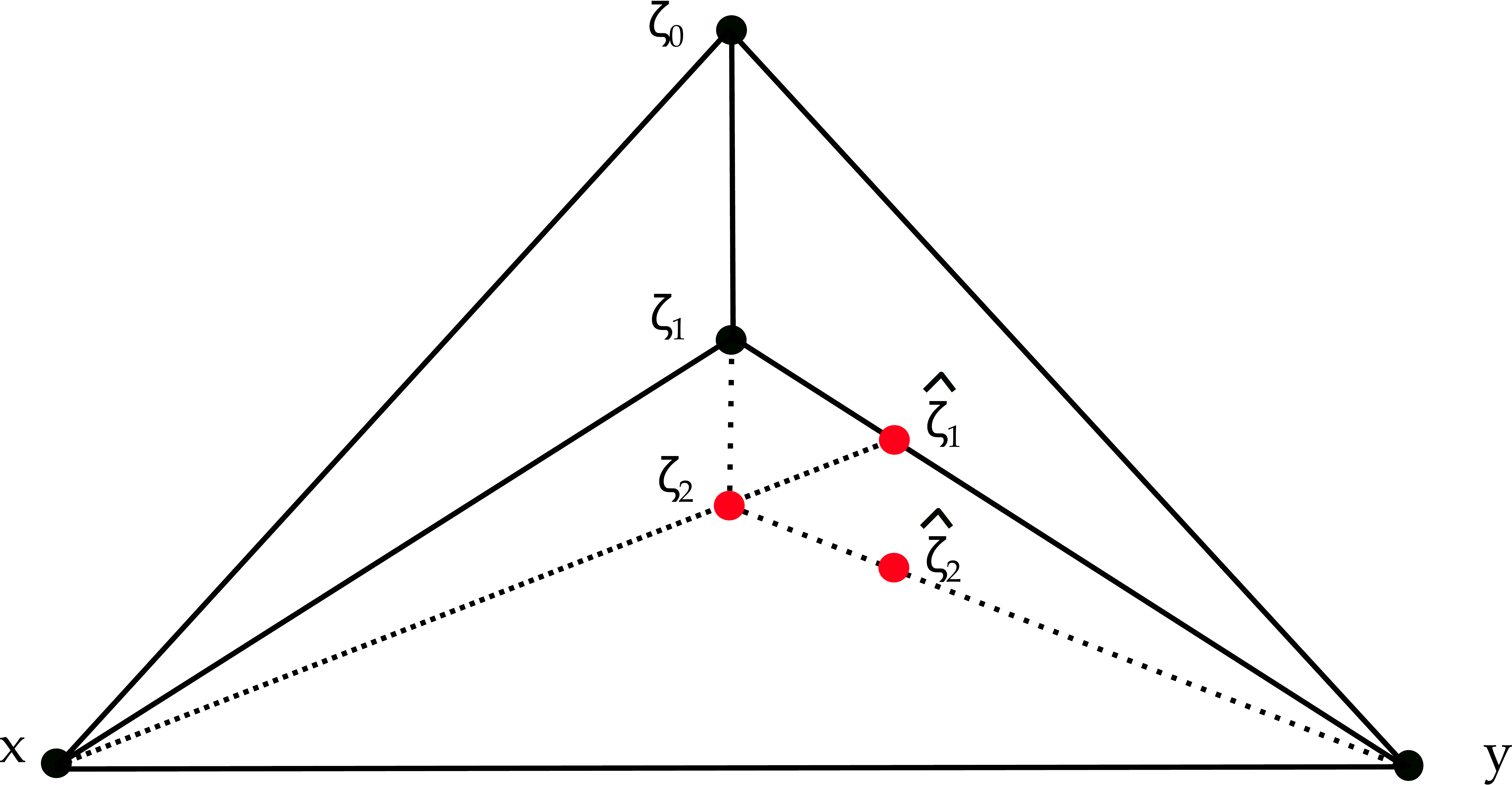} \quad\quad 
\includegraphics[width=7cm]{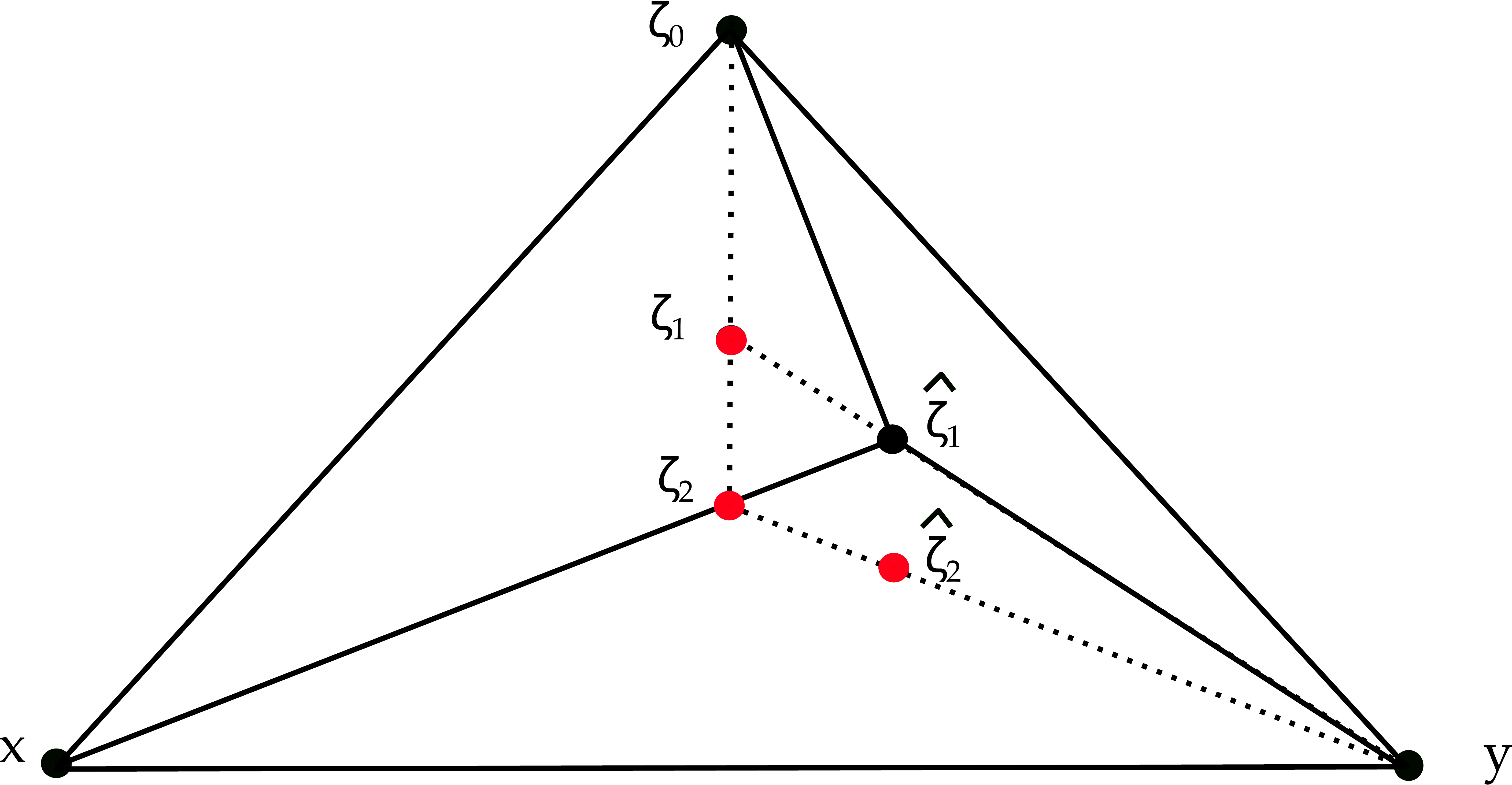} \\
\vspace{1cm}
 \includegraphics[width=7cm]{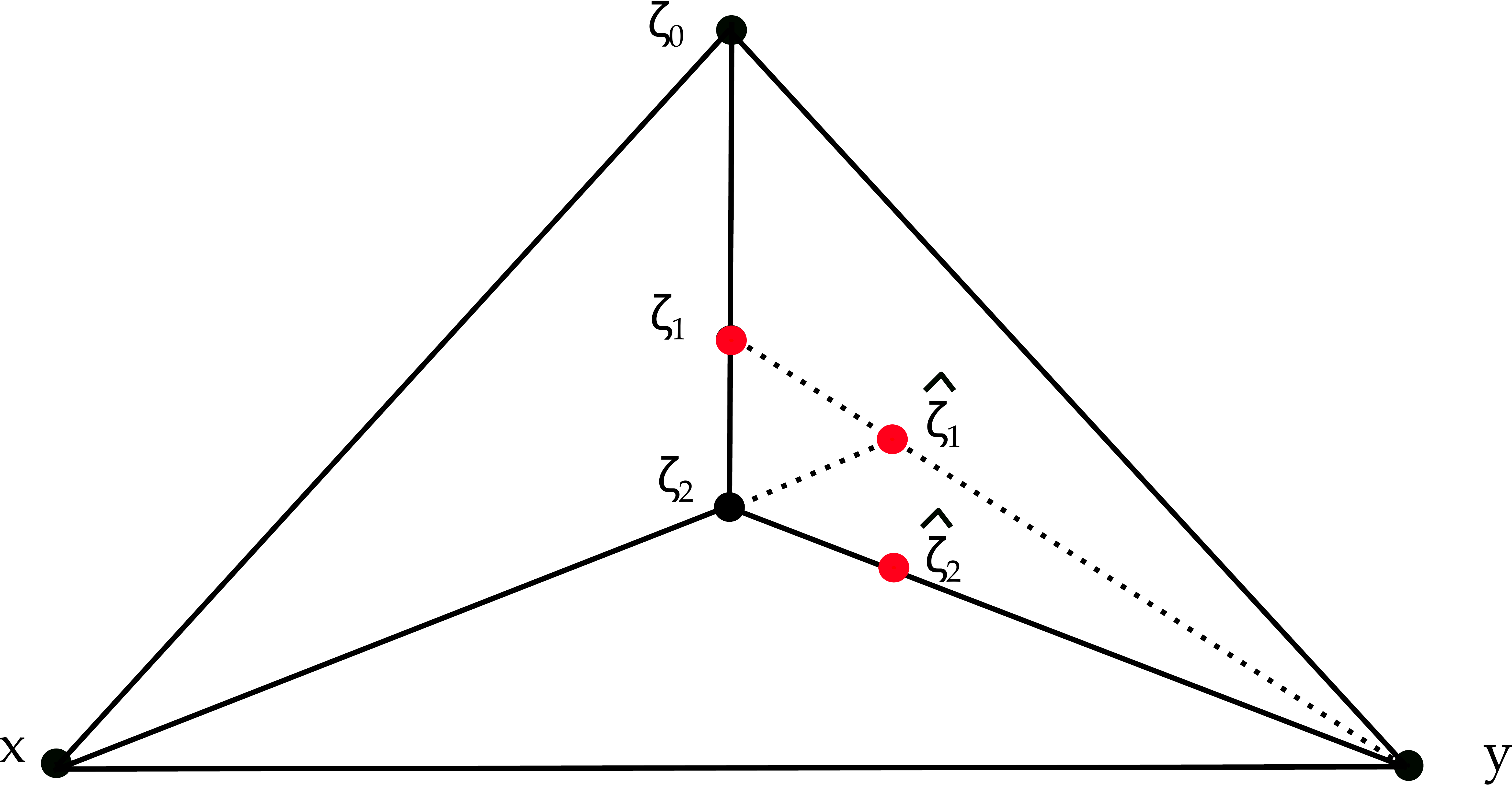}\quad \quad  
 \includegraphics[width=7cm]{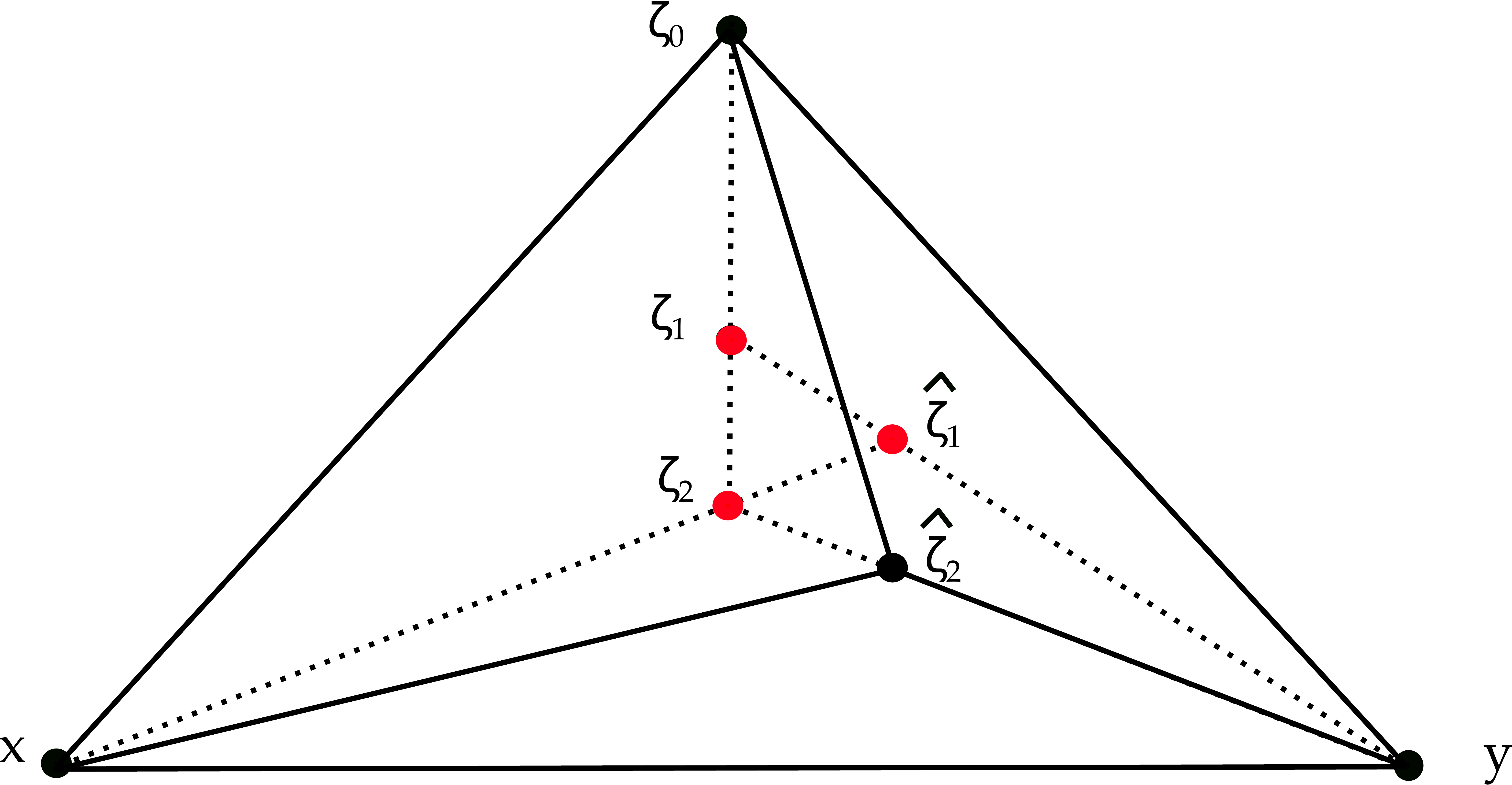} 
\end{center}\caption{The four first possibilities $\hbox{Res}_{\zeta_1}$, $\hbox{Res}_{\zeta_2}$, $\hbox{Res}_{\hat\zeta_2}$, and $\hbox{Res}_{\hat\zeta_1}$ in anti-clockwise ordering, respectively, for weighted blowups, shown in terms of top cones, i.e. the same projection as in figure \ref{fig:torictop}. 
For each case, we show the all of the cones introduced in the blow up. The points marked in red can still be used for further crepant blowups 
and are not used in the respective fans. This figure can be used to read off in which cones they are contained. \label{fig:StartingTriang}}
\end{figure}

In order to get a feeling for these methods, let us demonstrate which options we have for the first blowup. Introducing one
of the four coordinates $\zeta_1, \zeta_2, \hat{\zeta}_1, \hat{\zeta}_2$ corresponds to the four weighted blowups
\be\label{StartRes}
\ba
\hbox{Res}_{\zeta_1}:\quad  &([x,1],[y,1],[\zeta_0,1]; [\zeta_1,1]) \cr
\hbox{Res}_{\zeta_2}:\quad&([x,2],[y,2],[\zeta_0,1]; [\zeta_2,1]) \cr
\hbox{Res}_{\hat\zeta_1}:\quad&([x,1],[y,2],[\zeta_0,1]; [\hat{\zeta}_1,1])\cr
\hbox{Res}_{\hat\zeta_2}:\quad&([x,2],[y,3],[\zeta_0,1]; [\hat{\zeta}_2,1]) \,.
\ea
\ee
These blowups will subdivide the top cone $\langle v_x, v_y, v_{\zeta_0}\rangle$ in the way shown in 
figure \ref{fig:StartingTriang} in terms of cone diagrams. The alternative presentation in terms of fiber face diagrams is shown in figure \ref{fig:StartingFibFac}.

\begin{figure}
\begin{center}
\includegraphics[width=6cm]{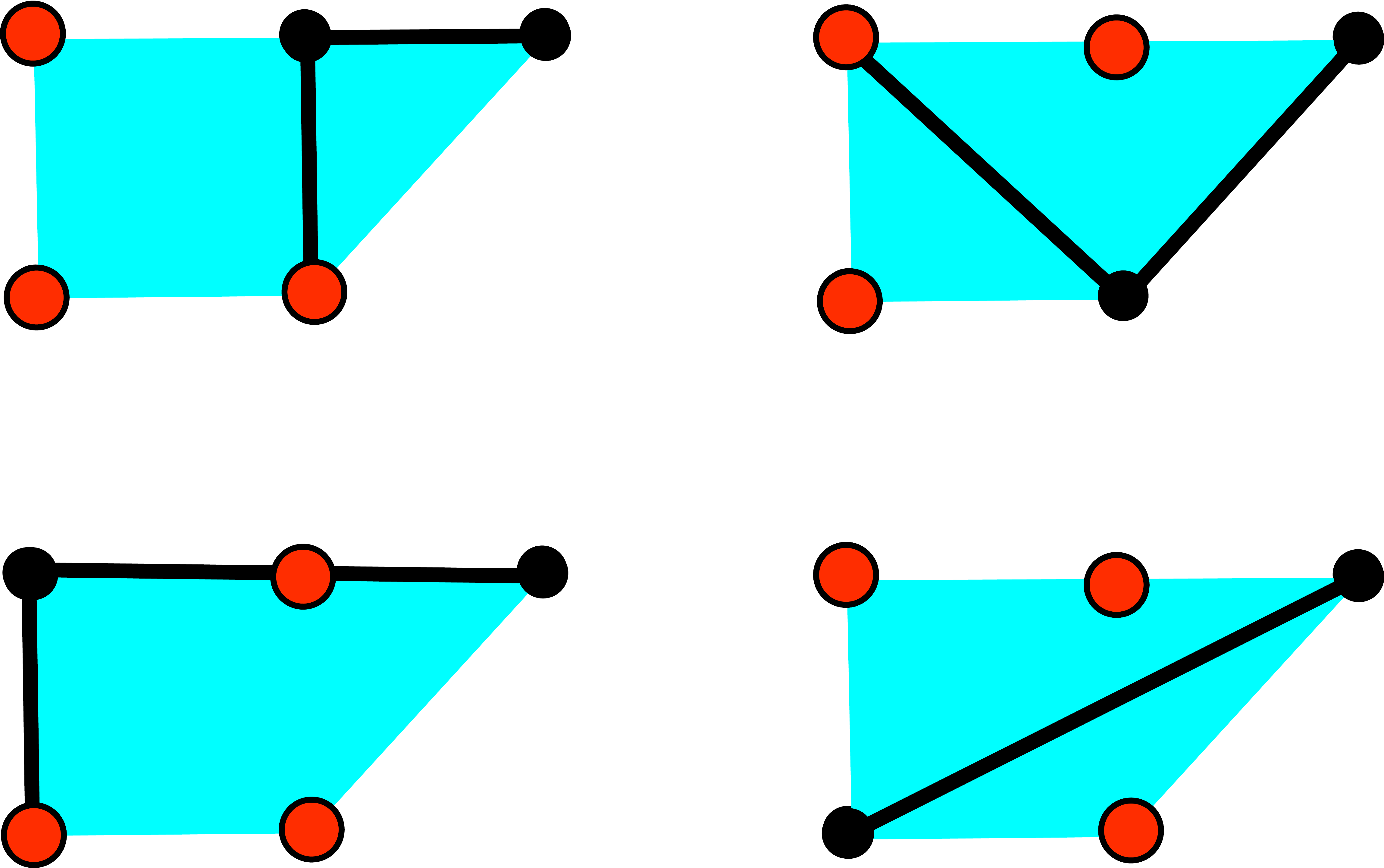} 
\end{center}
\caption{The fiber face presentation of the cone diagrams shown of figure \ref{fig:StartingTriang}.
The four first possibilities $\hbox{Res}_{\zeta_1}$, $\hbox{Res}_{\zeta_2}$, $\hbox{Res}_{\hat\zeta_2}$, and $\hbox{Res}_{\hat\zeta_1}$ in anti-clockwise ordering, respectively, for weighted blowups, shown in terms of  top cone diagrams in figure \ref{fig:torictop}. 
For each case, we show all of the cones introduced in the blow up in terms of black lines. The points used in the triangulation are marked in black. 
The points marked in red can still be used for further blowups 
and are not used in the respective fans.\label{fig:StartingFibFac}}
\end{figure}


\subsection{Fiber Faces and weighted blowups for Box Graphs}

We are now in the position to determine an explicit weighted blowup for each box graph in the network of small resolutions (or Coulomb phase analysis) \cite{Hayashi:2013lra, Hayashi:2014kca}, detailed in section \ref{sec:Box} and shown in figure \ref{fig:SU5AF}. 
The only exception to this is the graph corresponding to (11, IV) (and by reversing the order of the simple roots (6,I)), which we will discuss in the next section. This analysis provides a global construction of each box graph for $\mathfrak{su}(5)$ with ${\bf 5}$ and ${\bf 10}$ matter, confirming the flops performed in patches in \cite{Hayashi:2013lra}. The main advantage of the present approach is that it will have a natural generalization \cite{ABSSN}.

\begin{figure}
\centering
\includegraphics[width=14cm]{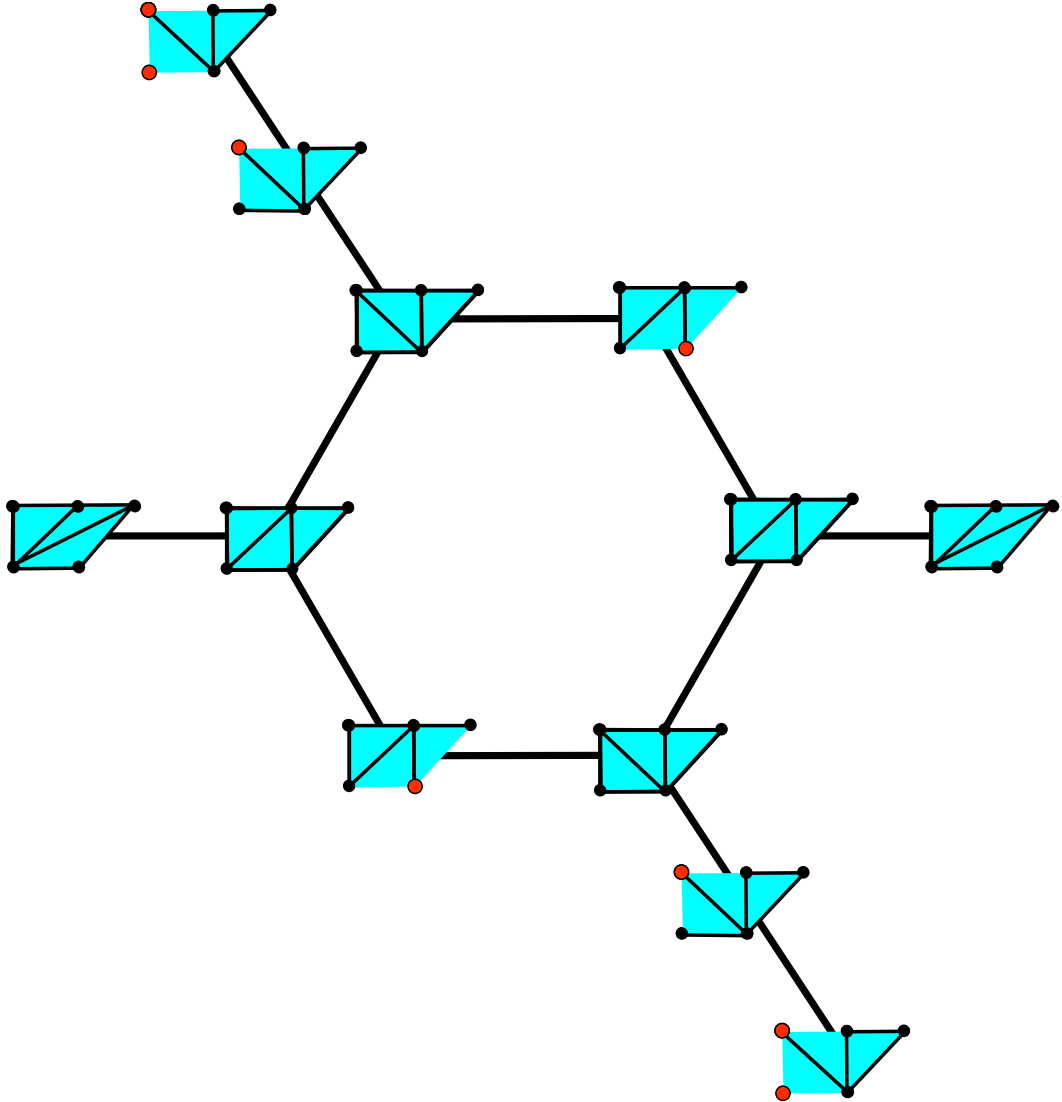}
\caption{Fiber face diagrams for the $SU(5)$ model with ${\bf 5}$ and ${\bf 10}$ representation, i.e. resolutions of the singular $I_5$ model with codimension 2 loci corresponding to $I_6$ and $I_1^*$ fibers.
Note that the codimension three fibers for these models are not necessarily of Kodaira type but monodromy reduced \cite{Hayashi:2014kca}. Each fiber face appears twice, and they only differ by reordering of the simple roots, i.e. on the right hand half of the network, the $\alpha_i$ are associted to the $\zeta_j, \hat{\zeta}_k$ with orientation that is clockwise, in the other half anti-clockwise. 
The fiber face diagram encodes the explicit weighted blowup sequence for a resolution. The lines connecting different fiber faces correspond to flops, so that this image corresponds to the box graph diagram figure \ref{fig:SU5AF}.\label{fig:SU5AFRes}}
\end{figure}

As explained in Section \ref{eq:ptforbus}, weighted crepant blowups of \eqref{TateSU5} can be found by successive refinements of 
the top cone $\langle v_x, v_y, v_{\zeta_0}\rangle$, using the four vectors \eqref{eq:viI5}. 
The sequence of blowups is determined from the fiber face diagram, which captures the essential information for the singularity resolution of the triangulation of the top cone $\langle v_x, v_y, v_{\zeta_0} \rangle$. 

The complete set of fiber faces and the network of flops among them is shown in figure \ref{fig:SU5AFRes}.
There are two situations that can arise:
\begin{itemize}
\item Standard toric resolutions correspond to {\it finely triangulated} fiber faces, where all points are black and connected by black lines,  i.e. are used in the triangulation. 
\item {\it Partially triangulated} fiber faces which contain red nodes correspond to partial toric resolutions, where the vertices corresponding to the red points have not been used in the triangulation. These are further resolved by algebraic blowups involving sections, that are not points in the triangulation, as we shall discuss momentarily. As discussed in Section \ref{sect:toricbu}, such phases may be realized as complete intersections of codimension two in toric varieties.
\end{itemize}
We now discuss these two situation in turn, highlighting the simplicity and generalizability of our approach:

If we successively introduce all four of the vectors $v_i$ in \eqref{eq:viI5}, we will obtain a resolution of \eqref{TateSU5}. As
the weight system of the corresponding toric variety is determined by the lattice vectors generating the one-dimensional cones alone,
we can write it down without specifying the order of resolutions we are performing:
\be\label{eq:I5topweightssystem}
\begin{array}{ccccccc}
 x & y & \zeta_0 & \zeta_1 & \zeta_2 & \hat{\zeta}_1 & \hat{\zeta}_2\\
 \hline
1 &1 &        1&       -1&       0 &    0          &  0  \\
2 &2 &        1&        0&       -1 &    0          &  0  \\
1 &2 &        1&       0&       0 &    -1          &  0  \\
2 &3 &        1&       0&       0 &    0          &  -1  \\
\end{array}\, .
\ee
This already determines the structure of the resolved Tate form
\be\label{eq:resolved}
y \left(\hat{\zeta}_1 \hat{\zeta}_2y + b_1 x + b_3 \hat{\zeta}_1 \zeta _1 \zeta _0^2\right)
=    \zeta _1 \zeta _2 \left(
\zeta _2\hat{\zeta}_2  x^3 
+b_2 \zeta _0 x^2
+ \zeta _1  \hat{\zeta}_1  b_4 \zeta _0^3x
+ \hat{\zeta}_1^2 \zeta _1^2 \zeta _0^5 b_6  
\right)
\ee

Different sequences of weighted blowups using all four of $v_{\zeta_1}, v_{\zeta_2}, v_{\hat{\zeta}_1}, v_{\hat{\zeta}_2}$ 
correspond to  fine 
triangulations of the point configuration shown in 
figure \ref{fig:torictop}, i.e. a triangulation, which uses all points.
Even though there are $4!=24$ different sequences of weighted
blowups, there are only $3$ inequivalent triangulations corresponding to three different phases. In order to find all
phases, we clearly have to use a more general strategy. 

The remaining cases correspond to partially triangulated fiber face diagrams, containing red nodes, i.e. points that are not used in the triangulation. 
Let us start with the observation that partial resolutions of \eqref{TateSU5} (or, equivalently, blowdowns of \eqref{eq:resolved})
can be described by simply deleting the absent coordinates from \eqref{eq:resolved} and \eqref{eq:I5topweightssystem}, and we also need to remove the $\C^*$-actions corresponding to these coordinates from \eqref{eq:I5topweightssystem}.
As we will see, in each of these cases, there is a small resolution which either turns the Tate model into a 
complete intersection, such as explained in section \ref{subsec:CI}, or into a determinantal variety, which will 
be discussed in section \ref{sec:DetBlowup}.

The fiber face diagrams of section \ref{sec:TCFF} are particularly well suited for the description of such resolutions, which go beyond the finely triangulated diagrams corresponding to standard toric resolutions. 

To show that the partially triangulated fiber face resolutions admit a description in terms of complete intersections, we write the resolved Tate model in two particularly interesting factored ways:
\be\label{yyhatP}
y \hat{y} = \zeta_1 \zeta_2 P \, ,
\ee
as well as in the alternatively factored form
\be\label{xWS}
x W = \hat{\zeta}_1 S \,.
\ee
Here, we have introduced the notation
\be
\ba\label{eq:defsPWSyhat}
\hat{y} &= \hat{\zeta}_1 \hat{\zeta}_2y + b_1 x + b_3 \hat{\zeta}_1 \zeta _1 \zeta _0^2\cr
P &= \zeta _2\hat{\zeta}_2  x^3 
+b_2 \zeta _0 x^2
+ \zeta _1  \hat{\zeta}_1  b_4 \zeta _0^3x
+ \hat{\zeta}_1^2 \zeta _1^2 \zeta _0^5 b_6  \cr
W&= \zeta _1 \zeta _2 \left(b_4 \hat{\zeta}_1 \zeta _1 \zeta _0^3+b_2 \zeta _0
   x+\hat{\zeta} _2 \zeta _2 x^2\right)-b_1 y \cr
S& = -b_6 \hat{\zeta} _1 \zeta _1^3 \zeta _2 \zeta _0^5+b_3 \zeta _1 \zeta
   _0^2 y+\hat{\zeta} _2 y^2\,.
\ea
\ee
It is clear now that any resolution sequence starting with $\hbox{Res}_{\zeta_1}$, $\hbox{Res}_{\zeta_2}$ or $\hbox{Res}_{\hat{\zeta_1}}$ has a partially resolved form given by either (\ref{yyhatP}) (with either $\zeta_2$ or $\zeta_1$ set to zero) or (\ref{xWS}). In all these cases, the equation takes the form of a conifold, and thus there is an alternative resolution sequence, which involves either $\hat{y}$ and $P$, or $W$ and $S$, which is not a toric resolution. The resolutions of this type correspond to fiber face diagrams which contain red nodes (i.e. resolutions where some elementary vertices are not used in the triangulation process). 

We will detail this process and the correspondence to the box graphs in the following for $SU(5)$, and in \cite{ABSSN} in general. The summary of the results for $SU(5)$ can be found in table \ref{tab:THETABLE}, which shows triplets of box graphs, fiber faces and algebraic resolutions. Note that by reversal of the ordering of the assignment of the exceptional sections to the simple roots each resolution corresponds to two box graphs, as detailed in the table. In the following we simply list only one half, associated to one ordering of the simple roots. 

\begin{table}\centering
$
\begin{array}{|c|c|c|c|c|}\hline
\# &\hbox{Box Graph} & \hbox{Fiber Face} & \hbox{Weighted Blowups}& \hbox{$\mathfrak{e}_6$ Fiber}\cr\hline
&&&&\cr
\hbox{(4,III)}&  \multirow{2}{*}{\includegraphics[width=2cm]{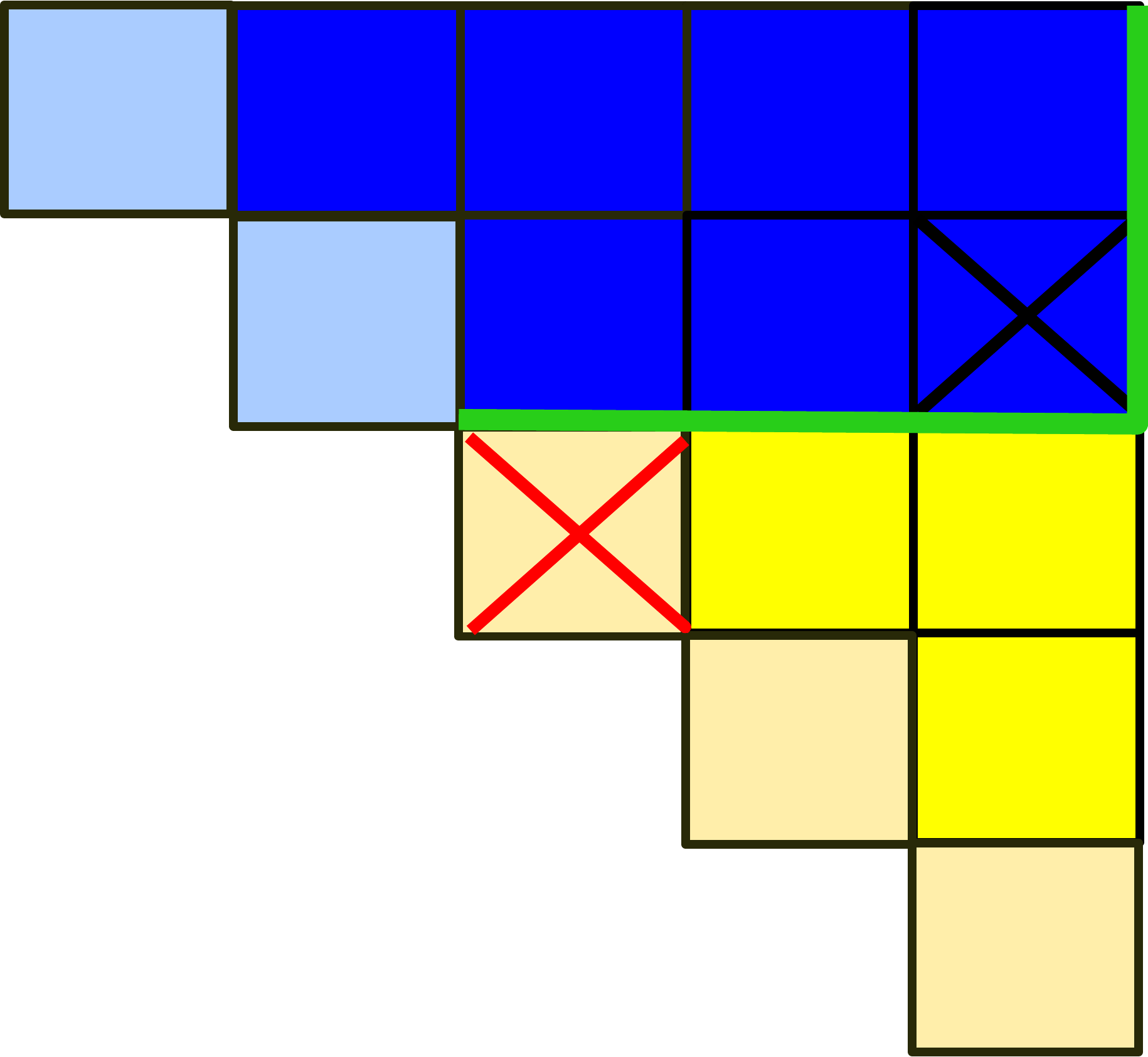}}
&\multirow{2}{*}{\includegraphics[width=2cm]{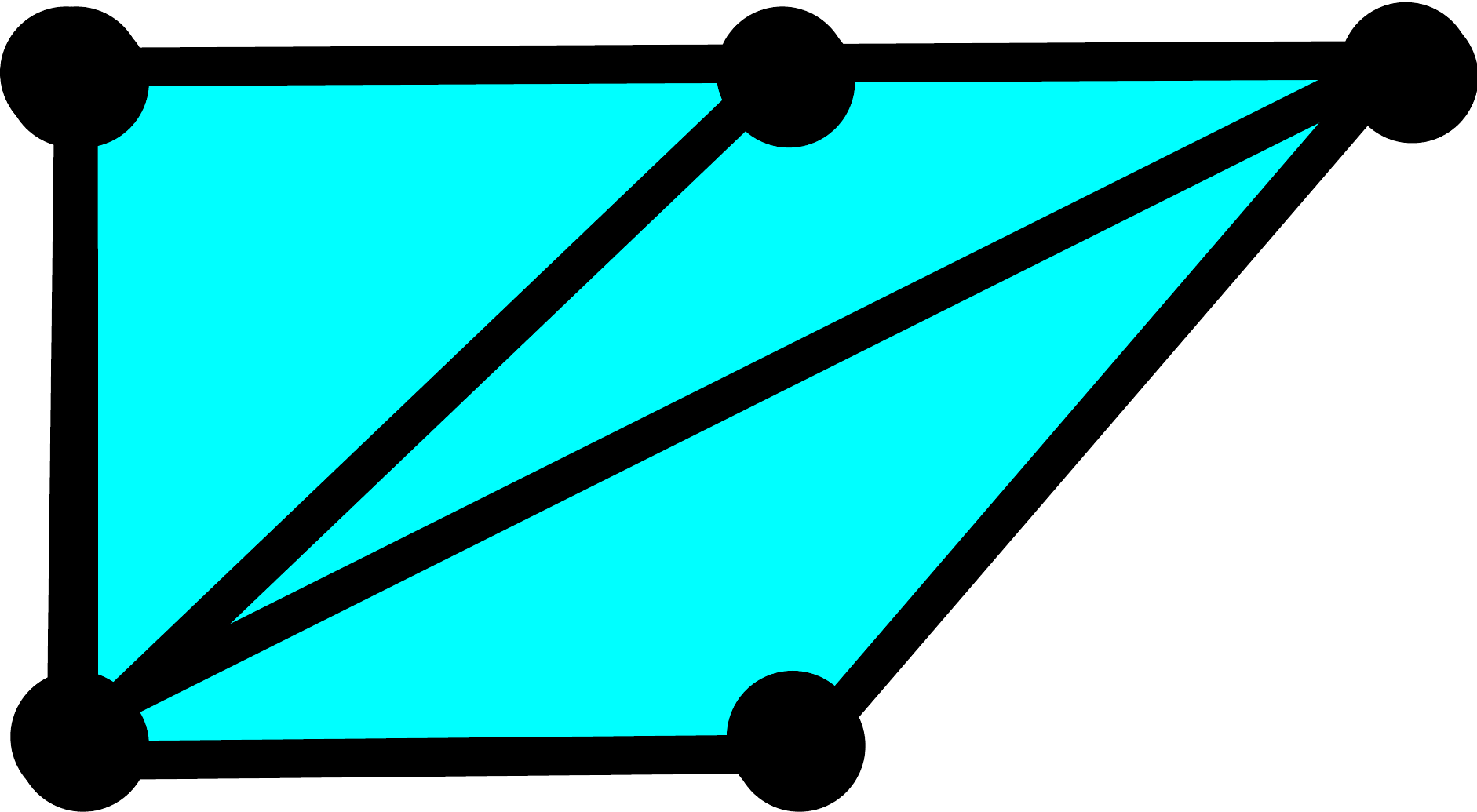}}
& ([{x,2}], [y, 3], [\zeta_0, 1] ; [\hat\zeta_2, 1]) 
&  \multirow{3}{*}{\includegraphics[width=3cm]{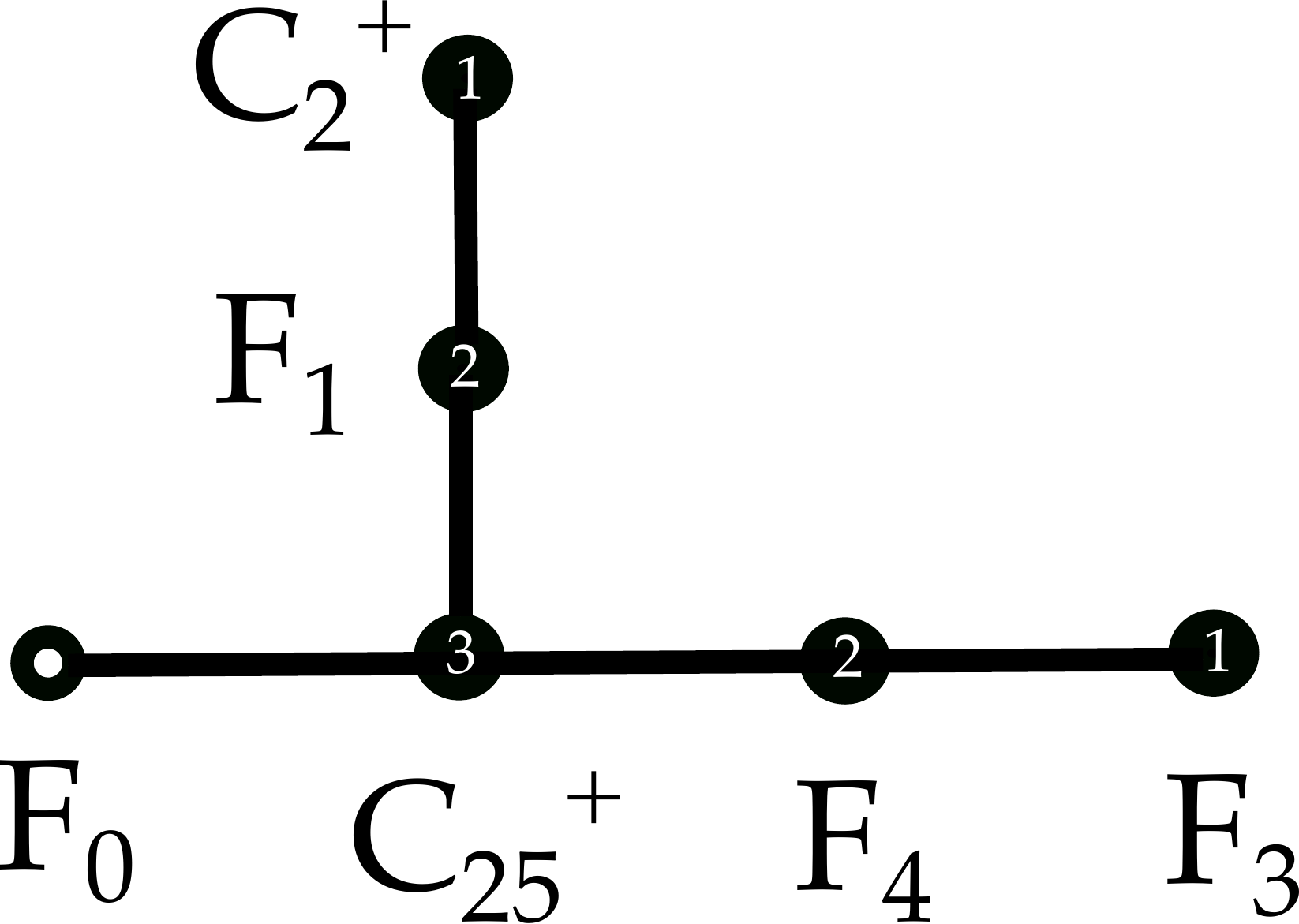}}
\cr
\hbox{Phase } 12&&& ([{y,1}], [\zeta_0, 1], [\hat\zeta_2, 1] ; [\hat\zeta_1, 2])&\cr
[\hbox{(13,II)}&&& ([{x,2}], [\zeta_0, 1], [\hat\zeta_2, 2] ; [\zeta_2, 3])&\cr
\hbox{Phase } 11]&&& ([\zeta_0, 1], [\zeta_2, 1] ; [\zeta_1, 2])
&\cr
&&&&\cr\hline
&&&&\cr
\hbox{(7,III)}&  \multirow{2}{*}{\includegraphics[width=2cm]{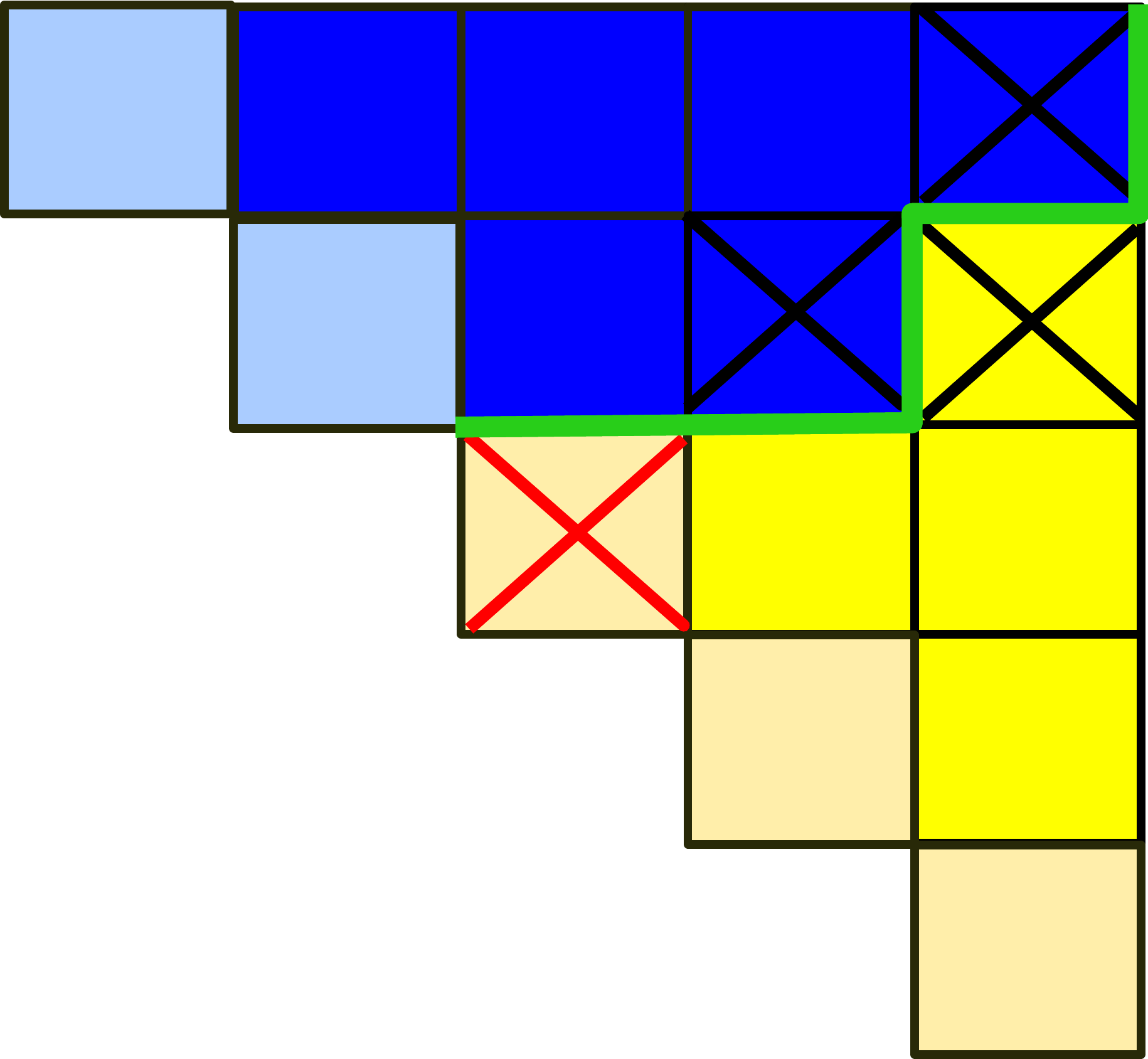}}
&\multirow{2}{*}{\includegraphics[width=2cm]{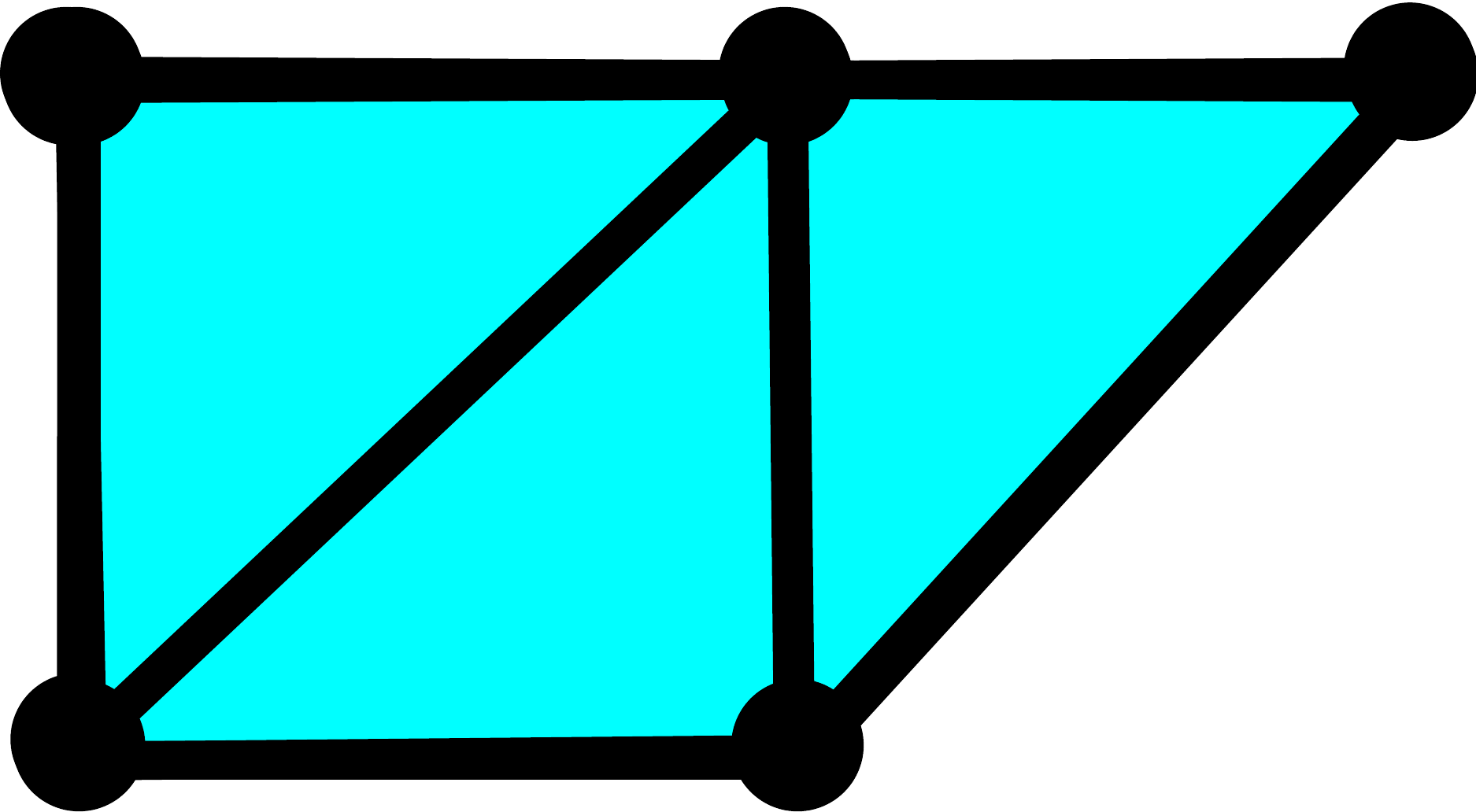}}
& (x, y, \zeta_0; \zeta_1) 
&  \multirow{3}{*}{\includegraphics[width=3cm]{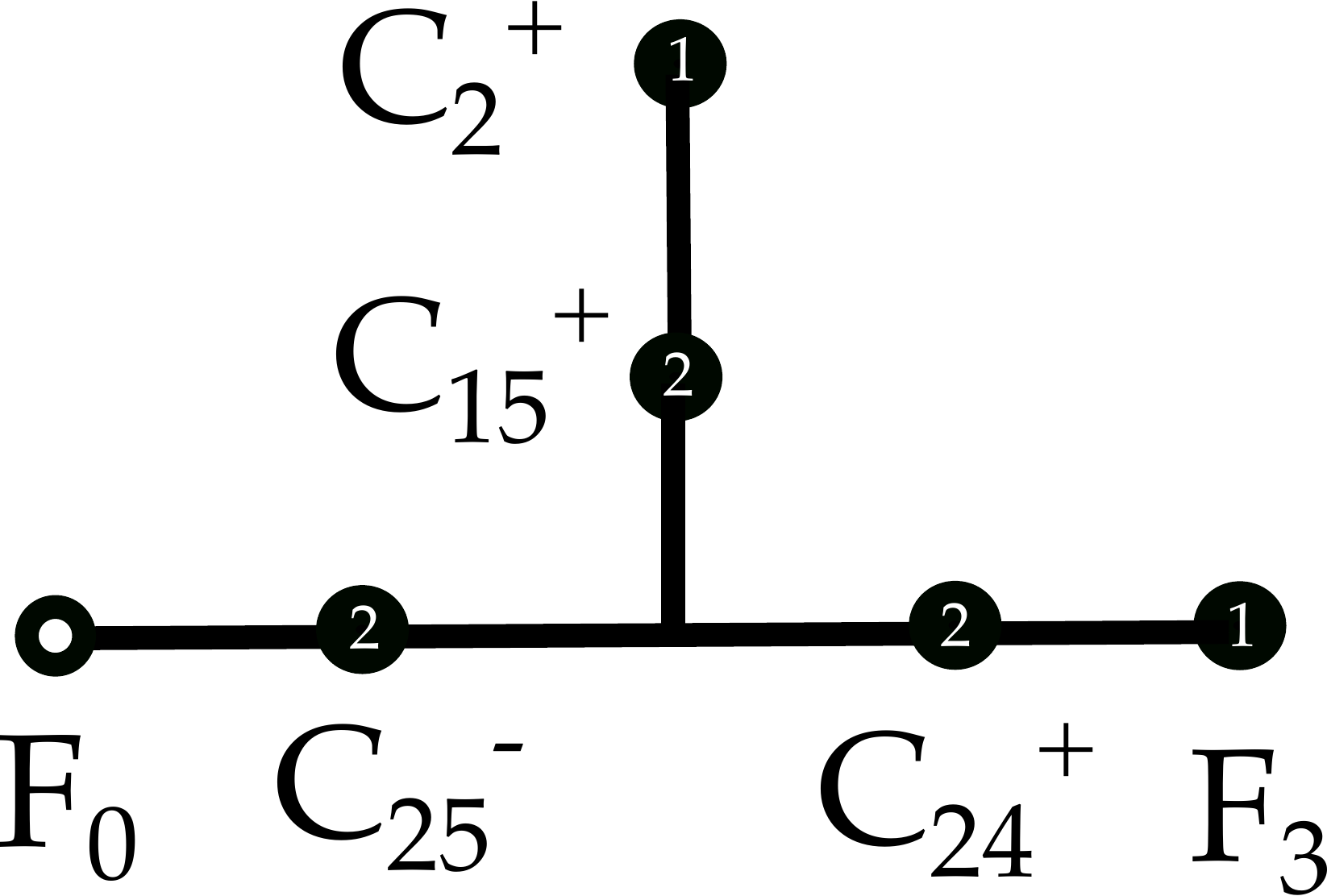}}\cr
\hbox{Phase }11&&& (x, y, \zeta_1; \zeta_2)&\cr
[\hbox{(10,II)}&&& (y, \zeta_1; \hat\zeta_1)&\cr
\hbox{Phase }2]&&&(y, \zeta_2; \hat\zeta_2)&\cr
&&&&\cr\hline
&&&&\cr
\hbox{(9,III)}&  \multirow{2}{*}{\includegraphics[width=2cm]{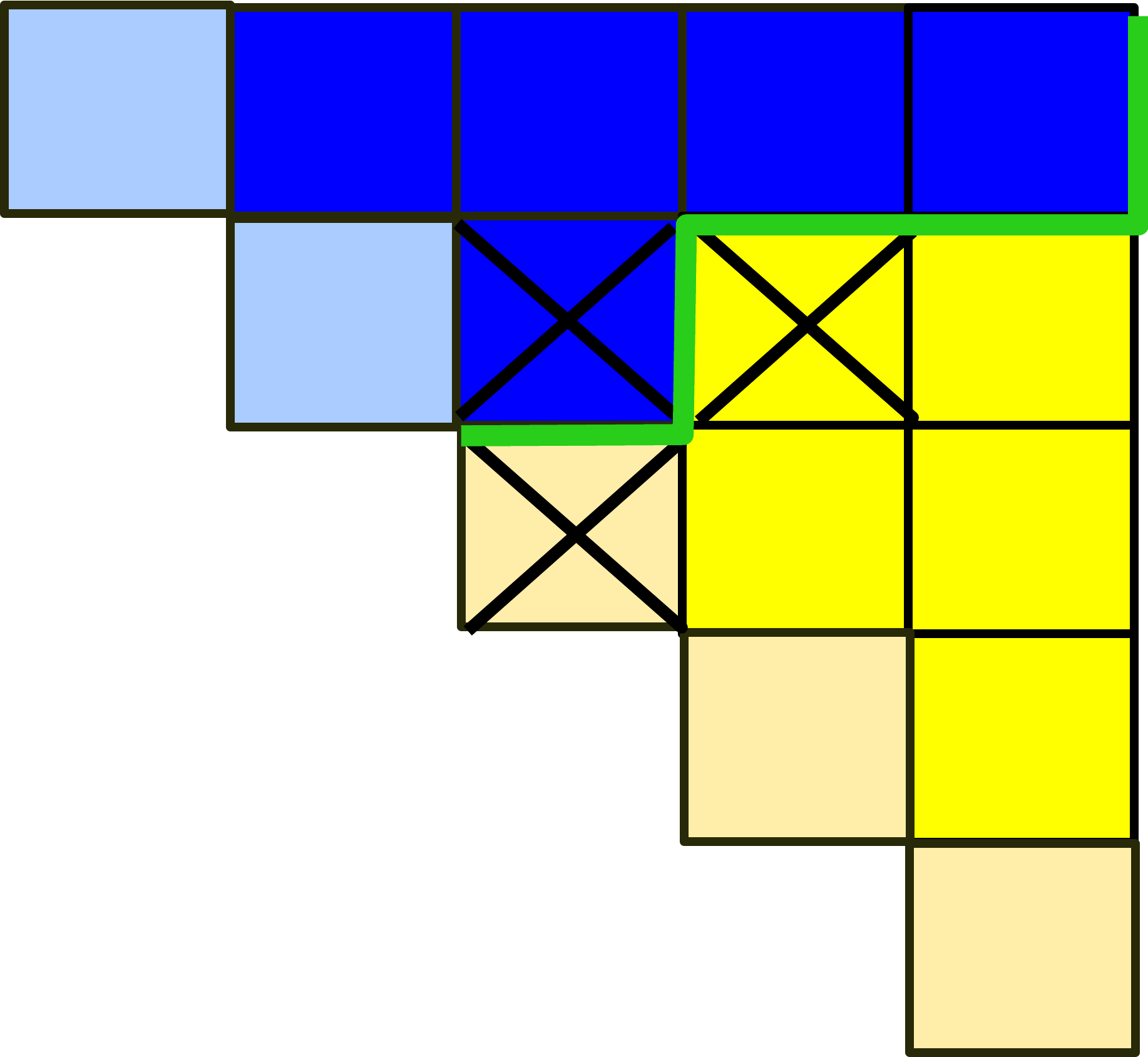}}
&\multirow{2}{*}{\includegraphics[width=2cm]{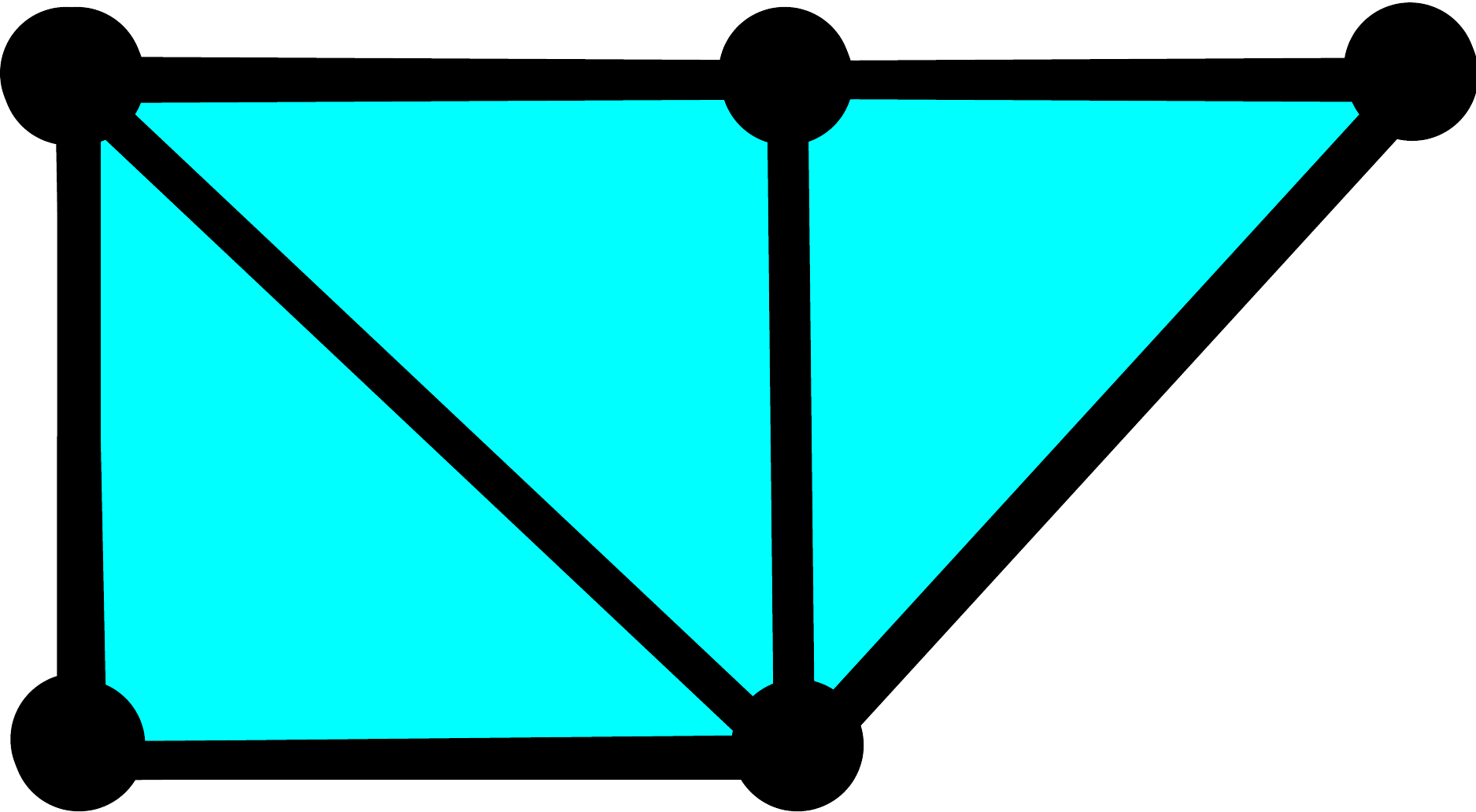}}
&([{x,1}], [y, 2], [\zeta_0, 1] ; [\hat\zeta_1, 1])
&  \multirow{3}{*}{\includegraphics[width=3.5cm]{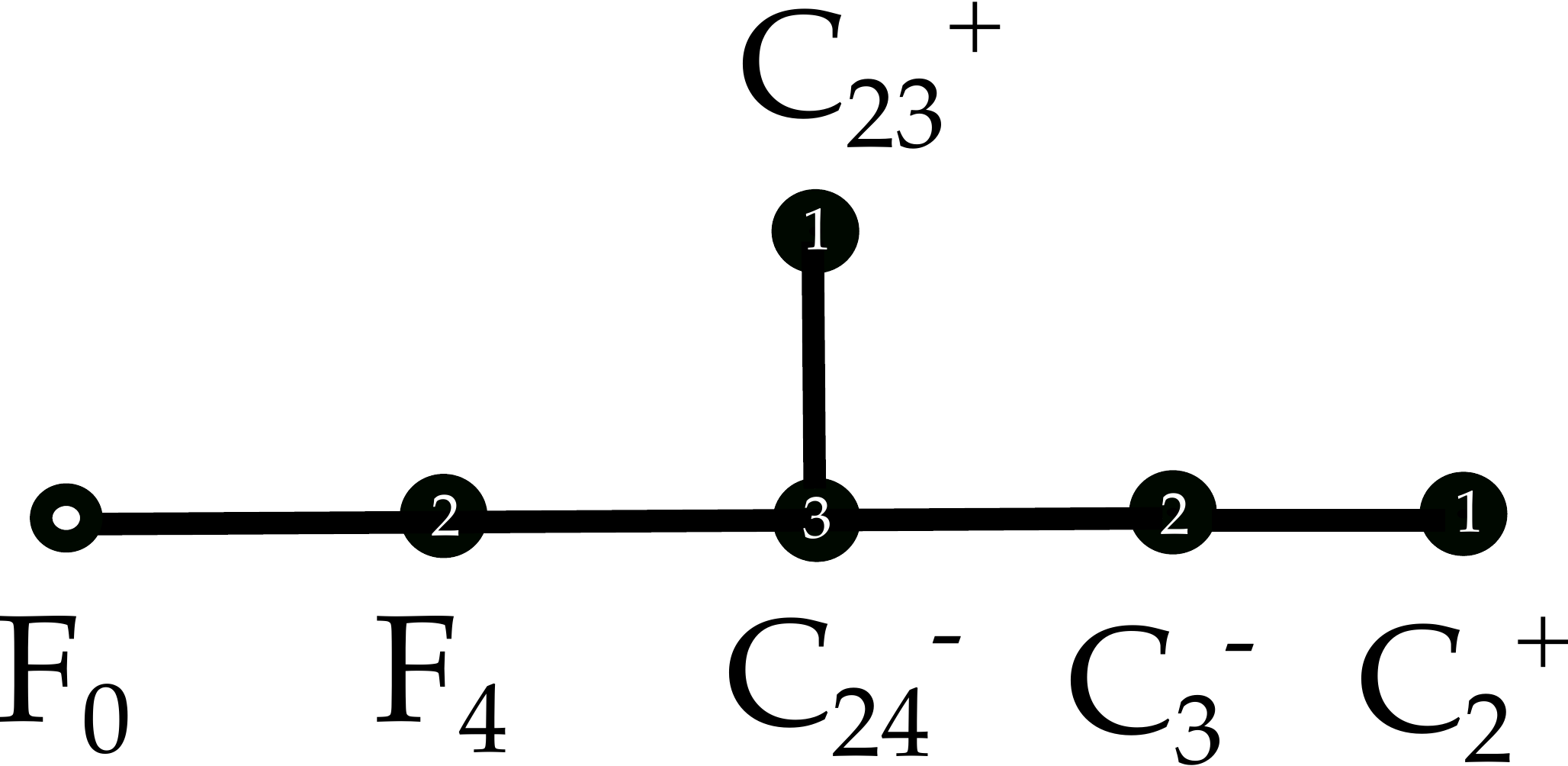}}\cr
\hbox{Phase }9&&&([{x,1}], [\zeta_0, 1], [\hat\zeta_1, 1] ; [\zeta_1, 2])&\cr
[\hbox{(8, II)}&&&([{x,1}], [y, 1], [\hat\zeta_1, 1] ; [\hat\zeta_2, 1])&\cr
\hbox{Phase }4]&&& ([x, 1],[\hat\zeta_1, 1]; [\zeta_2, 1] )&\cr
&&&&\cr\hline
&&&&\cr
\hbox{(9, II)}&  \multirow{2}{*}{\includegraphics[width=2cm]{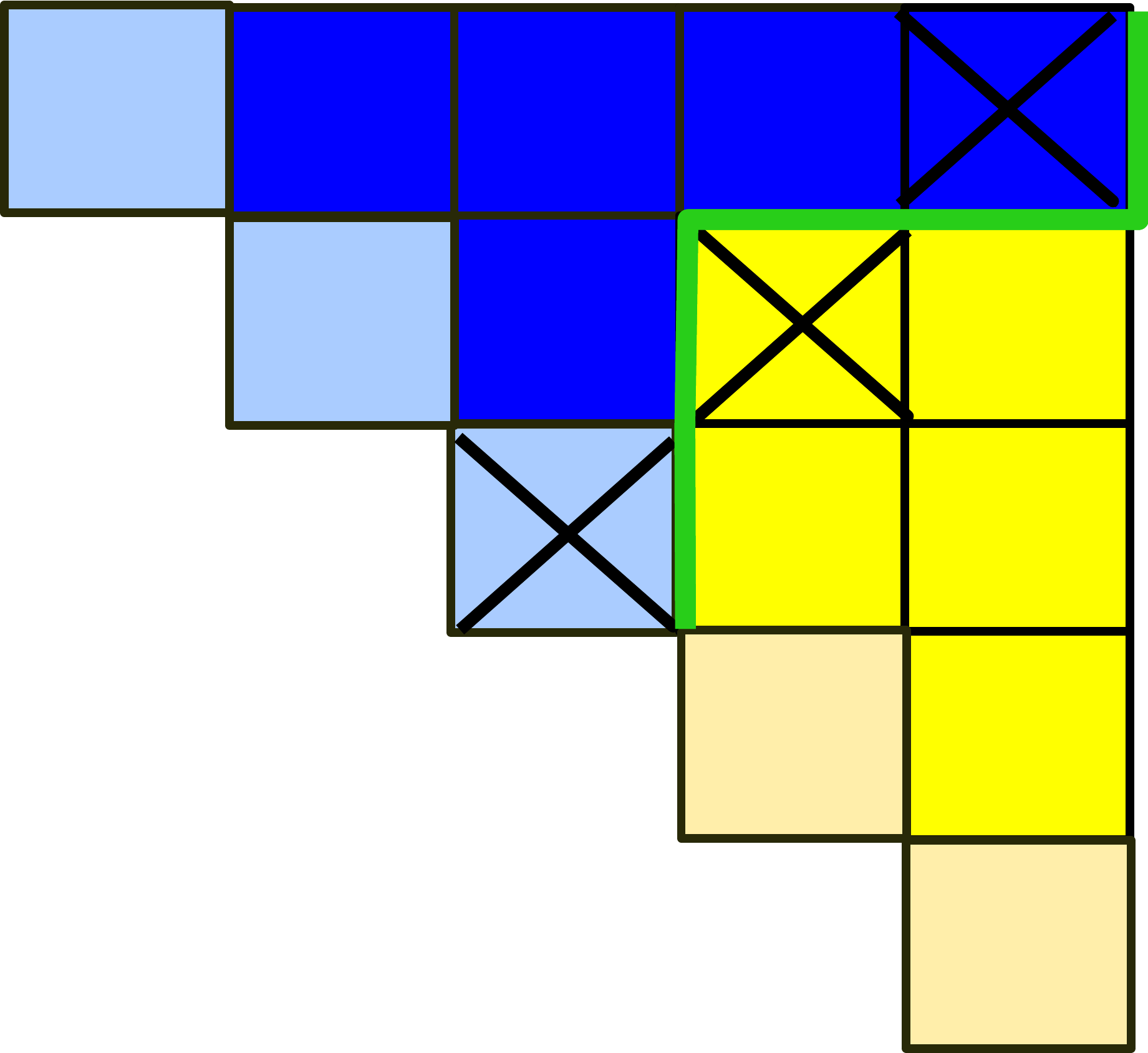}}
&\multirow{2}{*}{\includegraphics[width=2cm]{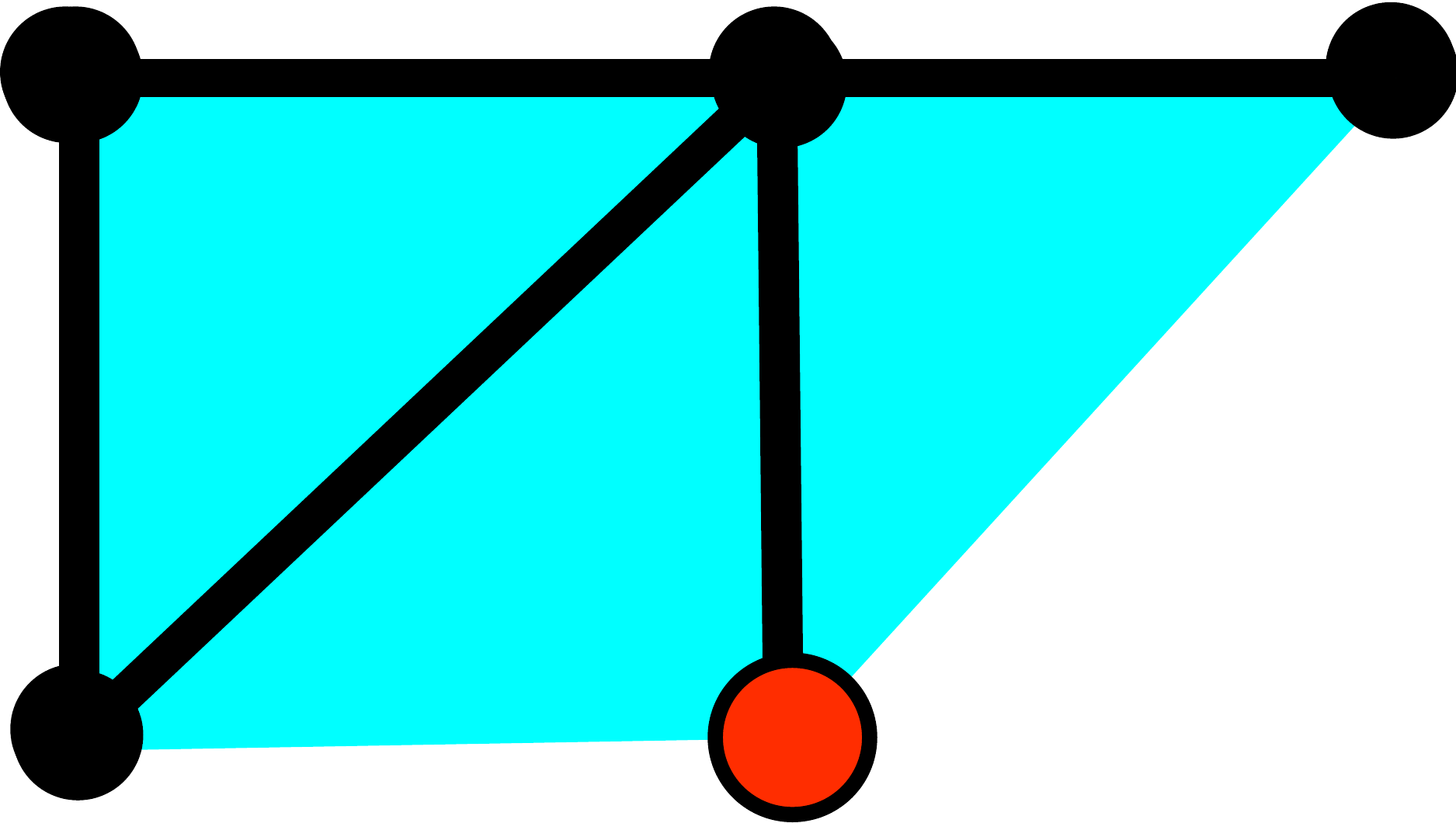}}
& (x, y, \zeta_0; \zeta_1) 
&  \multirow{3}{*}{\includegraphics[width=3.5cm]{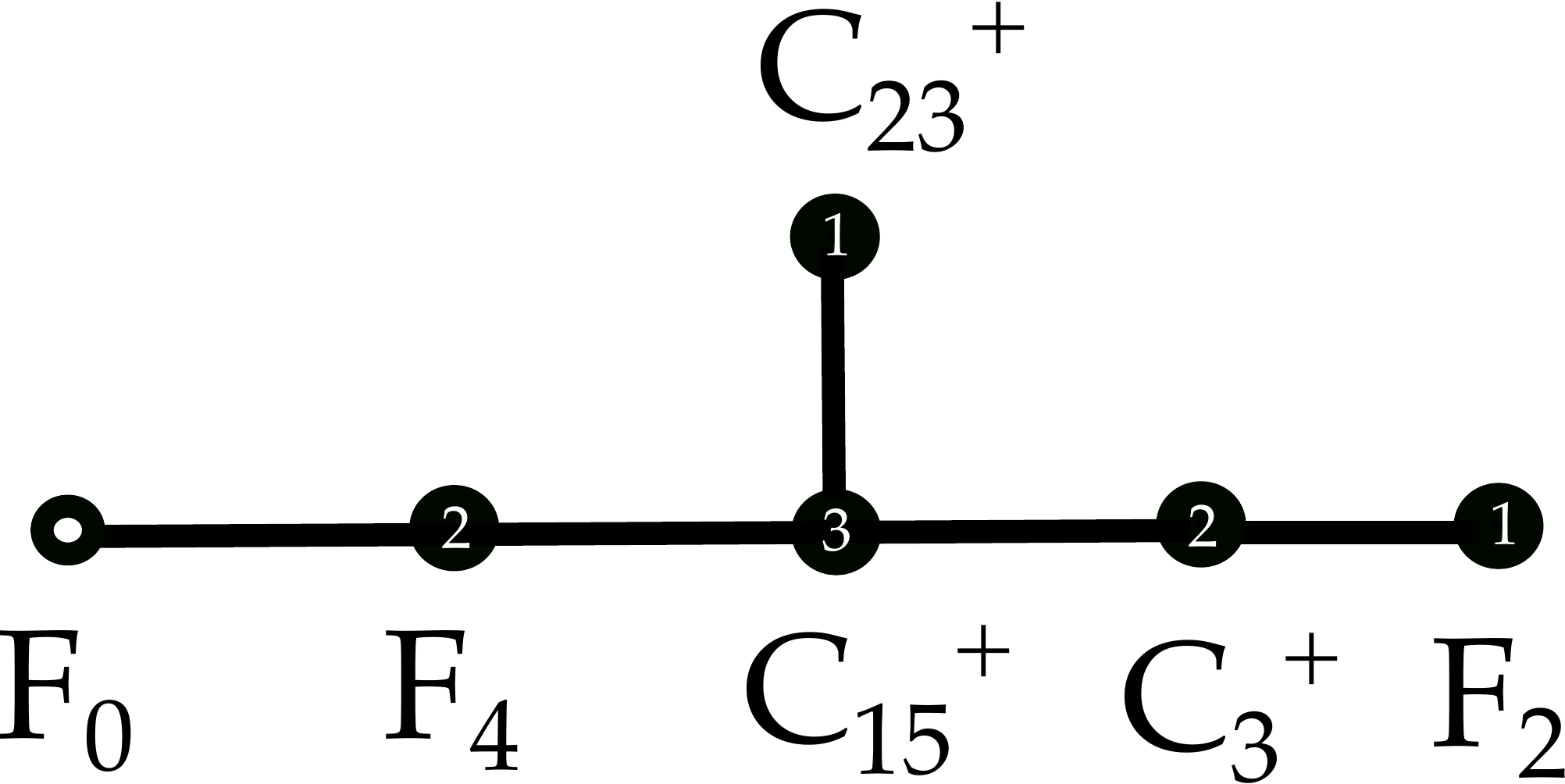}}
\cr
\hbox{Phase } 10&&& (x, y, \zeta_1; \zeta_2)&\cr
[\hbox{(8, III)}&&& (y, \zeta_1; \hat\zeta_1)&\cr
\hbox{Phase } 3]&&& (\hat{y}, P; \delta)&\cr
&&&&\cr\hline
&&&&\cr
\hbox{(11, III)}&  \multirow{2}{*}{\includegraphics[width=2cm]{SU5AF5.pdf}}
&\multirow{2}{*}{\includegraphics[width=2cm]{SU5AFT5.pdf}}
&([{x,1}], [y, 2], [\zeta_0, 1] ; [\hat\zeta_1, 1])
&  \multirow{3}{*}{\includegraphics[width=3.5cm]{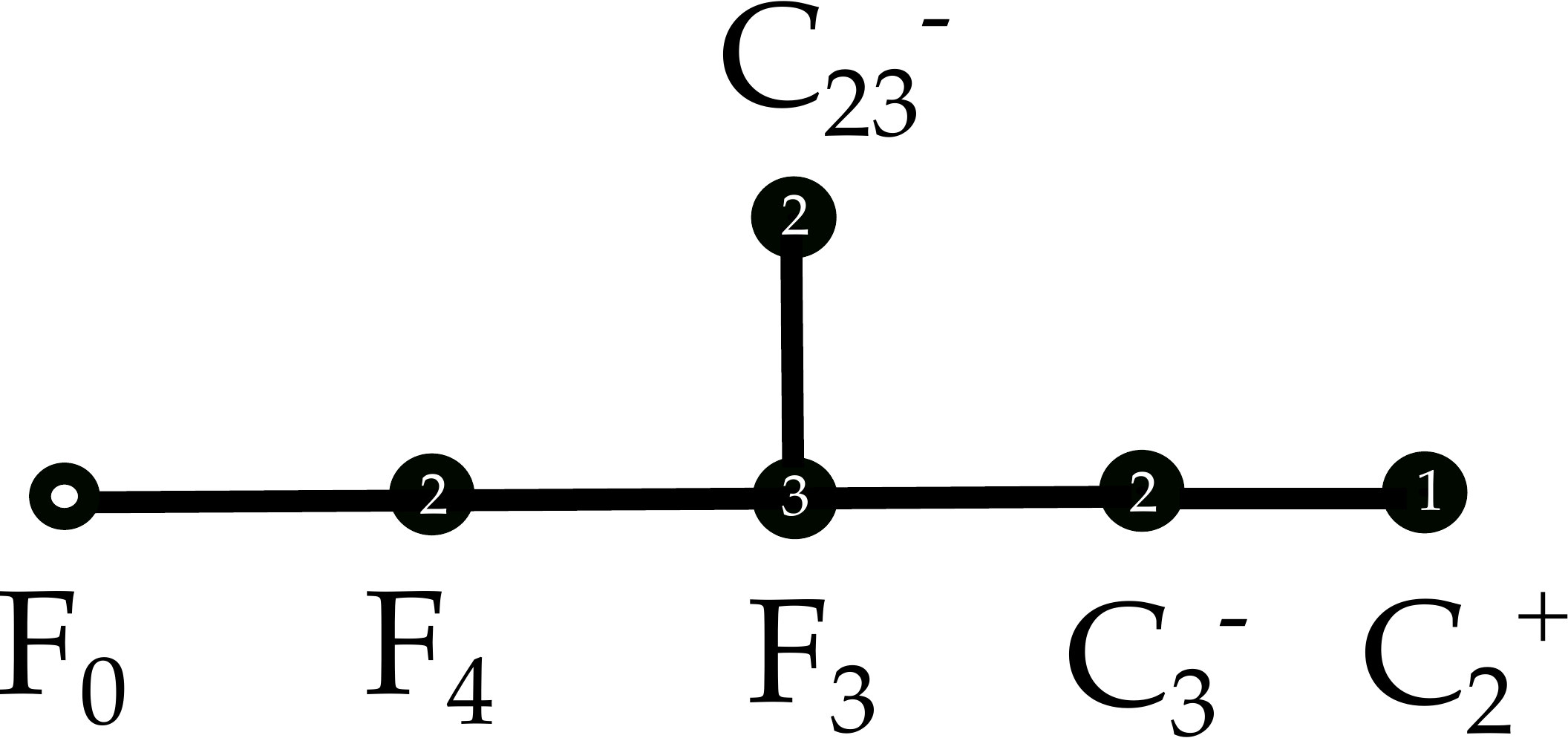}}
\cr
\hbox{Phase }8&&&([{x,1}], [\zeta_0, 1], [\hat\zeta_1, 1] ; [\zeta_1, 2])&\cr
[\hbox{(6, II)}&&&([{x,1}], [y, 1], [\hat\zeta_1, 1] ; [\hat\zeta_2, 1])&\cr
\hbox{Phase }5]&&& ([W, \hat{\zeta}_1; \delta])&\cr
&&&&\cr\hline
&&&&\cr
\hbox{(11, IV)}&  \multirow{2}{*}{\includegraphics[width=2cm]{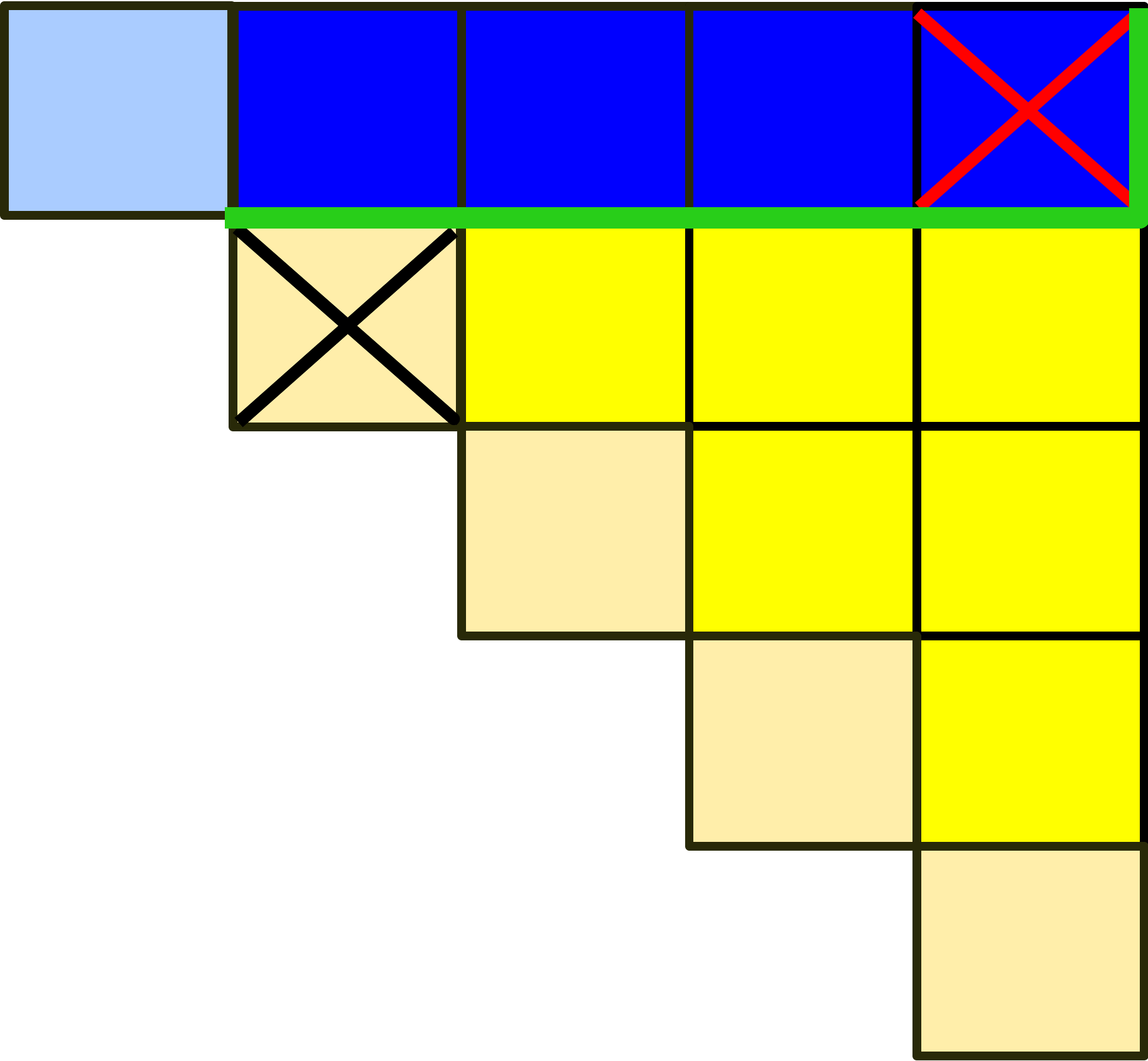}}
&\multirow{2}{*}{\includegraphics[width=2cm]{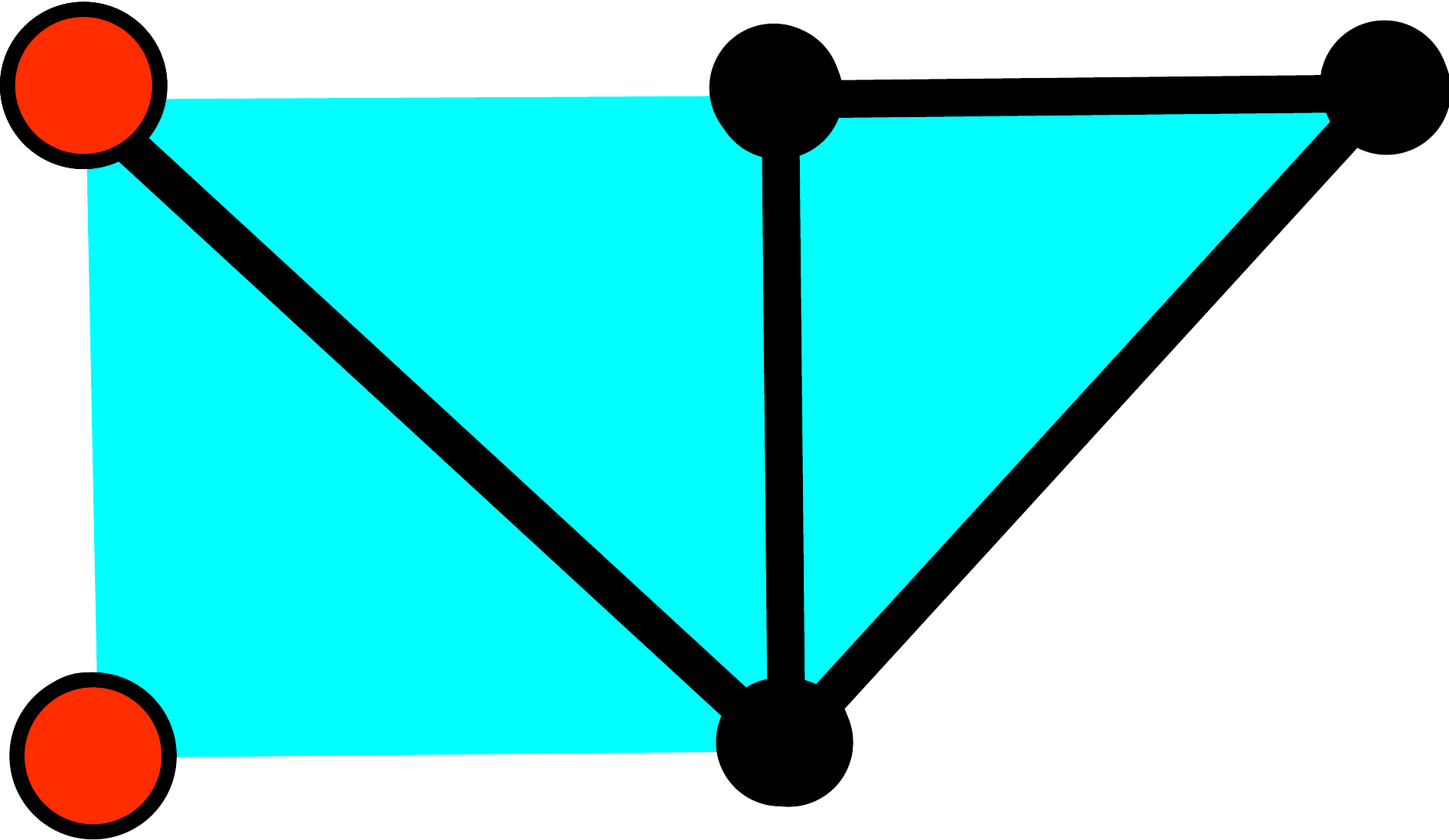}}
&([{x,1}], [y, 2], [\zeta_0, 1] ; [\hat\zeta_1, 1])
&  \multirow{3}{*}{\includegraphics[width=3.5cm]{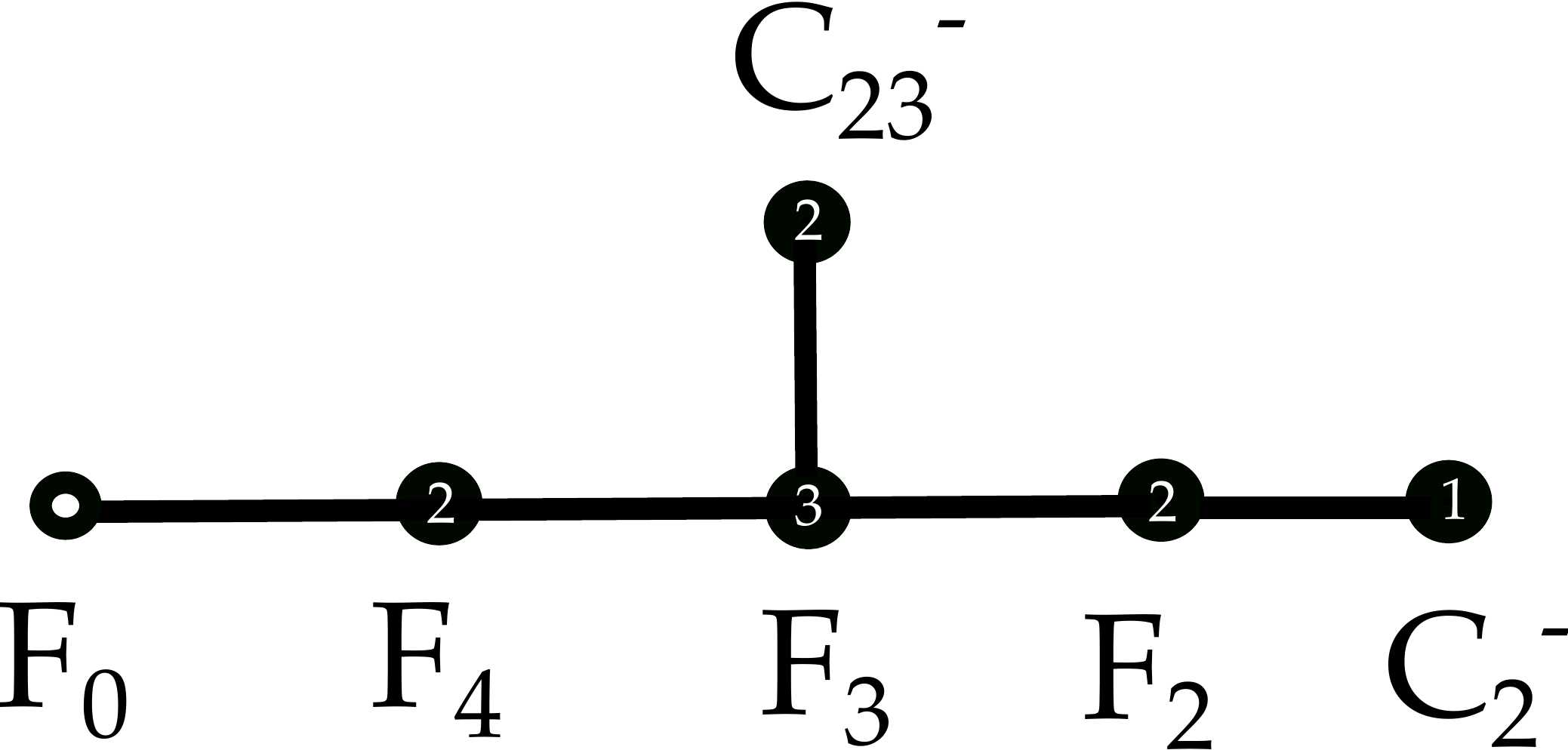}}
\cr
\hbox{Phase }7&&&([{x,1}], [\zeta_0, 1], [\hat\zeta_1, 1] ; [\zeta_1, 2])&\cr
[\hbox{(6, I)}&&&\hbox{Determinantal Blowup}&\cr
\hbox{Phase }6]&&&\hbox{Section \ref{sec:DetBlowup}}&\cr
&&&&\cr\hline
\end{array}
$
\caption{ Correspondence of box graphs, fiber faces, and algebraic resolutions for  $SU(5)$ with ${\bf 5}$ and ${\bf 10}$ representation, and codimension 3 monodromy reduced $\mathfrak{e}_6$ fibers in agreement with \cite{Hayashi:2014kca}. The labels for box graphs are as in (\ref{5Split}) and (\ref{10Split}). The Coulomb phase labels are as in \cite{Hayashi:2013lra}. In parenthesis we write the phases, which are obtained by the same resolution by choosing the inverted labeling of the roots of $\mathfrak{su}(5)$. 
The sections $\hat{y}, P, S,  W$ are defined in (\ref{eq:defsPWSyhat}). The first three resolutions are toric (realized as weighted blowups), the fourth standard algebraic, and the last is determinantal. 
 \label{tab:THETABLE}}
\end{table}

\subsubsection*{Box Graph (4, III)}
The corresponding fiber face diagram is  \includegraphics[width=1cm]{SU5AFT1.pdf}. 
Note that this is a fine triangulation, and thus corresponds to a toric resolution discussed in \cite{Krause:2011xj}. As we argued in general, these toric triangulations have an algebraic realization in terms of weighted blowups. We shall now present these here. 

The first step to reach \includegraphics[width=1cm]{SU5AFT1.pdf} is completely determined to be the starting resolution $\hbox{Res}_{\hat\zeta_1}$ in (\ref{StartRes})
\be
\hbox{Res}_{\hat\zeta_1}: \qquad ([{x,2}], [y, 3], [\zeta_0, 1] ; [\hat\zeta_2, 1]) \,,
\ee
which yields the corresponding triangulation in figure \ref{fig:StartingTriang}. 
This yields the fiber face  \includegraphics[width=1cm]{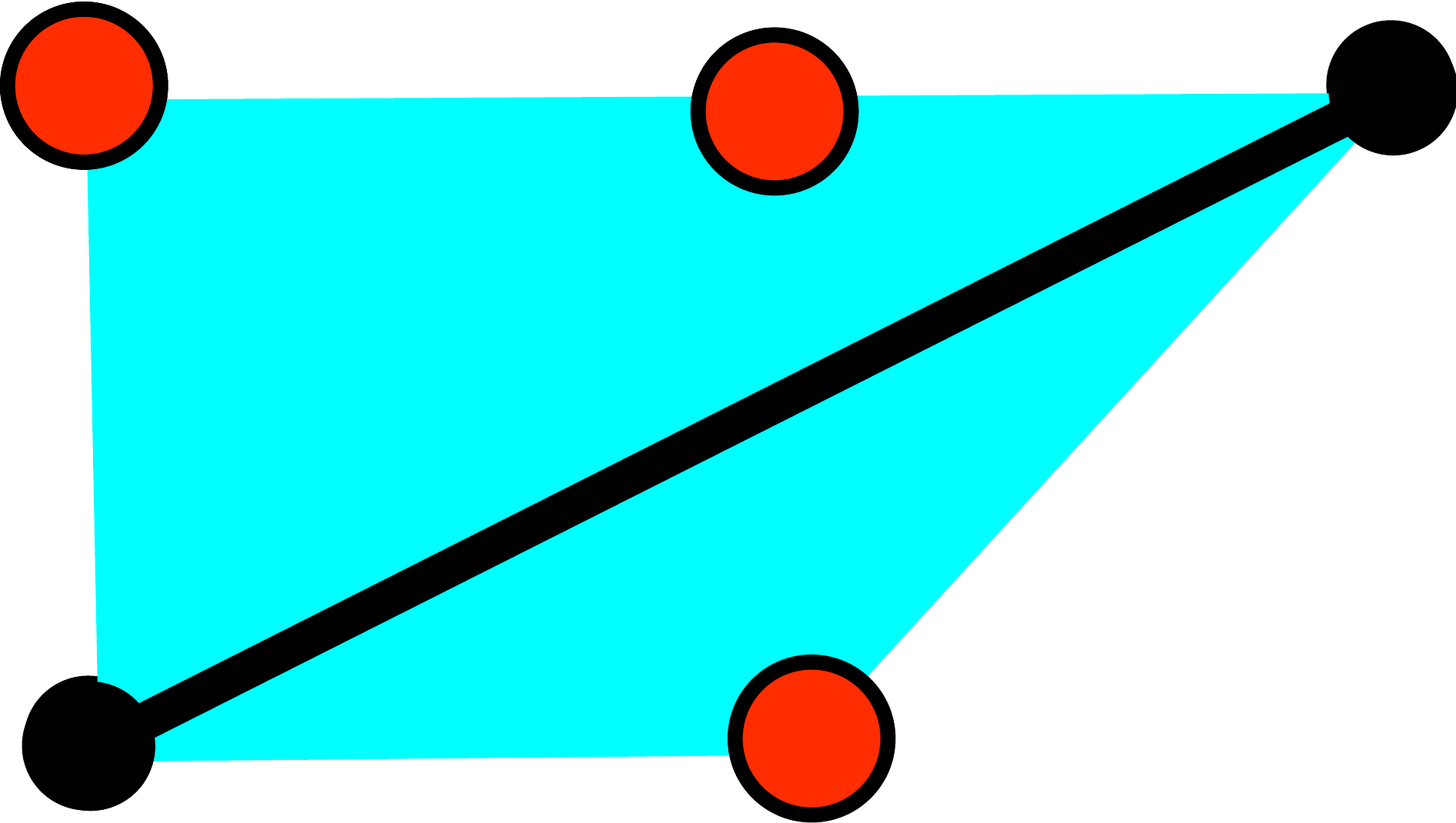}. 
In fact the following triangulations are completely determined and we arrive at the sequence
\be 
\includegraphics[width=11cm]{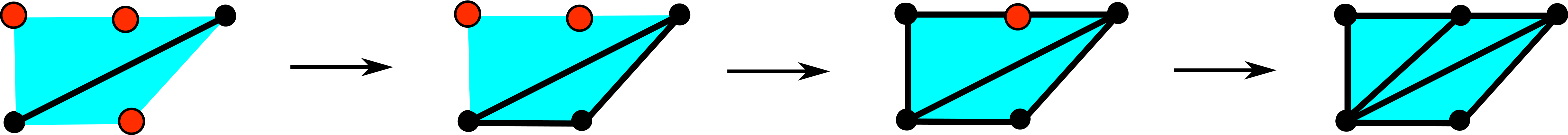}
\ee
The map between each of these fiber face diagrams is a triangulation of the top cone, and thereby a weighted blowup following our general discussion. E.g. the second step corresponds to subdividing the cone $\langle y, \zeta_0, \hat{\zeta}_1\rangle$ by $\hat{\zeta}_1$. From this we can determine the scalings (\ref{eq:scalingE}) and thereby the weights for the blowup to be 
$([{y,1}], [\zeta_0, 3], [\hat\zeta_2, 1] ; [\hat\zeta_1, 2])$. 

It remains to show that this reproduces the box graph splitting (4, III). We can either apply the weighted blowups, or use a slightly more elegant method, which will be proven in \cite{ABSSN}. Here we provide the explicit resolutions for reference in the appendix \ref{app:Blowups}. 


\subsubsection*{Box Graph (7, III) and (9, III) and (9, II)}

These are obtained as a standard resolution with unit weights: first we resolve in codimension one, which corresponds to introducing the subdivisions of the cone using $\zeta_1$ and $\zeta_2$:
\be
(x, y, \zeta_0; \zeta_1) \,,\qquad (x, y, \zeta_1; \zeta_2) \,.
\ee
We obtain the factored form (\ref{yyhatP}), with $\hat{\zeta}_i=1$ after these blowups. There are three distinct small resolutions of 
\be
y \hat{y} = \zeta_1 \zeta_2 P \,,
\ee
which correspond either to the fine triangulation \includegraphics[width=1cm]{SU5AFT2.pdf}, with blowups
\be
(y, \zeta_1; \hat\zeta_1)\,,\quad (y, \zeta_2; \hat\zeta_2)\,,
\ee
or the fine triangulation \includegraphics[width=1cm]{SU5AFT3.pdf}, which can be reached by 
\be
(y, \zeta_2; \hat\zeta_2)\,,\quad (y, \zeta_1; \hat\zeta_1)\,.
\ee
These two resolutions were studied algebraically in \cite{Lawrie:2012gg, Hayashi:2013lra} and from since these are fine triangulations and thus standard toric resolutions, they are exactly also those discussed in \cite{Krause:2011xj}. The two cases correspond to the phases (7, III) and (9, III) respectively. 

Finally, we  can use $P$ in the resolution, which implies that the Tate model becomes a complete intersection
\be
(y, \zeta_1; \hat\zeta_1)\,,\quad (\hat{y}, P; \delta) \,.
\ee
This corresponds to the triangulation \includegraphics[width=1cm]{SU5AFT4.pdf}, where the red node indicates that the node $\hat\zeta_2$ is not part of the triangulation, but the non-toric resolution with $(y, P;\delta)$ was applied. This resolution was studied from algebraic resolutions in \cite{MS, EY}, and corresponds to the box graph (9, II).

\subsubsection*{Box Graph (9, III) and (11, III)}

There is an alternative resolution that results in the fiber corresponding to the box graph (9,III), which is in fact more amenable to the flop from (9, III) to (11, III). This furthermore prepares the flop to (11, IV) which is the subject of the next section. 
The sequence of blowups is 
\be\label{FibFac789}
\includegraphics[width=12cm]{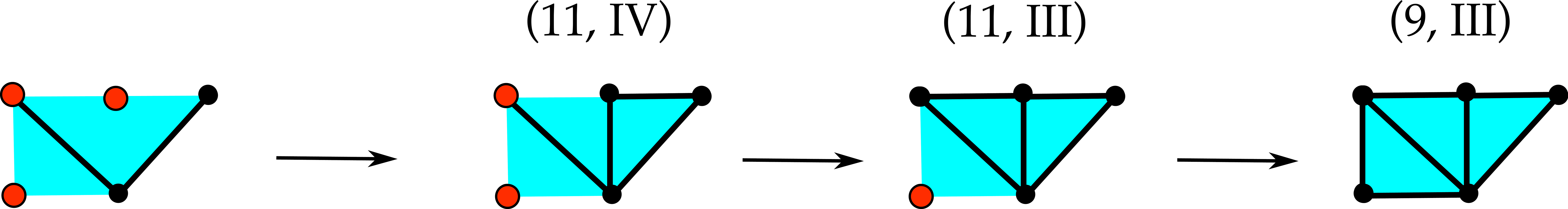}
\ee
Applying all of these results in the fine triangulation corresponding to (9, III), which we discussed already. Again each arrow corresponds to a weighted blowup, which we have listed in table \ref{tab:THETABLE}. Applying only the first three, results in a partial triangulation, which has the factored form (\ref{xWS}) with $\hat\zeta_2$ set to 1 (as we have not used this in the triangulation),
\be
x W = \hat{\zeta}_1 S \,,
\ee
We can now apply the blowup $(W, \hat{\zeta}_1; \delta)$, which is a algebraic blowup not realized purely in terms of homogeneous coordinates, with $W$, $S$ as in (\ref{eq:defsPWSyhat}). This yields the phase characterized by the box graph (11, III). The details of this resolution are provided for the reader's convenience in appendix \ref{app:Blowups}.
 The last case will be discussed in the next section and is also based on the above equation (\ref{FibFac789}). 

\begin{figure}
\centering
\includegraphics[width=5cm]{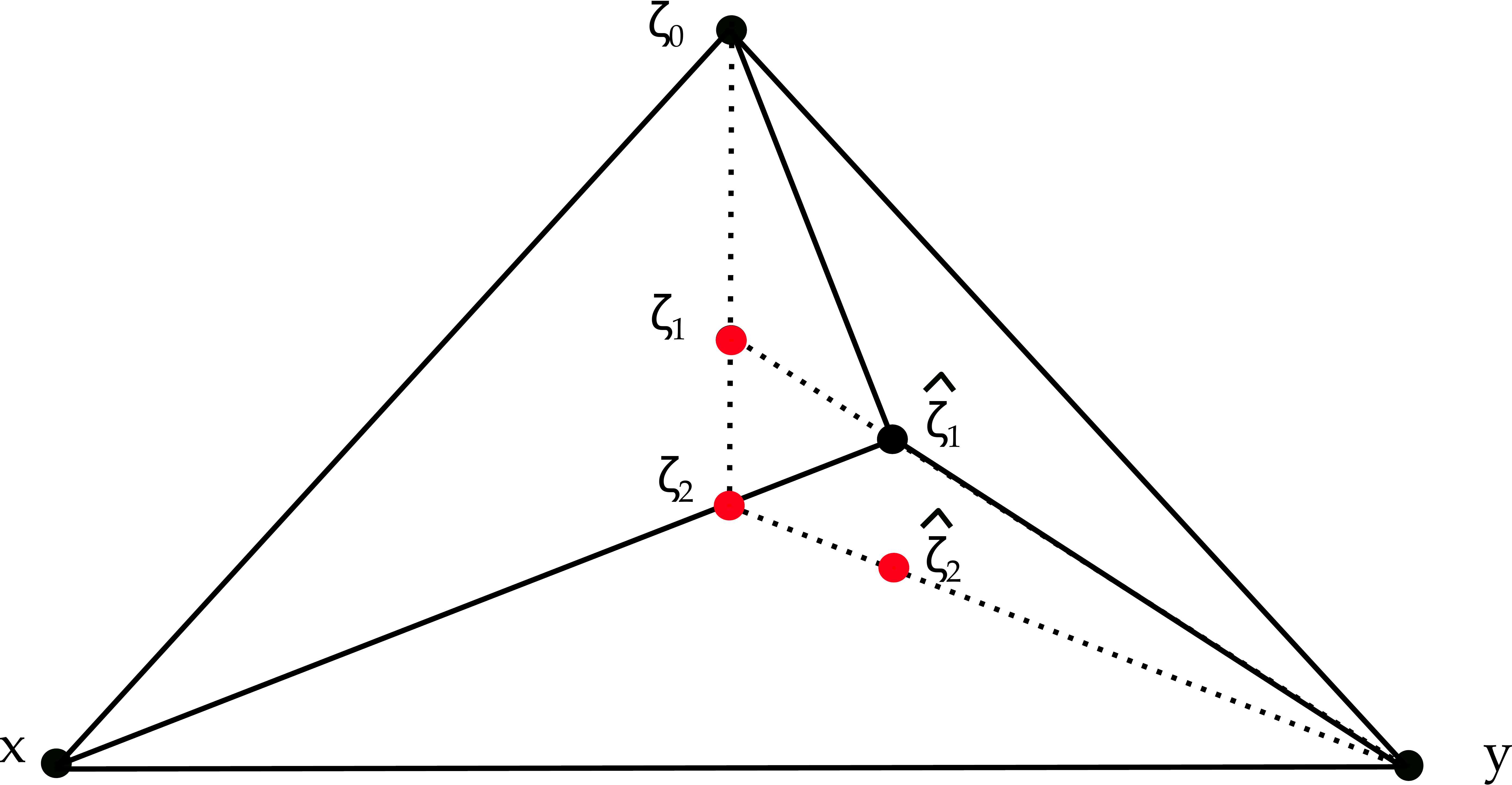}
\includegraphics[width=5cm]{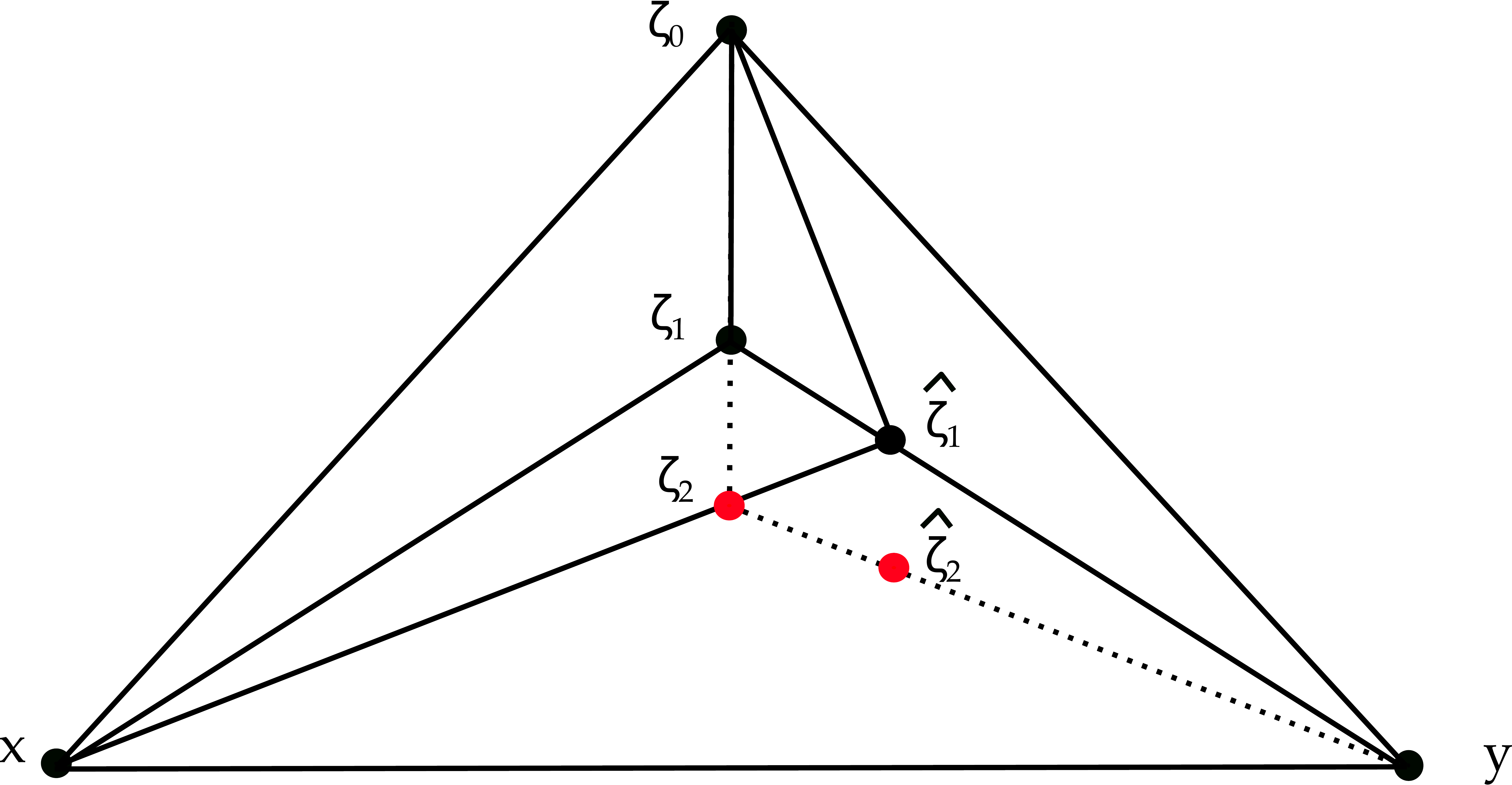}\\\smallskip
\includegraphics[width=5cm]{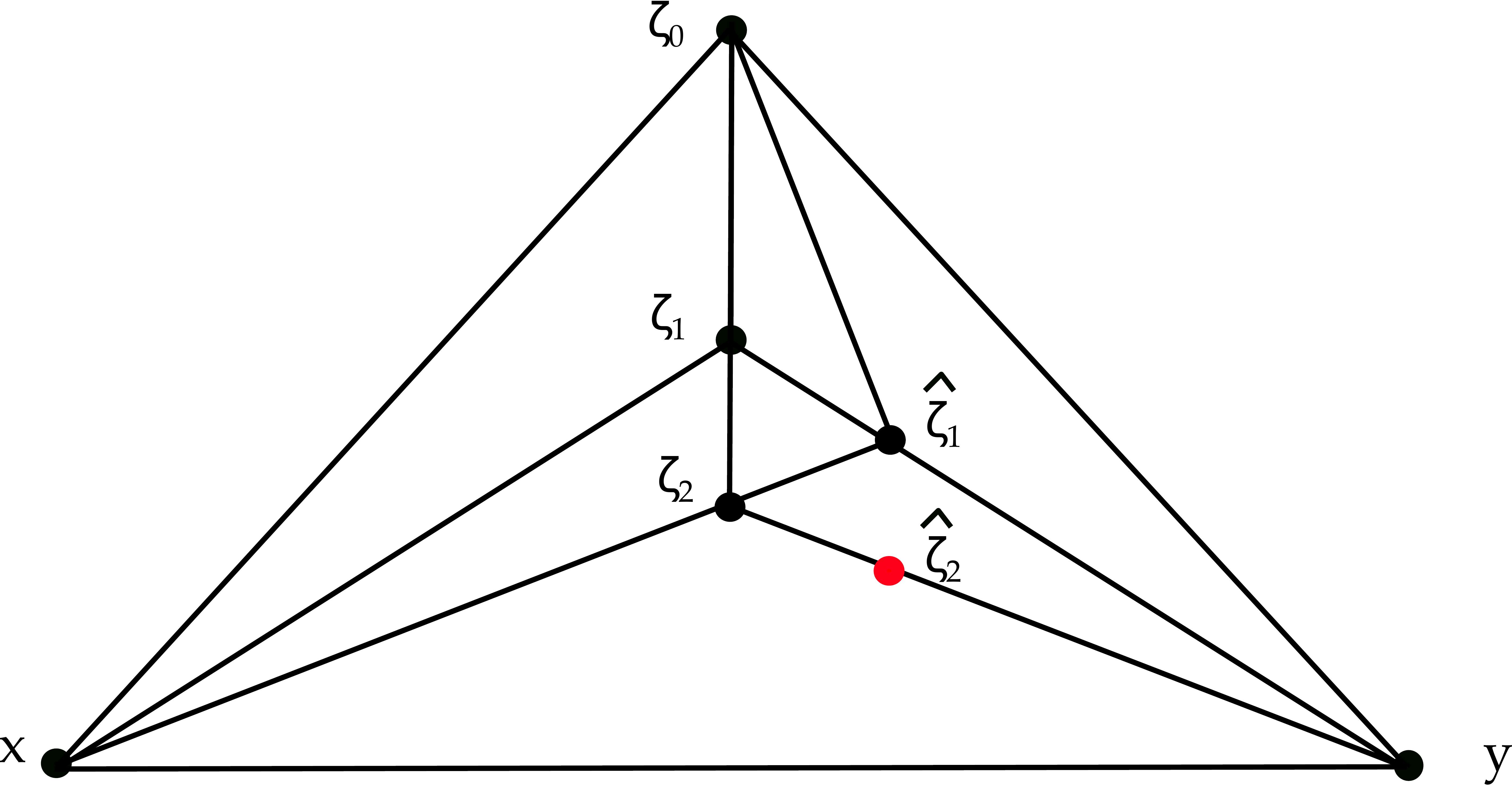}
\includegraphics[width=5cm]{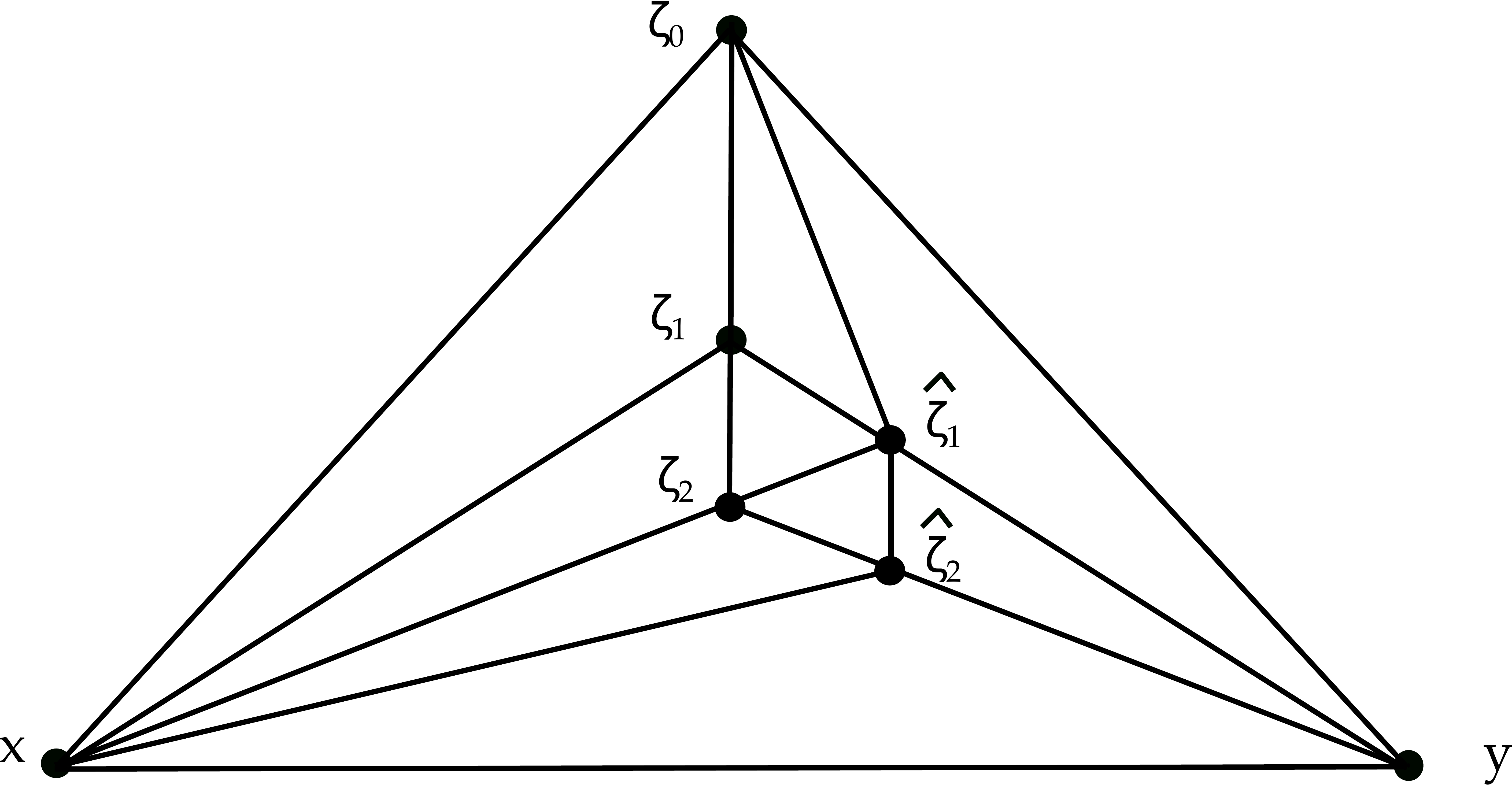}
\caption{Top cones for resolutions leading to (9, III) (applying all four), (11, III) (first three) and (11, IV) (first two). These are the cone diagrams 
corresponding to the subdivision of the fiber faces in (\ref{FibFac789}).   
\label{fig:TopCon789}}
\end{figure}

\subsection{Flops and Codimension 3 Fibers}

The box graphs have a simple realization of flops as single box sign changes. The algebraic resolutions that we constructed based on the fiber faces allow us to equally simply spot the flops. The flops among the fiber faces with fine triangulations is completely standard and explained around figure \ref{fig:toricflop}. The more interesting cases are the partially triangulated fiber face, where we have already seen the partial resolutions take one of the two forms (\ref{yyhatP}) or (\ref{xWS}), which also make the flop transitions manifest. 

Finally, we can confirm by direct constructions the monodromy-reduce $\mathfrak{e}_6$ fibers obtained from the box graphs in \cite{Hayashi:2014kca}. Note that the monodromy-reduction arises due to the absence of an extra section, we refer the reader for details to \cite{Hayashi:2014kca}.
For each of the resolutions we can compute the splitting of the $b_1=0$ locus as we pass to the codimension 3 locus $b_1=b_2=0$ along the discriminant component $z=0$. The fibers splittings and intersections follow by straight forward intersection computations, and are shown alongside the resolution sequences in table \ref{tab:THETABLE}. As already argued in \cite{Hayashi:2014kca} the possible monodromy-reduced $\mathfrak{e}_6$ fibers are obtained by deleting nodes of the Kodaira fiber $IV^*$, yielding non-Kodaira fibers in codimension 3, which have multiplicities as shown within the nodes in the table. 


\section{Determinantal Blowups}
\label{sec:DetBlowup}

In this section, we show how to obtain the resolution associated to the box graph $(11,{\rm IV})$ (and thus by reversal of the ordering of simple roots, the box graph (6, I)) by a series of successive blowups. After a sequence
of weighted algebraic blowups, we reach  a singular space which sits in between phase $(11,{\rm IV})$ and
$(11,{\rm III})$.  Whereas a further standard algebraic blowup realizes $(11,{\rm III})$, we need to blow up
the ambient space along a determinantal ideal to reach phase $(11,{\rm IV})$. This means that the Tate model 
corresponding to this phace is not a complete intersection.
The box graphs $(11,{\rm III})$ and  $(11,{\rm IV})$ are connected by a flop along a {\bf 5} curve, $C_2$, 
which is above the codimension two locus $\zeta_0=0 $ and 
\be
P_5 = b_1^2 b_6 + b_2 b_3^3 -b_1 b_3 b_4 = 0 \,,
\ee
which is less manifest in the Tate formulation, and thus makes the construction of this phase more challenging. 

\subsection{Setup and determination of singular locus}

First we introduce the subdivisions corresponding to a weighted blowup introducing $\hat{\zeta}_1$ and $\zeta_1$ introduced already in the last section
\be
([x,1],[y,2],[\zeta_0,1]; [\hat{\zeta_1},1])\, , \quad \quad ([x,1],[\hat{\zeta}_1,1],[\zeta_0,1]; [\zeta_1,2]) \,.
\ee
After these two steps, the partially resolved Tate model takes the factored form
\be
x W = \hat{\zeta}_1 S \, ,
\ee
where of course $\hat{\zeta}_2$ and $\zeta_2$ have to be set to unity in the expressions for $W$ and $S$ in \eqref{eq:defsPWSyhat}.
After these two blowups, we have induced the triangulation \includegraphics[width=1cm]{SU5AFT6.pdf} on the fiber face, i.e. the second step in the sequence (\ref{FibFac789}). 

Instead of continuing with a refinement of the cones of the corresponding fan as in (\ref{FibFac789}), we now promote $W$ to a new coordinate 
$\omega$ and do the crepant blowup 
\be
([\omega,1],[\hat{\zeta}_1,1] ;[\delta,1])\, .
\ee
After this blowup, the geometry is a complete intersection
\be
\ba
x \omega & = \hat{\zeta}_1 \left( -b_6 \hat{\zeta} _1 \zeta _1^3  \zeta _0^5 \delta +b_3 \zeta _1 \zeta
   _0^2 y+ y^2 \right)  \\
\delta \omega & = \zeta _1  \left(b_4 \hat{\zeta}_1 \zeta _1 \zeta _0^3 \delta +b_2 \zeta _0
   x+  x^2\right)-b_1 y \label{inbetween8and7}
\ea
\ee
This is not a completely resolve space yet as there are singularities remaining. 
Let us see this explicitely in order to guide
us to the resolution realzing phase $(11,{\rm IV})$. To do so, we go to a chart $\C^4$ (for the fiber coordinates) spanned by $x,y,\delta$ 
and $\omega$ (or, equivalently, $\hat{\zeta}_1$ and compute the Jacobian matrix. As we already anticipate that we will find the singularity 
over $x=y=\delta=0$ we evaluate it there. 
The two homogenous coordinates $\zeta_0$ and $\zeta_1$ cannot vanish simultaneously with $x$ and $y$, we can set them to unity. 
The Jacobian matrix then gives
\begin{equation}\label{singmatrix}
  {\rm rk} \left(\begin{array}{cc}
 \omega & b_2 \\
b_3 \hat{\zeta}_1& -b_1 \\
 - b_6 \hat{\zeta}_1^2 &  b_4 \hat{\zeta}_1 +\omega \\
 0 & 0 
 \end{array}\right) = 1 \,,
\end{equation}
as a condition for a singularity over $x=y=\delta=0$. We can rewrite this condition as
\be
\ba
\label{minorsjac8to7}
 b_1 w + b_2 b_3 \hat{\zeta}_1 = & 0\nn \\
\hat{\zeta_1} \left(  \omega b_3  + b_3 b_4 \hat{\zeta}_1 - b_1 b_6 \hat{\zeta}_1 \right)= & 0\nn \\
 \omega^2 + \omega \hat{\zeta}_1 b_4 +b_2 b_6 \hat{\zeta}_1^2 = & 0 \,.
\ea
\ee
As $\omega$ and $\hat{\zeta}_1$ cannot vanish simultaneously, these equations can 
only have a common solution if $\hat{\zeta}_1= 0$ and $b_1=0$ or
all the resultants vanish, which implies that
\begin{equation}
b_1^2 b_6 + b_2 b_3^3 -b_1 b_3 b_4 = 0 \,,
\end{equation}
thus confirming that the curve is indeed a ${\bf 5}$ curve. 
Besides this relation, $[\omega:\hat{\zeta}_1]$ are fixed by the above conditions, 
so that we find a singularity at codimension $5-2=3$ after the blow down. We have hence learned that while \eqref{inbetween8and7}
is smooth in codimension two (codimension  one over the base), it is still singular in codimesion three (two in the base). There
are two singular strata located along the ${\bf 10}$ and ${\bf 5}$ matter curves.

The singularities over $x=y=\delta=0$ may be easily resolved by performing the blowup $([x,1],[y,1],[\delta,1]; [\hat{\zeta_2},1])$. 
This will change the anticanonical class of the ambient space by $2D_{\hat{\zeta} _2}$, so that we obtain a crepant resolution after computing
the proper transform of \eqref{inbetween8and7}. It is not hard to see that this again realizes the box graph resolution $(11,{\rm III})$. To find 
$(11,{\rm IV})$, we hence need to use a different resolution.
The resolution we are looking for must be similar to a (partially) flopped version of the resolution at $x=y=\delta=0$. 

\subsection{Resolution and Determinantal variety}

We now construct the resolution that realizes the box graph (11, IV). 
To find the alternate locus to resolving $x=y=\delta=0$, we can resolve along, note that we can rewrite \eqref{inbetween8and7} as
\begin{equation}\label{inbetween8and7asmatrix}
\left(\begin{array}{ccc}
      -\omega & -(y + b_3 \zeta_1 \zeta_0^2)\hat{\zeta} _1  & b_6 \zeta_1^3 \zeta_0^5 \hat{\zeta} _1^2 \\
       -\zeta_1(x + b_2 \zeta_0) & b_1 & -(\omega + b_4 \zeta_1^2 \zeta_0^3 \hat{\zeta} _1 )
      \end{array}\right)
\left(\begin{array}{c}
       x\\ y \\ \delta
      \end{array}
\right) \, \equiv R \left(\begin{array}{c}
      x\\ y\\ \delta
      \end{array}\right)\,  = \, 0 \, .
\end{equation}
Observe that e.g. the divisor $y=0$ is not irreducible. It splits into $x = \delta =0 $ 
or as the simultanous solution of \eqref{inbetween8and7asmatrix} with 
\begin{equation}
\omega(\omega + b_4 \zeta_1^2 \zeta_0^3 \hat{\zeta} _1) + \zeta_1(x + b_2 \zeta_0)b_6 \zeta_1^3 \zeta_0^5\hat{\zeta} _1^2   = 0 \, .
\end{equation}
Similarly, $\delta=0$ splits in two components and $x=0$ into three components.

Let us define new coordinates $\rho_x$, $\rho_y$ and $\rho_\delta$ as the determinants of the $2\times 2$ matrices which are obtained 
from $R$ by deleting the columns corresponding to $x, y, \delta$ (with an extra sign for $\rho_y$), i.e.
\be
\ba
 \rho_x &= (\omega + b_4 \zeta_1^2 \zeta_0^3 \hat{\zeta} _1)(y + b_3 \zeta_1 \zeta_0^2)\hat{\zeta}_1 - b_1 b_6 \zeta_1^3 \zeta_0^5\hat{\zeta} _1^2 \cr
 \rho_y &= - \omega(\omega + b_4 \zeta_1^2 \zeta_0^3 \hat{\zeta} _1) - \zeta_1(x + b_2 \zeta_0)b_6 \zeta_1^3 \zeta_0^5\hat{\zeta} _1^2  \cr
 \rho_\delta &= -b_1\omega - \zeta_1(x + b_2 \zeta_0)(y + b_3 \zeta_1 \zeta_0^2)\hat{\zeta} _1\,.
\ea
\ee
In this way of presentation, the extra component of the divisor $x=0$ occurs because $\rho_x$ has $\hat{\zeta}_1$ as a factor.
In these coordinates, the singularities are at $x=y=\delta=0$ and $\rho_x=\rho_y=\rho_\delta=0$, which can equivalently be described as the intersection of $x=y=\delta=0$ with the determinantal variety defined by ${\rm rk}\, R=1$. As observed already, this fixes $[\omega:\hat{\zeta}_1]$ and forces $P=0$ or $b_1=0$ in the base, so that we are on top of either the ${\bf 5}$ or the ${\bf 10}$ matter curve.
We hence have a situation which is analogous to the conifold: there are Cartier divisors which split into several irreducible components (Weil divisors)
and we can resolve the singularity if we blow up along either one of them. 

The singular geometry (\ref{inbetween8and7asmatrix}) can be resolved in two ways, which are related by a flop transition 
\begin{itemize}
\item (11, III): \qquad  Blowup at $x=y=\delta=0$
\item  (11, IV): \qquad Blowup at complete intersection of \eqref{inbetween8and7asmatrix} with $x = 0$ and $\rho_x=0$. 
\end{itemize}
In order to implement the blowup along $x=\rho_x=0$, we start with the following observations: by construction the coordinates $\rho$ satisfy
\begin{equation}
 R \left(\begin{array}{c}
       \rho_x \\ \rho_y \\ \rho_\delta
      \end{array}\right)\,  = \, 0 \, .
\end{equation}
Furthemore, the ideal generated by \eqref{inbetween8and7asmatrix} contains the polynomials:
\be
\ba
\label{zvsrho}
 x \rho_y &- y \rho_x \cr
 y \rho_\delta &- \delta \rho_y \cr
 \delta \rho_x &- x \rho_\delta \, .
\ea
\ee
The meaning of this is not difficult to see: both the vectors $(x,y,\delta)$ and $(\rho_x,\rho_y,\rho_\delta)$
are orthogonal to both row vectors of $R$. Hence they must be parallel, as expressed in \eqref{zvsrho} above.
As we have discussed above, there is a singularity when both vectors vanish.
To perform the resolution, we hence introduce an auxiliary $\P^1$ with coordinates $[\xi_1:\xi_2]$ subject to the relations
\be
\ba\label{resolutionto7}
 x \xi_1 &= \rho_x \xi_2 \nn\\
 y \xi_1 &= \rho_y \xi_2 \nn\\ 
 \delta \xi_1 &= \rho_\delta \xi_2 \, .
\ea
\ee
so that $[\xi_1:\xi_2]$ measures the proportionality constant of the two parallel vectors $(x,y,\delta)$ and $(\rho_x,\rho_y,\rho_\delta)$. 

We now show that \eqref{resolutionto7} and \eqref{inbetween8and7asmatrix} indeed describe a crepant resolution of the singularity at $x=y=\delta=\rho_x=\rho_y=\rho_\delta=0$.  To see this, first note that the relations \eqref{resolutionto7} uniquely specify a point on the auxiliary $\P^1$ except when we are at the locus of the former singularity. Hence we have pasted in a $\P^1$ at the codimension three locus $x=y=\delta=\rho_x=\rho_y=\rho_\delta=0$. This means that we not only have a resolution, but that it 
is also small (i.e. there is no exceptional divisor), from which it follows that we have a crepant resolution as well.

The weight system of the ambient space is now
\be\label{Scaling87}
  \begin{array}{ccccccccc}
  x & y &\zeta_0& \zeta_1 & \hat{\zeta} _1 & \delta & \omega & \xi_1 & \xi_2 \cr 
    \hline
 1&1&1&-1&0&0&1&1&0 \cr
 0&1&0&1&-1&0&1&1&0  \cr
 0&0&0&0&1&-1&1&2&0  \cr
 0&0&0&0&0& 0&0&1&1
 \end{array}
\ee

The fiber components of the resolved phase just obtained can be written as intersections of \eqref{resolutionto7} and \eqref{inbetween8and7asmatrix} 
with
\be
\ba
F_0:\qquad & \zeta_0 = 0\cr
F_1:\qquad &\delta  =  \rho_\delta = 0\cr
F_2:\qquad &  \delta  = \xi_2 = 0 \cr
F_3:\qquad & \hat{\zeta}_1 =  x = 0 \cr
F_4:\qquad  & \zeta_1= 0 \,.
\ea
\ee

Writing its defining relations out explicitely, we find that $F_1$ is given by the (non-independent)
equations
\be
\ba\label{eq:F4phase7}
\delta & = \rho_\delta = 0 \\
0& = \zeta_1(x + b_2 \zeta_0)(y + b_3 \zeta_1 \zeta_0^2)\hat{\zeta} _1 - b_1\omega\ \\
x \xi_1 &= \left((\omega + b_4 \zeta_1^2 \zeta_0^3 \hat{\zeta} _1)(y + b_3 \zeta_1 \zeta_0^2)\hat{\zeta}_1 - b_1 b_6 \zeta_1^3 \zeta_0^5\hat{\zeta} _1^2\right) \xi_2 \\
y \xi_1 &= - \left(\omega(\omega + b_4 \zeta_1^2 \zeta_0^3 \hat{\zeta} _1) + \zeta_1(x + b_2 \zeta_0)b_6 \zeta_1^3 \zeta_0^5\hat{\zeta} _1^2 \right)\xi_2 \\ 
x\omega &= - y \hat{\zeta}_1 (y+b_3 \zeta_1 \zeta_0^2) \\
y b_1 & = x \zeta_1(x+b_2 \zeta_0) 
\ea
\ee

Our first task is to show that the splitting of the fiber components over the ${\bf 10}$ curve is as expected for
phase $(11,{\rm IV})$. For this, it is sufficient to consider the component $F_1$, which is expected to split into four components.
Over $b_1=0$, \eqref{eq:F4phase7} splits into the four irreducible components
\be
\ba
\zeta_1   &= 0 : \qquad \left\{ 
 \ba
        x \xi_1 &= \xi_2 \hat{\zeta}_1 y \omega =0 \cr
	 y \xi_1 &= -\omega^2 \xi_2 \\
	x\omega &= -y^2 \hat{\zeta}_1 
	\ea
	\right.
\cr	
\hat{\zeta}_1&= 0 :\qquad  
\left\{ 
\ba  x  &= 0\cr
 y\xi_1 & = - \omega^2 \xi_2\cr
\ea
\right.
\cr
x+ b_2 \zeta_0  & = 0 :\qquad
\left\{ 
\ba 
 x \xi_1  & = (\omega + b_4 \zeta_1^2 \zeta_0^3 \hat{\zeta} _1)(y + b_3 \zeta_1 \zeta_0^2)\hat{\zeta}_1  \xi_2 \\
		    y \xi_1 &  = - \omega(\omega + b_4 \zeta_1^2 \zeta_0^3 \hat{\zeta} _1)\xi_2 \\
		 x\omega & = -y \hat{\zeta}_1(y+b_3 \zeta_1 \zeta_0^2) 
\ea 
\right. \cr
 y+b_3 \zeta_1 \zeta_0^2 & = 0 
  :\qquad
\left\{ 
\ba 
  x & = 0 \\
 y \xi_1 & = - \left(\omega(\omega + b_4 \zeta_1^2 \zeta_0^3 \hat{\zeta} _1) + b_2 b_6 \zeta_1^4 \zeta_0^6 \hat{\zeta} _1^2  \right) \xi_2 \,.
 \ea
 \right.
\ea
\ee
where $\delta=0$ is understood for all of them. Note that in some cases, not all equations defining the ideal corresponding to the
fiber component are independent. We recognize these as $F_4$, $F_3$ restricted to $b_1=0$, as well as the two curves $C_{1,5}^+$ and $C_{2,3}^-$,
which is the splitting we expect over the ${\bf 10}$ matter curve from the box graphs discussed in Section \ref{sect:singfib10}.

Finally, let us see that we have obtained the expected splitting over the ${\bf 5 }$ matter curve.
Over $P=0$ in the base, there exist $x,y,\omega$ simultaneously solving $\rho_\delta=\rho_y=\rho_x=0$,
as well as $R (x,y, 0 ) = 0 $. On the other hand we can also simultaneously
solve $x=y=\delta = \rho_\delta=\rho_y=\rho_x=0$ using only $\omega$ when $P=0$ as noticed before. 
Correspondingly, $F_1$ splits into two irreducible components defined by the intersection of
\eqref{eq:F4phase7} with
\begin{eqnarray}
C_1^+: & \xi_1 = 0 \nn\\
C_2^- :& x = y = 0 \, .
\end{eqnarray}
From these splittings it follows that we have indeed realized a global three/fourfold description of phase $(11,{\rm IV})$. Although we have
no proof that this phase cannot be realized as a complete intersection, we find it amusing that we have to venture out of well-charted
territory to realize this `outlying' phase. 

Finally, we can confirm the splitting along codimension three, which yields one of the $\mathfrak{e}_6$ monodromy reduced fibers. Setting $b_2=0$ in addition, we observe the splitting
\be
C_{1,5} \quad \rightarrow \quad F_2 + C_2^- + F_3 + C_{2,3}^- \,.
\ee
Again, using the projective relations, the intersections are readily obtained to be as shown in table \ref{tab:THETABLE}.



\section*{Acknowledgments}

We wish to thank Andres Collinucci, Hirotaka Hayashi, Craig Lawrie and Roberto Valandro for discussions and Taizan Watari for sparking the idea for
the cone diagrams used in this work. This work is supported in part by the STFC grant ST/J002798/1. SSN thanks BA 
for an upgrade from LHR to NRT while completing this paper. 


\appendix


\section{Toric Primer}
\label{sect:toric101}

In this section we describe some basics of toric geometry which are needed for our discussion in this paper. We focus 
on the construction of toric varieties using fans, see \cite{danilov, Fultontoric, cox2011toric, Kreuzer:2006ax, Bouchard:2007ik} 
for a more thorough treatment.

The starting point for our discussion is a lattice $N$ which we conventionally choose as $\Z^n$. A rational strongly convex polyhedral cone $\sigma$
in $N\otimes \R$ is a cone generated by a finite number of primitive\footnote{In this context, primitive means that it is the closest lattice point
to the origin in its direction.} lattice points which does not contain any linear subspace of $N\otimes \R$ (except the point). A collection of such cones
forms a fan $\Sigma$ if it contains the face of each cone and the intersection between any two cones is a face of each. 

From a given fan $\Sigma$, a toric variety can be constructed in several (equivalent) ways. For our purposes, the most convenient construction is
the one using homogeneous coordinates. A $n$-dimensional toric variety $T_\Sigma$ can be described as
\begin{equation}
 T = \left( \C^{n+k} \setminus Z \right) / \left( (\C^*)^k \times G \right) \, ,
\end{equation}
for a subset $Z$ of $\C^{n+k}$ and a finite group $G$. The action of each $C^*$ on $\C^{n+k}$ can be described by a system of weights:
\begin{equation}
 \left( z_1, z_2, \cdots, z_n \right) \mapsto  \left( z_1 \lambda^{s_1},  z_2 \lambda^{s_2} \cdots,  z_n \lambda^{s_n} \right) \, .
\end{equation}
The data used above can be recovered from the fan as follows. Every one-dimensional cone $\rho_i$ is generated by a primitive lattice 
vector $v_i$ (as we have to frequently refer to these lattice vectors, we will simply call them generators in the following).
Assigning a coordinate $z_i$ to each one-dimensional cone, there is a corresponding $\C^*$ action with weights 
$s_i$ for every linear equation of the form
\begin{equation}
 \sum s_i v_i = 0 \, .
\end{equation}
This means that a fan with $n+k$ one-dimensional cones sitting in a lattice $N$ of (real) dimension $n$ describes a toric variety of complex
dimension $n+k - (n+k-n)=n$. We frequently display the weights of homogeneous coordinates in the form
\be
\begin{array}{cccc}
 z_1 & z_2 & \cdots & z_n \\
 \hline
 s_1 & s_2 & \cdots & s_N
\end{array}\, .
\ee

The exceptional set $Z$, which is equivalent to the Stanley-Reisner (SR) ideal in the ring of homogenous coordinates, 
is defined such that a collection of coordinates $z_I$ can only vanish simultaneously if the corresponding cones share a common cone in $\Sigma$. We write such relations as $[z_a,z_b,\cdots]$ for $a,b \in I$. We will not be interested in cases with a non-trivial group $G$, so we omit its description from the discussion, see e.g. \cite{danilov, Fultontoric, cox2011toric, Kreuzer:2006ax, Bouchard:2007ik} for a nice exposition.

A toric variety $T_\Sigma$ only has orbifold singularities if all cones are simplicial\footnote{A $p$-dimensional cone is 
simplicial if it can be generated by $p$ vectors.}. A simplicial $p$-dimensional cone $\sigma$ with generators $v_1, \cdots , v_p$ leads to
a singularity at codimension $p$ in $T_\Sigma$ if its generators fail to span the restriction of the lattice $N$ to their supporting hyperplane.
The singularity is then located at the locus $z_1= \cdots = z_p = 0$. $T_\Sigma$ is hence smooth if all cones are simplicial and the generators 
of all of the $n$-dimensional cones span $N$. 

A fan gives rise to a compact toric variety $T_\Sigma$ if the union of all cones spans $N\otimes \R$. 

The vanishing loci of the homogeneous coordinates $z_i$ define toric (Weil) divisors $D_i$. As these Divisors
can only vanish simultaneously if they are in a common cone, we may think of higher-dimensional cones as corresponding
to algebraic subvarieties of higher codimension. 

The toric divisors obey the linear relations
\begin{equation}
 \sum_i \langle v_i,m\rangle D_i = 0 
\end{equation}
for every $m$ in the dual lattice (usually called the $M$-lattice). This means that the class of any
divisor $D$ corresponding to the vanishing locus of a polynomial is specified by the weights $s_i$ of $P$ 
under the $\C^*$ actions. 

To compute intersection numbers between divisors, we can first use the SR ideal to see if the intersection can be non-vanishing. 
Non-zero intersection numbers can be computed by using that, for a collection of $n$ different $v_i, i\in I$ spanning 
an $n$-dimensional cone $\sigma$ in $\Sigma$
\begin{equation}
 \prod_I D_i = 1/{\rm Vol}(\sigma)
\end{equation}
Here, ${\rm Vol}(\sigma)$ is the lattice volume of the cone $\sigma$, which is given by the determinant of its generators.

A Weil divisor $D = c_i D_i$ is also Cartier if there is a piecewise linear (on each cone) support function $\psi_D$ satisfying
\begin{equation}
\psi_D|_{\sigma} =\, <m_{\sigma}, v_i >\, = - c_i  \, ,
\end{equation}
for each cone $\sigma$ generated by $\{ v_i, i \in I\}$. If all cones of $\Sigma$ are simplicial, all toric divisors are 
also $\mathbb{Q}$-Cartier, so that such a support function exists. The cone of ample curves (or, equivalently, the K\"ahler cone) 
contains the set of divisors for which $\psi_D$ is strongly convex. This means that $\psi_D|_{\sigma} > -c_j$ 
for all one-dimensional cones \emph{not} in $\sigma$. If the open cone of ample curves is non-empty, we can find a 
line bundle which is very ample and hence defines an embedding of $T_\Sigma$ into projective space.


\section{Details of weighted blowups}
\label{app:Blowups}

\subsection*{Box Graph (4, III) or (13, II)}

After the weighted blowups,with proper transform for 
$([x, n_x], [y, n_y], [z, n_z]; [\zeta, n_{\zeta}])$ given by $\zeta^{n_\zeta-(n_x + n_y + n_z)}$, and each term scaling as 
$x \rightarrow x \zeta^{n_x/n_\zeta}$ etc,
the equation is exactly (\ref{eq:resolved}), as the corresponding fiber face has a fine triangulation.
The projective relations are
\be
\ba
{[}\zeta _1 \zeta _2 x,\hat{\zeta }_1 y,\zeta _0 \zeta _1
   \zeta _2 \hat{\zeta }_1] &\cr
   [y,\zeta _0 \zeta _1 \zeta_2,\zeta _1 \zeta _2 \hat{\zeta }_2]&\cr
   [x,\zeta _0\zeta _1,\hat{\zeta }_2]& \cr
   [\zeta _0,\zeta_2]
\ea
\ee
Using the projective relations we can determine the splittings of the Cartan divisors $\zeta_i=0$ and $\hat\zeta_i=0$. 
Along $b_1=0$, which is the ${\bf 10}$ locus, the curves dual to the divisors $\zeta_0=0$ and $\hat\zeta_2=0$ split into two three components, respectively. Computation of the intersections with {\tt Smooth} \cite{Smooth} yields that this is precisely the splitting for the box graph 4 respectively 13  of the {\bf 10} matter. Likewise along $P= b_1^2 b_6 - b_1 b_3 b_4 + b_2 b_3^2$ again $\hat\zeta_2=0$ splits into two components, consistently with the box graph III or II respectively. 

\subsection*{Box Graph (9, III) or (8, II)}

After the weighted blowups we again obtain the equation (\ref{eq:resolved}), as the corresponding fiber face has a fine triangulation.
The projective relations are
\be
\ba
{[}\zeta _1 \zeta _2 \hat{\zeta }_2 x,\hat{\zeta }_2
   y,\zeta _0 \zeta _1]&\cr
   [\zeta _2 \hat{\zeta }_2
   x,\zeta _0,\zeta _2 \hat{\zeta }_1 \hat{\zeta
   }_2]&\cr 
   [\zeta _2 x,y,\zeta _2 \hat{\zeta
   }_1]&\cr 
   [x,\hat{\zeta }_1]
   \ea
\ee
Along $b_1=0$, which is the ${\bf 10}$ locus, the curves dual to the divisors  $\zeta_2=0$ and $\hat\zeta_1=0$ split into two and three components respectively. 
The precise charges from  {\tt Smooth} \cite{Smooth} yields that this is precisely the splitting for the box graph 9 and 8, respectively. Along 
$P= b_1^2 b_6 - b_1 b_3 b_4 + b_2 b_3^2$ again $\hat\zeta_2=0$ splits into two components, consistently with the box graph III or II respectively. 
Thus showing that these are (9, III) and (8, II), depending on which ordering of the simple roots we choose. 

\subsection*{Box Graph (11, III) or (6, II)}

Finally, we get to a resolution which corresponds to a partially triangulated fiber face. 
The weighted blowups give rise to an equation of the form 
\be
x W = \hat\zeta_1 S 
\ee
with $W$ and $S$ as in (\ref{eq:defsPWSyhat}). 
Instead of continuing with $(x, \hat\zeta_1; \zeta_2)$, which would yield the previous box graph resolution (9, III), we instead take the  resolution
\be
(W, \hat\zeta_1; \delta) \,.
\ee
The equation then takes the form
\be
\ba
x \omega &=  \hat\zeta_1 (\hat\zeta_2 y^2 + b_3 \zeta_1 \zeta_0^2 y - b_6\delta \hat\zeta_1 \zeta_1^3  \zeta_0^5) \cr
\delta\omega &= \zeta_1 (b_4 \hat\zeta_1 \delta \zeta_1 \zeta_0^3 + b_2 \zeta_0 x + \zeta_2 x^2 ) - b_1 y \,.
\ea
\ee
The projective relations are then
\be
\ba
{[}\zeta _1  \hat{\zeta }_2 x,\hat{\zeta }_2
   y,\zeta _0 \zeta _1]&\cr
   [ \hat{\zeta }_2
   x,\zeta _0, \hat{\zeta }_1 \delta \hat{\zeta
   }_2]&\cr 
   [ x,y,\hat{\zeta}_1\delta]&\cr 
  [\omega, \hat{\zeta}_1]
   \ea
\ee
As this resolution has not appeared anywhere so far in the literature, we will provide some more details. 
The curves associated to the simple roots are (we choose one of the orderings here, corresponding to (6, II), however trivially, the reverse ordering will give rise to the other resolution (11, III))
\be
\ba
F_0 : &\qquad \zeta_0=0\qquad  \left\{ \ba
	x \omega  &= \hat\zeta_1 \hat\zeta_2 y^2  \cr
	\delta \omega& = - b_1 y + x^2 \hat\zeta_2 \zeta_1 
	\ea  \right. \cr 
F_1 : &\qquad 
 	\zeta_1=0\qquad 
	\left\{ \ba 
	x \omega &= \hat\zeta_1 \hat\zeta_2 y^2 \cr 
	\delta \omega& =- b_1 y 
	\ea  \right. \cr 
F_2 : &\qquad 
	\hat\zeta_1 = 0\qquad 
	\left\{ \ba 
	x&=0 \cr
	\delta \omega& = - b_1 y 
	\ea  \right. \cr 
F_3 : &\qquad 
	\hat\zeta_2=0\qquad 
	\left\{ \ba 
	x \omega&= \hat\zeta_1 ( b_3 \zeta_1 \zeta_0^2 y - b_6\delta \hat\zeta_1 \zeta_1^3  \zeta_0^5)  \cr
	\delta \omega & = \zeta_1 (b_4 \hat\zeta_1 \delta \zeta_1 \zeta_0^3 + b_2 \zeta_0 x ) - b_1 y 
	\ea  \right. \cr 
F_4 : &\qquad 
	\delta=0 \qquad 
	\left\{ \ba 
	x \omega & = y \hat\zeta_1 (\hat\zeta_2 y + b_3 \zeta_1 \zeta_0^2) \cr
	0 &= \zeta_1 x (b_2 \zeta_0 + x \hat\zeta_2) - b_1 y 
	\ea  \right. 
\ea
\ee
Along $b_1=0$, only $F_4$ splits, using the projective relations, 
\be
\delta=b_1=0 \qquad \left\{ 
\ba
\zeta_1= x \omega - y^2 \hat\zeta_2 \hat\zeta_2&=0 \cr
x= \hat\zeta_1&=0 \cr
x = \zeta_2 y + b_3  \zeta_1 \zeta_0^2 &=0  \cr
(b_2 \zeta_0 + x \hat\zeta_2) = x \omega - y \hat\zeta_1 (\hat\zeta_2 y + b_3 \zeta_1 \zeta_0^2) &=0 
\ea
\right. 
\ee
which precisely correspond to the splitting 
\be
F_4 \quad \rightarrow F_1 + F_2 + C_{3,4}^++ C_{1,5}^-  \,.
\ee
Along the ${\bf 5}$ locus  $P=0$ it is $F_3$ that splits into 
two components. Using the projective relations the fiber intersections are $I_1^*$ and $I_6$ respectively. 
And thus confirming that these realize the box graphs  (11, III) or (6, II). 

Finally we can also determine the splitting and fiber along the codimension 3 locus  $b_1=b_2=0$, where 
from the above the further splitting is of $C_{1,5}^-$ 
\be
\ba
b_1 =b_2 =\delta= x \hat\zeta_2 =   x \omega - y \hat\zeta_1 (\hat\zeta_2 y + b_3 \zeta_1 \zeta_0^2) =0
\ea
\ee
which has three components and $F_3$, which split as
\be
\ba
C_{1,5}^- \quad &\rightarrow \quad F_2 + C_{3,4}^+ + C_{3}^+ \cr 
F_3 \quad &\rightarrow \quad C_3^+ + C_4^- \,.
\ea
\ee
Again these splittings are consistent with the box graphs. Intersections of the fiber components using the projective relations, yield precisely the monodromy reduced $\mathfrak{e}_6$ fiber shown in table \ref{tab:THETABLE}. 



\providecommand{\href}[2]{#2}\begingroup\raggedright\endgroup


\end{document}